\documentclass[journal, twocolumn]{IEEEtran}
\usepackage[affil-it]{authblk}  
\usepackage[utf8]{inputenc}
\usepackage[a4paper, total={5.6in, 8.7in}]{geometry}
\usepackage{cite}
\usepackage{titlesec}
\usepackage{graphicx}
\usepackage[font=small,labelfont=bf]{caption}
\usepackage{amsthm}
\usepackage{amsmath}
\usepackage{enumerate}
\usepackage{subcaption}
\usepackage{fancyhdr}
\usepackage{titling}
\usepackage{amssymb}
\usepackage{comment}
\usepackage{xcolor}
\usepackage{url}
\usepackage{booktabs}
\usepackage{tabularx}

\newtheorem{theorem}{Theorem}[section]

\newtheorem{lemma}[theorem]{Lemma}
\newtheorem{definition}[theorem]{Definition}
\titleformat{\section}{\normalfont\large\bfseries}{\thesection}{1em}{\MakeUppercase}
\titleformat{\subsection}{\normalfont\small\bfseries}{\thesubsection}{1em}{\MakeUppercase}
\titleformat{\subsubsection}{\normalfont\small\bfseries}{\thesubsubsection}{1em}{\MakeUppercase}

\renewcommand\thesection{\arabic{section}}
\renewcommand\thesubsection{\thesection.\arabic{subsection}}
\renewcommand\thesubsubsection{\thesubsection.\arabic{subsubsection}}

\begin{document}


\title{From Microbes to Methane: AI-Based Predictive Modeling of Feed Additive Efficacy in Dairy Cows}

\makeatletter
\renewcommand*{\@fnsymbol}[1]{\the#1}
\makeatother

\author{%
    Yaniv~Altshuler\textsuperscript{1,3} \quad %
    Tzruya~Calv\~{a}o~Chebach\textsuperscript{2,3} \quad
    Shalom~Cohen\textsuperscript{3} \\
    \quad \IEEEmembership{yanival@mit.edu} \qquad \IEEEmembership{tzruyac@mail.tau.ac.il} \quad \qquad \IEEEmembership{dr.shalom@metha.ai}
    \thanks{Massachusetts Institute of Technology}%
    \thanks{Tel Aviv University}%
    \thanks{Metha Artificial Intelligence}%
}

\date{}

\maketitle

\begin{abstract}

In an era of increasing pressure to achieve sustainable agriculture, the optimization of livestock feed for enhancing yield and minimizing environmental impact is a paramount objective. This study presents a pioneering approach towards this goal, using rumen microbiome data to predict the efficacy of feed additives in dairy cattle.

We collected an extensive dataset that includes methane emissions from 2,190 Holstein cows distributed across 34 distinct sites. The cows were divided into control and experimental groups in a double-blind, unbiased manner, accounting for variables such as age, days in lactation, and average milk yield. The experimental groups were administered one of four leading commercial feed additives: Agolin, Kexxtone, Allimax, and Relyon. Methane emissions were measured individually both before the administration of additives and over a subsequent 12-week period. To develop our predictive model for additive efficacy, rumen microbiome samples were collected from 510 cows from the same herds prior to the study's onset. These samples underwent deep metagenomic shotgun sequencing, yielding an average of 15.7 million reads per sample. Utilizing innovative artificial intelligence techniques we successfully estimated the efficacy of these feed additives across different farms. The model's robustness was further confirmed through validation with independent cohorts, affirming its generalizability and reliability.

Our results underscore the transformative capability of using targeted feed additive strategies to both optimize dairy yield and milk composition, and to significantly reduce methane emissions. Specifically, our predictive model demonstrates a scenario where its application could guide the assignment of additives to farms where they are most effective. In doing so, we could achieve an average potential reduction of over 27\% in overall emissions.

\end{abstract}

\section{Introduction}

Ruminants, thanks to the intricate symbiotic relationship with their resident microbiota, have the unique ability to breakdown complex polysaccharides like cellulose and hemicellulose, which constitute the primary components of their plant-based diet. This process is facilitated by the host animal's provision of a stable environment, facilitating continuous mixing, deconstruction, and fermentation of ingested plant material. This, in turn, results in the production of short-chain fatty acids which serve as a digestible energy source for the host animal \cite{metha-intro1,metha-intro2}.

The assembly and development of the rumen microbiota is a multifactorial process, influenced by several host and environmental factors. These include the host's age, diet \cite{metha-intro4}, genetic makeup \cite{metha-intro5}, and herd origin \cite{metha-intro6}, all of which play a pivotal role in defining the microbiota's compositional layout \cite{metha-intro3}. Moreover, the stochastic colonization events of the rumen during early life stages can leave lasting imprints on the ruminant microbiome's structure \cite{metha-intro4}.

While this symbiotic relationship allows ruminants to thrive on fibrous diets, it also has an environmental cost. The ruminant digestive process is a significant contributor to the emission of methane, a potent greenhouse gas, which accounts for about 14\% of total greenhouse emissions and has a global warming potential 28 times higher than carbon dioxide ($CO_2$). Notably, livestock are estimated to contribute to nearly 30\% of all anthropogenic methane emissions \cite{metha-intro2}.

Efforts to mitigate the environmental impact of dairy farming have given rise to several innovative strategies. One such approach involves the utilization of microbial biomarkers to identify cows with high methane emission rates, thereby enabling targeted management strategies aimed at reducing methane emissions and fostering environmental sustainability \cite{metha-intro7}. A more refined strategy involves characterizing microbial gene abundances as proxies for methane emissions, focusing specifically on metabolic pathways expected to exhibit variation between low and high methane emitters \cite{metha-intro8}.

This study adopted an unbiased metagenomic approach to create a model that determines the most suitable feed additive customized for individual herds, allowing for precision application based on individual microbiome profiles. This methodology acknowledges the significant variation and multitude of contributing factors that lead to the diverse responses observed among ruminants. We capitalize on the rich biological information stored in the rumen microbiome, transforming the microbiome of a select few cows in each herd into a living 'sensor', and allowing us to predict the most effective feed additive tailored to that specific herd.

We devised a two-stage trial design targeting the prediction of the efficacy of methane-reducing additives using cows' microbiome data. The initial stage engages an unsupervised machine learning process, trained on a diverse dataset that includes microbiome samples from a wide spectrum of cows across various farms. The following stage makes use of a smaller subset of cows, whose methane emissions have been documented periodically, to implement supervised learning. This stage is aimed at constructing a predictive model that associates microbiome profiles with the effectiveness of feed additives.

By directly tapping into the microbiome, we bypass conventional variables such as weather and diet, which, though traditionally deemed critical, pose a challenge in establishing a clear, direct association with feed additive efficacy. Our approach presents an objective and comprehensive solution designed to effectively mitigate methane emissions in livestock farming.

The innovation of this approach lies not only in the use of the microbiome as a predictive tool, but also in our capacity to make sense of its complex raw data. The rumen microbiome, rich in diversity and complexity, presents a significant analytical challenge that until now has hindered its utility in such applications. To tackle this, we employed a data-driven approach powered by state-of-the-art artificial intelligence technology, creating a pioneering model that acknowledges the extensive variation and plethora of factors contributing to diverse responses observed among ruminants. By leveraging the power of the microbiome and artificial intelligence, our approach offers a promising avenue for future environmentally-conscious livestock management.

Our approach is noteworthy for its scalability, applicability beyond mere methane reduction, and potential for continuous improvement as more data accumulate. This manuscript elucidates our detailed trial design and deliberates upon its prospective influence on environmental sustainability, farm economy, and the expansive realm of precision agriculture.

\section{Materials and Methods}

\subsection{Microbiome Acquisition and Ruminal Fluid Sampling}

Rumen cannulation (RC) and stomach tubing (ST) stand as the two most prevalent techniques for the study of ruminal fermentation and microbial community composition in both large and small ruminants \cite{muizelaar2020rumen,nocek1997bovine}. While some researchers posited that samples acquired using ST may be less indicative of certain rumen parameters like pH, VFA concentrations, or bacterial communities compared to RC \cite{de2020comparison}, ST remains a valuable technique when exploring molar proportions of VFA, protozoa count, dH2, methane, and ammonia concentrations \cite{shen2012insertion,duffield2004comparison,wang2016sampling,geishauser1996comparison}.

For our study, we employed ST, specifically the passive ruminal fluid collection technique, for its distinct advantages. Cannulation, although precise, is both expensive and invasive. It typically necessitates a smaller cohort of animals due to its inherent complexity. In contrast, ST allowed us to collect ruminal fluid from a large cohort of intact animals. This not only induces less stress in the animals but also facilitates sampling in commercial dairies rather than solely experimental facilities.

Opting for stomach tubing as our primary technique for ruminal fluid extraction also aligns well with anticipated expansions to studies involving other ruminants. In research focused on small ruminants, a methodological comparison between surgical rumen cannulation and stomach tubing found the latter to be a more viable, safer, and pragmatic choice for sampling rumen contents in sheep and goats for the study of ruminal fermentation \cite{ramos2014use}. Notably, stomach tubing facilitates the collection of a diverse bacterial community and effectively mirrors most results garnered from cannula-based sampling.

The utilization of stomach tubing for sample collection is particularly conducive to the comprehensive, data-driven approach we adopted in our study. Recent research \cite{hagey2022rumen} illustrated that, as anticipated, stomach tubing garners more diverse and varied microbial samples than those derived from cannulated animals. This heightened diversity is invaluable for our analytics. A richer microbial profile provides a broader spectrum of data, enabling us to capture a more nuanced and holistic understanding of the complex interactions and processes occurring within the rumen. Crucially, despite this increase in diversity, stomach tubing samples remain highly representative of the fermentation processes and the methanogenic microbiota present in the rumen. This ensures our analyses are grounded in a more complete representation of the rumen environment, maximizing the reliability and precision of our results.

In our ST sampling method, we utilized a 300-cm long polyvinyl chloride orogastric tube (2.9 cm O.D. and 2.5 cm I.D.) with 4 holes perforated at its distal 30 cm. During each sampling event, the cow's head was restrained, and the tube, when attached to a 50 cm speculum, was gently inserted through the esophagus into the rumen. Once in place, the tube remained lower than the cow's head, allowing ruminal fluid to passively accumulate. We discarded the initial 10 ml of fluid to mitigate potential saliva contamination, after which we collected 50 ml of ruminal fluid in a sterile conical tube for further analyses. To prevent cross-contamination between samples, both the tube and the speculum were meticulously rinsed and bleached after each use.

\subsection{Sample Processing and Sequencing}

The rumen fluid samples collected were instantaneously frozen on-site with dry-ice and dispatched the very same day to a storage facility, where they were securely stored at a temperature of -80°C.

When ready for shipment, the frozen rumen fluid samples were brought to room temperature by defrosting them in ZYMO DNA/RNA Shield buffer (ZR-R1100-250). This process safeguards the samples from degradation during transport. They were then transported at room temperature to a specialized DNA service facility.

The DNA was extracted from the rumen fluid samples using the ZymoBIOMICS 96 MagBead DNA Kit (Zymo Research, Cat. no. D4308), in strict adherence to the provided manufacturer's protocol.

Subsequently, library preparation for sequencing was executed strictly adhering to the guidelines provided by the Illumina DNA Prep (Illumina, Cat. no. 20060059).
The assembled libraries were then subjected to paired-end sequencing with 2x150 base reads employing an Illumina NovaSeq 6000 instrument. For this purpose, the NovaSeq 6000 S4 Reagent Kit v1.5 (300 cycles) (Illumina, Cat. no. 20028312) was used.

Notably, all processes, inclusive of DNA extraction, library preparation, and sequencing, were conducted at ZYMO Research based in Freiburg, Germany.

Upon receiving the results of the sequencing process, the integrity of the raw FASTQ sequences was assessed by employing the FASTQC software for quality control, and subsequently, the data was fine-tuned by BBDUK with a set of customized parameters for optimal refinement.

\begin{figure*}[htbp]
    \centering
    \begin{subfigure}{0.3\textwidth}  
        \centering
        \includegraphics[width=\linewidth]{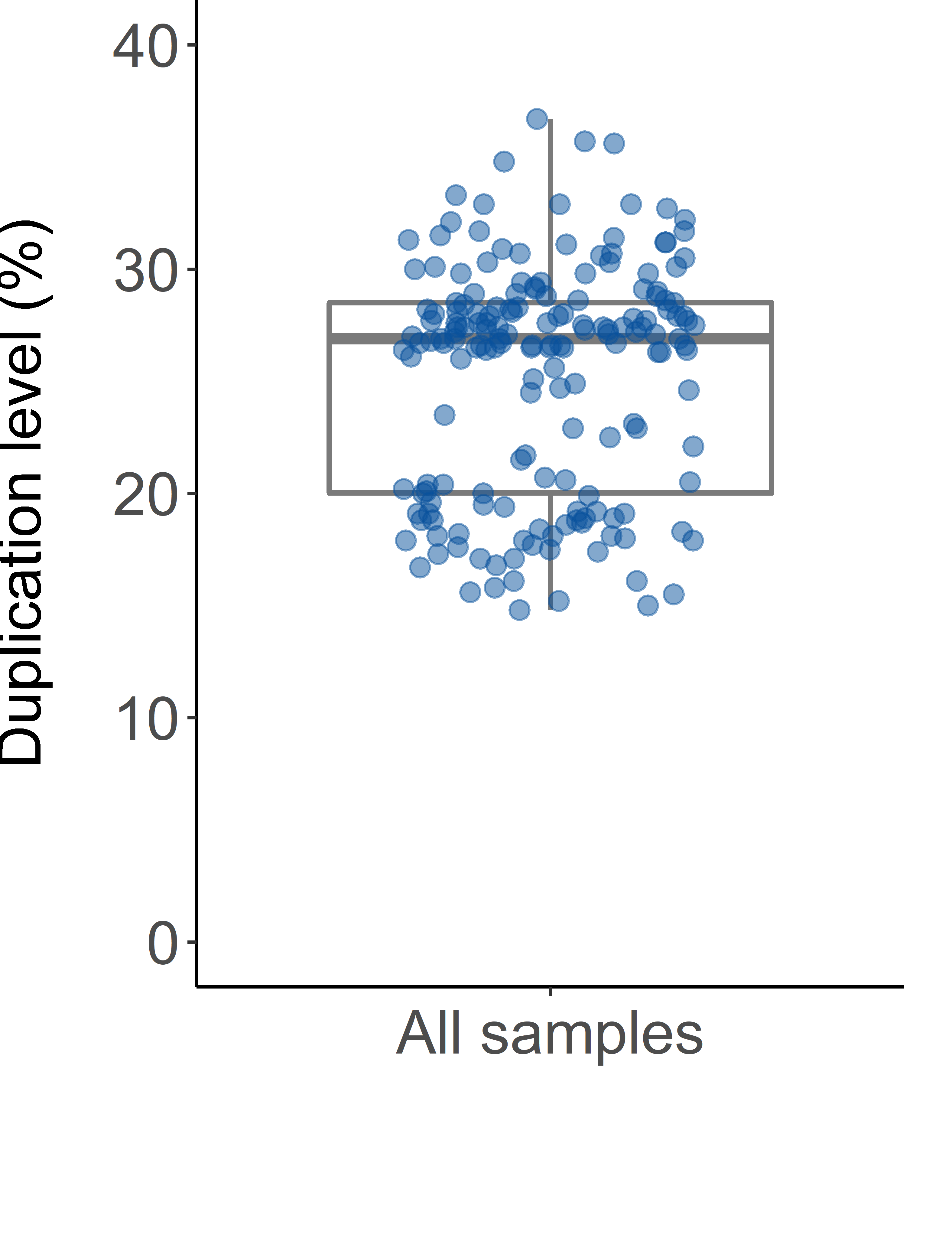}
    \end{subfigure}%
    \hspace{0.05\textwidth}  
    \begin{subfigure}{0.6\textwidth}  
        \centering
        \includegraphics[width=\linewidth]{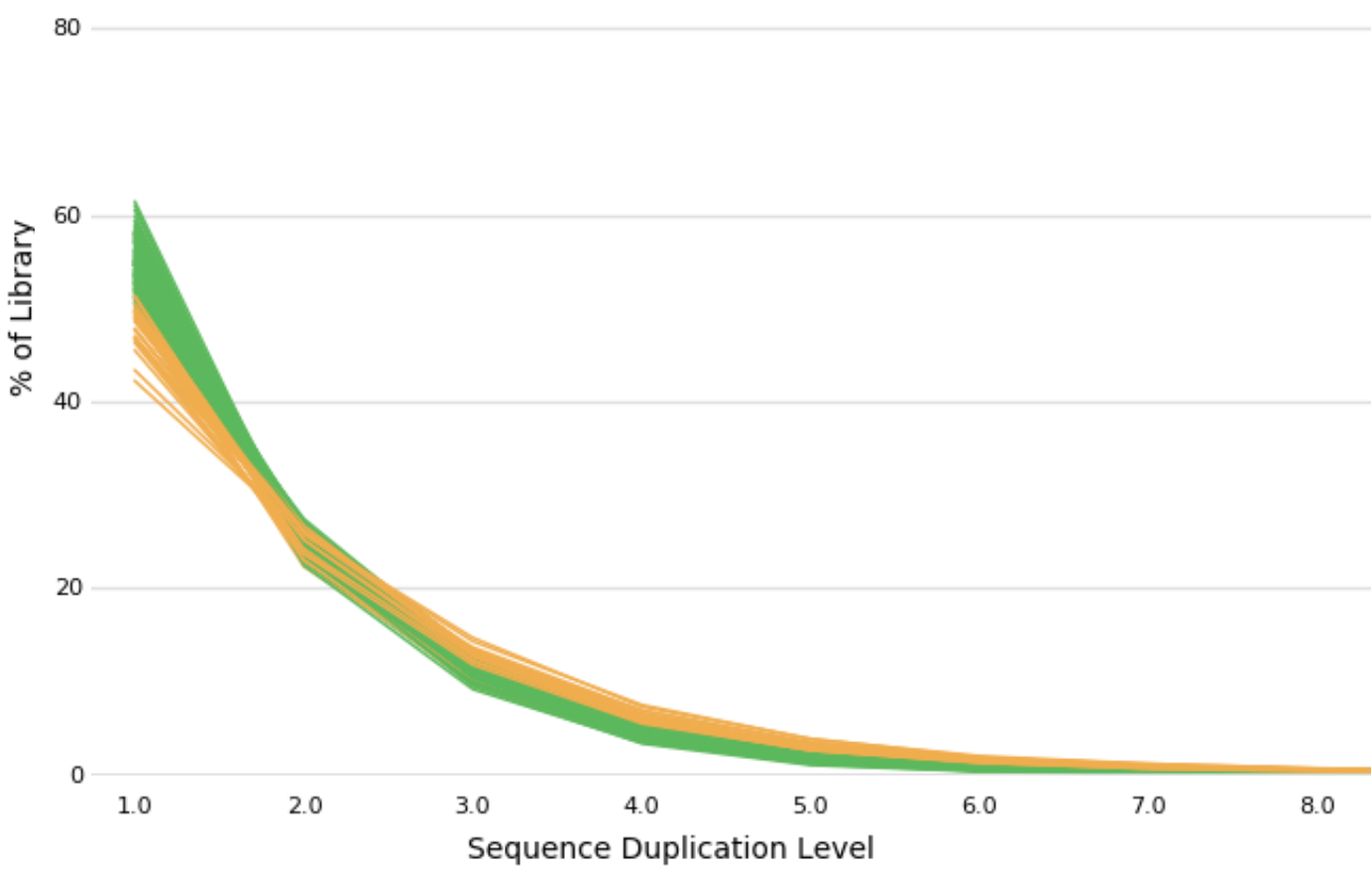}
    \end{subfigure}
    \caption{Analysis of sequence duplication levels, illustrating the extent of duplication across various sequencing processes. Notably, the majority of samples exhibit a low count of duplicates, signifying the high quality of the sequencing process.}
    \label{fig:duplication_level}
\end{figure*}

Specifically, Illumina adapters were meticulously eliminated from the 3' end of the sequence, and all reads that contained fewer than 100 bp were systematically discarded. This data filtering and refinement process resulted in an average yield of approximately 15.7 million reads per sample (see more detailed in Figure \ref{fig:reads_depth}). The average Phred Score for all samples was higher than 35, the adapter content was less than 0.5\% with duplication rate of 25\% (see Figure \ref{fig:duplication_level}). The complete sequencing quality control report is available online and will be provided by the authors upon request.

\begin{figure}[htbp]
    \centering
    \includegraphics[width=0.3\textwidth]{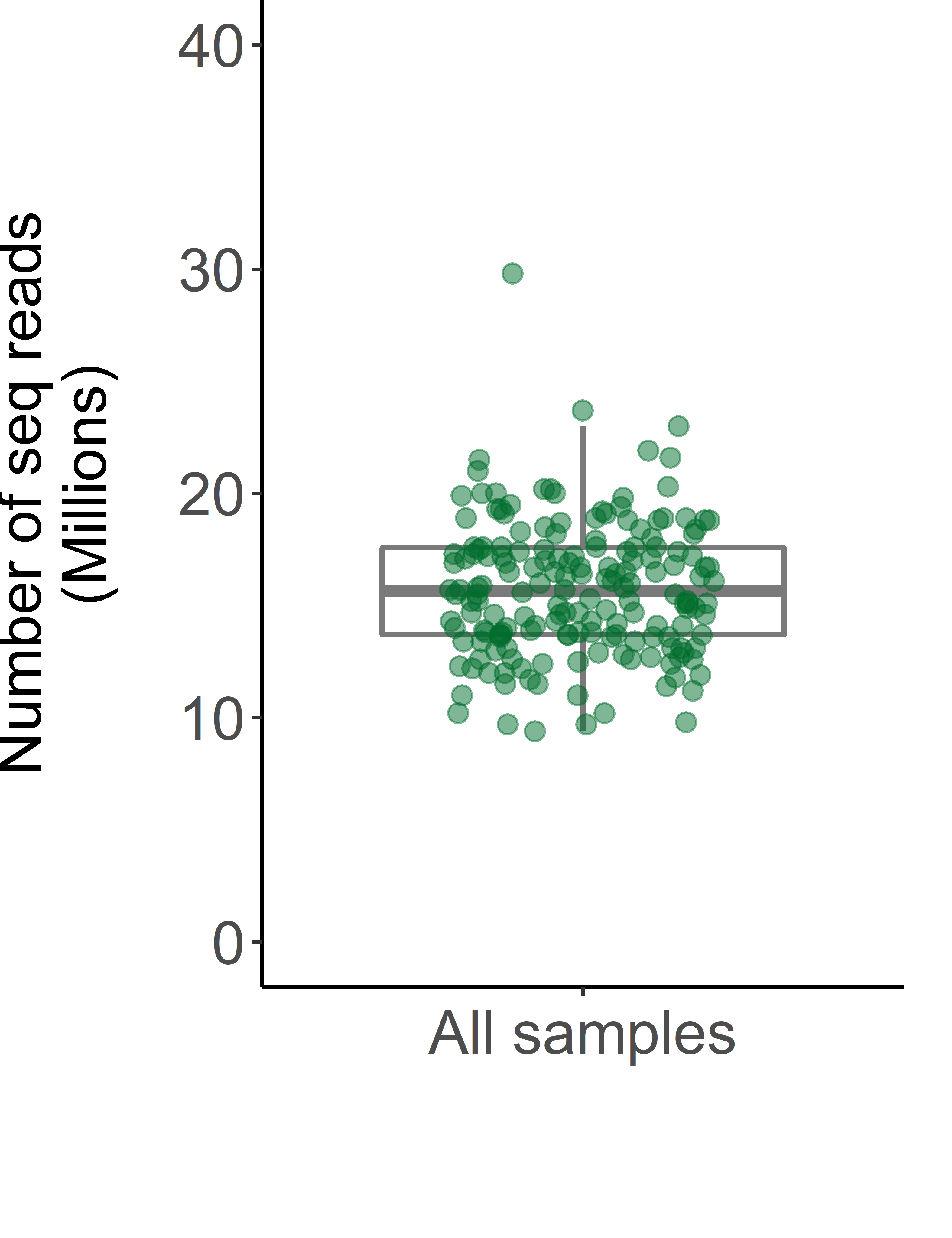}
    \caption{Evaluation of the depths of sequenced samples: an in-depth examination of the sampling depth across various sequencing processes.}
    \label{fig:reads_depth}
\end{figure}

\subsection{Animals}
In this study, we examined 2,190 Israeli Holstein cows. Currently, Israel has approximately 102,200 dairy cows, the vast majority of which are of the Israeli Holstein breed. Roughly 70\% of these cows are found in Kibbutz herds, which are large units within cooperatively owned and managed farms. The remaining 30\% are part of Moshav herds, which are smaller, family-owned farms.
According to the Israeli Herd Book annual report of 2022 \cite{weller2023genetic} the Israeli cow produced an average of 12,442 kg of milk (production/cow/305 days), of which 3.32\% is protein and 3.89\% is fat.
The annual milk yield per cow recorded here is among the highest globally. For comparison's sake, the per-cow milk production in the USA in 2022 stood at 10,839 kg, as referenced in \cite{milkProduction2023}.

The participating study sites housed between 400 and 900 cows each. No significant correlation was found between the farm size and the average efficacy of the feed additives. The microbiome sample from 10 cows used for prediction constituted at least 1.1\% of the total herd size. Given the absence of a significant correlation between farm size and the prediction engine's accuracy, we postulate that our method should be pertinent for sample sizes of at least 0.5\%. For example, this suggests that a sample from 15 cows could representatively predict the expected efficacy for a herd of up to 3,000 cows.

\subsection{Global applicability: overcoming geographical and nutritional variabilities}

While the methodology outlined in this study was initially implemented in commercial dairies across Israel, it possesses a broad applicability that extends well beyond this geographical boundary. This universality stems from several key factors that were integral to our research design and execution.

First, our trials were deliberately conducted across a diverse range of geographical settings within Israel, encompassing areas from arid deserts to cooler mountainous regions and more temperate plains. This variety in environmental conditions mirrors the broad spectrum of climatic zones found in many countries where extensive dairy farming is practiced. Hence, the efficacy of our methodology in these varied Israeli locales strongly suggests its potential adaptability and effectiveness in similar geographical contexts worldwide.

Secondly, our research accounted for various nutritional regimes implemented in these farms. The diet of dairy cows in Israel, much like in numerous other countries, includes by-products from the food industry, such as pulp, gluten meal, and citrus fruits. This aspect of our study is particularly significant as it demonstrates the adaptability of our methodology to different feeding practices and regimes, which are a common feature in global dairy farming.

Furthermore, while our primary focus lies in the realm of the microbiome and its mathematical interactions, it is important to consider the potential variability of microbiomes across different countries. However, we posit that regional differences in the rumen microbiome are unlikely to significantly impact the applicability of our method. This is because our approach hinges on a mathematical framework that is largely data-agnostic. The core of our methodology is the analysis of power-law and network dynamics within the microbiome, a process that is not inherently limited by geographical variations in microbiome composition.

In essence, the requirement of our methodology is straightforward: to sequence the microbiome and measure methane emissions. From these datasets, regardless of their geographic origin, our model is capable of making accurate predictions. The mathematical nature of this approach underlines its potential global applicability, as the fundamental principles of power-law dynamics and network analysis are universally applicable.

\subsection{Methane Detection and Quantification}

A multitude of methane sensing techniques exists in the market today, reflecting the diverse array of applications they serve - from safety monitoring in mining and natural gas industries, air quality surveillance in urban areas, to greenhouse gas emission tracking for climate research. Originally, these devices were not specifically conceived for agricultural settings, let alone for monitoring ruminant animals like cows. However, recognizing the crucial role of livestock in methane emissions, our study undertook a comprehensive evaluation of the available sensing technologies. After thorough scrutiny, which considered aspects such as accuracy, durability, suitability for large scale deployment, and adaptability to the unique conditions of a farm environment, we were able to select those sensors most apt for our specific needs in measuring bovine methane emissions.

\subsubsection{Existing Technologies for Methane Detection and Measurement}

The following presents an overview of the principal technological methodologies employed today in commercially available methane sensors and their potential suitability for ruminant enteric methane emissions.

\paragraph*{\textbf{Infrared Sensors:}}
These sensors measure methane concentration by detecting the specific wavelengths absorbed by methane. They tend to be reliable and require low maintenance. Some commercial examples include the ExplorIR-M 5\% $CO_2$ Sensor and the SGX Sensortech's IR Methane Sensor \cite{bishop2016intra}. While these sensors are quite accurate, their placement and exposure to environmental conditions could impact the readings in a free-range cattle environment.

\paragraph*{\textbf{Semiconductor Sensors:}}
These sensors measure methane by detecting the change in resistance of a semiconductor material exposed to different methane concentrations. These are usually less expensive than infrared sensors, but they tend to have a shorter lifespan and require more maintenance. Figaro's TGS2611 \cite{shah2023characterising} is an example of a semiconductor methane sensor. Their low-cost could be advantageous for wide-scale deployment across large cattle farms.

\paragraph*{\textbf{Catalytic Sensors:}}
These sensors measure methane concentration by detecting the heat produced when methane reacts with a catalyst. However, these sensors might not be ideal for methane measurement from cows due to their susceptibility to poisoning and their requirement for oxygen to function. An example of such a commercial sensor is the Honeywell XCD Methane Gas Detector. This fixed gas detector is designed to provide comprehensive monitoring of combustible gas levels in various environments and is known for its reliability and accuracy.

\paragraph*{\textbf{Electrochemical Sensors:}}
These sensors measure methane by detecting the current generated when methane is oxidized. While they are sensitive and compact, they tend to have a shorter lifespan than other sensor types. The ALTAIR Pro Single-Gas Detector \cite{roulston2018emissions} is an example of an electrochemical sensor, designed for worker safety in mind, with a primarily goal to monitor potentially harmful gases in confined spaces.

\paragraph*{\textbf{Photoacoustic Spectroscopy Sensors:}}
Instruments like the INNOVA 1412i Photoacoustic Gas Monitor \cite{qi2012study} use the principle of photoacoustic spectroscopy to measure methane emissions. These are highly accurate but can be more expensive and might be more suited to laboratory settings or small scale, intensive research studies.

\paragraph*{\textbf{Laser-based Sensors:}}
Sensors like the LI-COR's LI-7700 Open Path $CH_4$ Analyzer \cite{mcdermitt2011new} use laser technology to measure methane concentrations in the open air. These are highly accurate and can cover a large area, making them suitable for large farms, but they are often also significantly more expensive.

\paragraph*{\textbf{Sulfur Hexafluoride ($SF_6$) Tracer:}}
This technique is commonly used for measuring methane emissions in ruminants. It involves the animal inhaling a small quantity of $SF_6$, and the concentration of $SF_6$ and methane in the exhaled air is measured, allowing for the implicit calculation of the methane production rate of the animal \cite{mercadante2015relationship}. This technique is widely used in research settings due to its accuracy, but it requires specific equipment and technical expertise, making it less suitable for widespread commercial use. An example is the SF6 Sulfur Hexafluoride Gas Analyzer by Nova Analytical Systems, specifically designed for such applications.

\subsubsection{Assessment of Methane Emissions in Prior Ruminant Research}

In recent years, the quest to reduce emissions from enteric methane fermentation has garnered increasing attention. This has sparked significant efforts towards devising techniques that not only accurately represent in-field situations but also minimize the disturbance to the animals \cite{ribeiro2020comparison,danielsson2017methane}.

While conventional industrial methane sensors have been repurposed for measuring ruminant methane emissions (see infra-red-based sensing in \cite{bishop2016intra}, laser-based studies in \cite{dengel2011methane}, and a combination of photoacoustic spectroscopy and infrared sensors in \cite{place2011construction}), there has also been progress in developing methods specifically tailored for bovine applications.
A comprehensive comparison of these techniques is elaborated in \cite{garnsworthy2019comparison}. Below, we provide a succinct summary of the predominant methodologies, highlighting their respective advantages and drawbacks:

\paragraph*{\textbf{Respiration Chambers:}}
The breathing or calorimetric chamber has been the traditional benchmark for measuring $CH_4$ emissions from ruminants in various settings \cite{bhatta2007measurement}. This method's chief aim is to quantify the energy generated through an animal's regular metabolic processes. Such chambers play a pivotal role in exploring strategies to curtail $CH_4$ emissions. They function by monitoring the concentrations of gasses in the animal's exhaled air within a regulated environment. However, the use of the calorimetric chamber is generally confined to the analysis of a single animal due to construction costs and the need for specialized operational skills \cite{storm2012methods}.

\paragraph*{\textbf{Head chamber:}}
This method employs an airtight box, encircling the ruminant's head, with a curtain or sleeve around the neck to restrict air exchange between the internal and external atmospheres of the chamber \cite{bhatta2007measurement}. The box should be adequately sized to allow unhindered head movements and access to feed and water. Compared to the calorimetric chamber, the prime benefit of this approach lies in its cost-effectiveness (in comparison to the respiration chamber). Similar to the calorimetric chamber, measurements must be performed individually on trained animals.

\paragraph*{\textbf{Face chamber:}}
The face mask, akin to the calorimetric and head chambers, presents another approach for measuring $CH_4$ from ruminants \cite{johnson1995methane}. This method involves fitting a mask onto the animal's head to gather air exhaled through the airways. The animal requires a brief acclimation period to the equipment, typically spanning seven days, with six-minute sessions each day \cite{odai2010estimation}. During this time, the animal is not allowed to eat or drink, and the analyses are conducted similarly to those in an open calorimetric chamber.

\paragraph*{\textbf{Polyethylene tunnel:}}
This method utilizes a structure reminiscent of an agricultural greenhouse, erected on a pasture with dual layers of inflatable polyethylene walls and a large entrance. It serves as a simpler alternative to the calorimetric chamber in terms of operation and data collection. Inside this tunnel, air is consistently drawn in, allowing for continuous collection of air samples from an exhaust port for gas analysis or gas chromatography \cite{kebreab2006methane}. This method is typically employed to assess $CH_4$ emissions in areas of fresh forage, allowing animals to behave naturally and controlling selected forage within the confined tunnel space.
This technique's benefits include the animals' unrestricted movement within the tunnel and the relatively low acquisition and installation costs. However, it is impractical to control the tunnel's temperature during periods of high ambient temperature. Most studies using this method have focused on sheep due to pasture space constraints \cite{bhatta2007measurement}. Additionally, this technique is unsuitable for experiments evaluating various treatments.

\paragraph*{\textbf{Sulfur Hexafluoride ($SF_6$) Tracer:}}

The sulfur hexafluoride (SF6) method involves a small permeation capsule, essentially a metal tube with a porous plate at one end, filled with $SF_6$. Initially, the capsule is placed in a thermostatic water bath for a month before it is inserted into the animal's rumen. The animal is fitted with a halter that has a capillary tube connected to a PVC yoke. Over a specified duration, this apparatus collects exhaled gases. After a vacuum is applied, the sample is sent to the lab for gas chromatography analysis.
A valve in the PVC yoke ensures the collection of exhaled air at a steady rate. This collection system is calibrated to stop once the sample fills approximately half of the system's storage capacity, typically within 24 hours.
This method allows the animals to move freely and engage in normal grazing activities, removing the necessity for confining them in cages or barometric chambers. Nevertheless, the animals need training to acclimate to the equipment, and the PVC tubing requires daily replacement \cite{mercadante2015relationship}.

\paragraph*{\textbf{Automatic feeder technique (GreenFeed):}}
The GreenFeed technique operates by recognizing an electronic tag on the animal as it begins to feed. The system then measures the gases emitted every second during the feeding process, allowing for the monitoring of individual emission rates over time. Given that approximately 90\% of gases produced by ruminants are released through eructation via the mouth and nostrils, this system generates a highly reliable dataset for research on GHG reduction strategies \cite{hammond2016greenfeed}.
Upon insertion of its head into the feeder, the animal is identified via an electronic tag using radio frequency technology (RFID). A fan then activates to draw in the air exhaled through the animal's nostrils and mouth. Sensors within the equipment measure gas concentrations, the volume of emitted gas, and other environmental parameters.
Despite its advantages, GreenFeed presents considerable challenges that may limit its use. Its high cost can be prohibitive, especially in larger studies, making implementation unfeasible in many research centers. Additionally, the time required to acclimate animals, particularly Zebu and other native commercial breeds, to the equipment should be considered when planning studies utilizing this technique \cite{hristov2015use}.

\subsubsection{Selected sensor:}

Though several dedicated sensing mechanisms have been developed specifically for monitoring methane emissions from cows, such as respiration chambers, polyethylene tunnel, head chamber, face mask or an automatic feeder, these tools share common disadvantages. Primarily, their use results in an alteration of the cows' natural behavior, making the captured data less representative of the animals' day-to-day emission patterns. The disruption of normal behavior is due to the intrusive nature of these devices, which typically require direct contact with the animals or confinement within a restricted space.

These methods also lack scalability. In larger farms with hundreds or thousands of cows, the application of these techniques becomes a logistical challenge, limiting their utility in extensive real-world scenarios. The expenses associated with these techniques further dampen their practicality - the high costs involved in the construction, maintenance, and operation of these devices often make them economically unfeasible for most farms. Furthermore, their use typically demands trained cows which accustom to the devices, imposing an additional layer of complexity to the measurement process.

Given these limitations, we elected to leverage an industrial methane sensor for our study. Industrial sensors are known for their high degree of accuracy and sensitivity, essential features for reliable data collection. More importantly, their non-intrusive nature allows the cows to behave naturally, ensuring that the data gathered is reflective of standard methane emissions under typical conditions. This non-intrusiveness also means the cows require no special training or conditioning to tolerate the device. Being designed for industrial applications, these sensors are robust, cost-effective, and scalable, enabling their usage across larger herds without a significant uptick in operational complexity. Thus, by using an adapted industrial sensor, we hope to bypass many of the hurdles associated with dedicated cow methane sensors and collect reliable and representative data on methane emissions from ruminants.

\paragraph*{\textbf{SEM5000 by Geotech: Detailed Specifications and Primary Advantages:}}
The ATEX Gas Analyser Geotech SEM5000 is a robust, hand-held device specifically designed to measure methane concentrations. Built for use in challenging environments, this device has some key capabilities and advantages that make it well-suited for methane measurement in ruminants like cows.
The SEM5000 utilizes laser-based technology for detecting and quantifying methane levels with a range of 0 ppm to 100\% volume (laser based sensors were demonstrated to be of ideal use for methane measurements in ruminants \cite{rey2019comparison}). It boasts a rapid response time, delivering results in seconds. The device also has an in-built GPS, which allows for geo-tagging of measurements and the creation of gas concentration maps.
The key advantages of this sensor in methane measurement for cows are as follows:

\begin{itemize}
  \item \textbf{Non-invasive and very low stress method:} Unlike some techniques that require animal confinement or behaviour modification, the SEM5000 allows for measurements to be taken in a non-invasive manner, reducing stress on the animals and ensuring data gathered represents their natural behaviour.
  \item \textbf{Accuracy and precision:} The laser technology used in the SEM5000 delivers high-accuracy and precision readings, reducing the potential for errors and increasing the reliability of the data.
  \item \textbf{Portability and robustness:} Given its hand-held design and robust construction, the SEM5000 can be used in a variety of field conditions, making it a practical tool for monitoring methane emissions in grazing environments.
  \item \textbf{Scalability:} The SEM5000 allows for high-throughput data collection and can be used to measure methane emissions from a large number of animals over a short period, making it a more scalable solution than other techniques.
  \item \textbf{Cost:} Notably, the SEM5000 Methane Detector is a cost-effective choice, priced at nearly one-tenth the cost of the Greenfeed system, offering efficient and affordable monitoring of methane emissions.
\end{itemize}

\begin{figure}[htbp]
    \centering
    \includegraphics[width=0.3\textwidth]{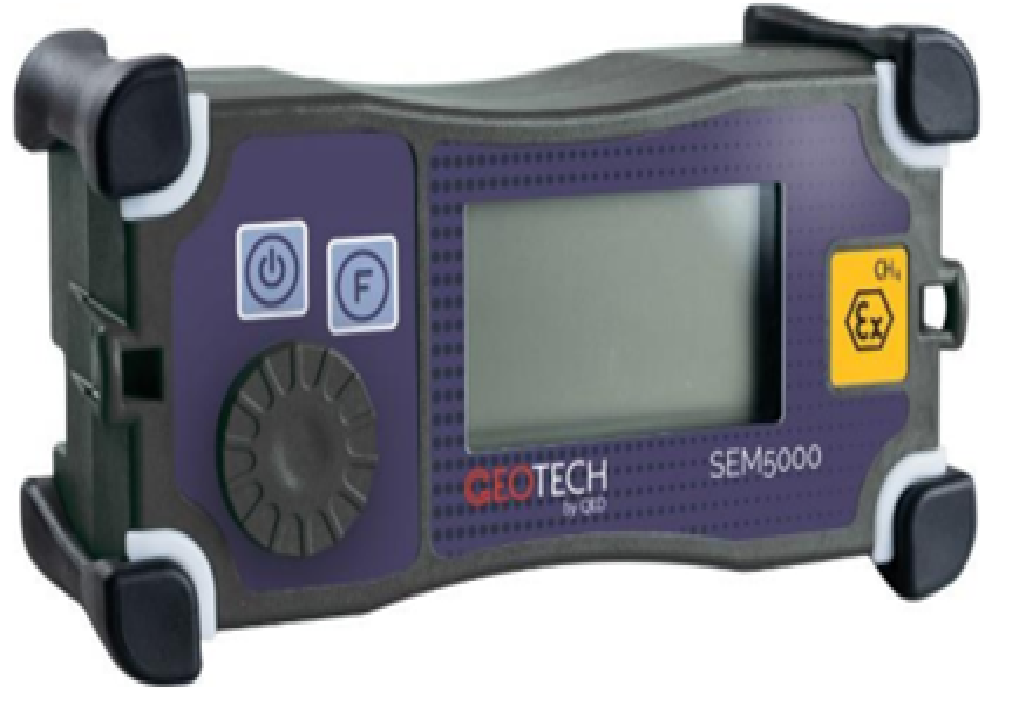}
    \caption{The Geotech SEM5000 Methane Detector, a hand-held device utilizing laser-based technology for highly accurate and rapid methane concentration measurements. Its compact design enhances portability and usability under various field conditions, and its fast response time coupled with high data throughput make it an ideal tool for extensive in-field ruminant measurements.}
    \label{fig:reads_depth}
\end{figure}

\begin{table}[htbp]
\centering
\begin{tabular}{|l|l|}
\hline
\textbf{Specification} & \textbf{Value} \\
\hline
Range & 0 to 10,000 ppm \\
\hline
Resolution & 0.1 ppm \\
\hline
Accuracy & 0.7 ppm \\
\hline
Technology & laser based\\
\hline
Response time & $<$ 2.5 sec. \\
\hline
Rate & 2 sec. per reading \\
\hline
Battery life & 10 hours \\
\hline
Flow & 1 litter per Minute \\
\hline
Weight & 1.3 kg \\
\hline
\end{tabular}
\caption{Main technical specifications of the Geotech SEM5000 methane detector.}
\label{tab:my_label}
\end{table}

\subsubsection{Comparative Performance Evaluation of the SEM5000 Sensor}
In order to ascertain the reliability and robustness of the SEM5000 sensor for ruminant methane measurement, we devised a comprehensive study. Our objective was to show that, although less costly and more scalable than other commonly used technologies, the SEM5000 can deliver equivalent levels of accuracy. For comparative benchmarking, we selected the Li-Cor LI-7810 laser-based \cite{xu2018chamber} system and the GreenFeed system \cite{hammond2016greenfeed} - two established technologies in the field. Li-Cor LI-7810, while robust and accurate \cite{korben2021estimation,johannesson2022forest,takano2021spatial}, is considerably more expensive, and GreenFeed, though regarded as the gold standard \cite{hristov2015use,hammond2015methane,huhtanen2019enteric}, presents limitations in scalability, usability and cost.

\paragraph*{\textbf{Comparative analysis methodology:}}
Our comparison methodology involved conducting two distinct sets of tests. In the first set, we concurrently measured the same cows using the SEM5000 and GreenFeed devices. This allowed for a direct comparison between these two methods on the same animal subjects.
We compared the mean and median methane emissions from each cow as measured by both sensors. Given the heightened concern over high methane emissions in the context of mitigation, we also analyzed the average of the top 25\% of readings. To further validate the consistency between the two sensors, we employed the Mann-Whitney U Test \cite{nachar2008mann}, a non-parametric method designed to determine if two sets of readings originate from the same source.

In the second set of tests, we used the SEM5000 and Li-Cor 7810 devices in parallel over a period of four weeks, measuring 48 different cows. This extended period of observation, as well as the large number of specimens being measured, provided a comprehensive set of data to compare the performance of the SEM5000 sensor to the well-established Li-Cor 7810 laser-based system. For each cow, we determined the regression between measurements from the two sensors. Subsequently, we used the Bland–Altman method \cite{bland1986statistical}, a technique tailored for evaluating the concordance of two sensors measuring identical data. As a final step, we computed the Root Mean Square Error (RMSE) to quantify the differences between the two sensors.

The order of measurements was randomized in both sets of tests to minimize potential bias.

\paragraph*{\textbf{Results:}}
Our preliminary findings provide compelling evidence supporting the suitability of the SEM5000 sensor for ruminant methane measurement. The comparative data suggest that the SEM5000 sensor achieves high levels of accuracy, comparable to that of the pricier Li-Cor and GreenFeed systems. A detailed analysis of these results is presented in the forthcoming figures, demonstrating the commendable performance of the SEM5000 sensor.

\begin{figure*}[htbp]
    \centering
    \includegraphics[width=\textwidth]{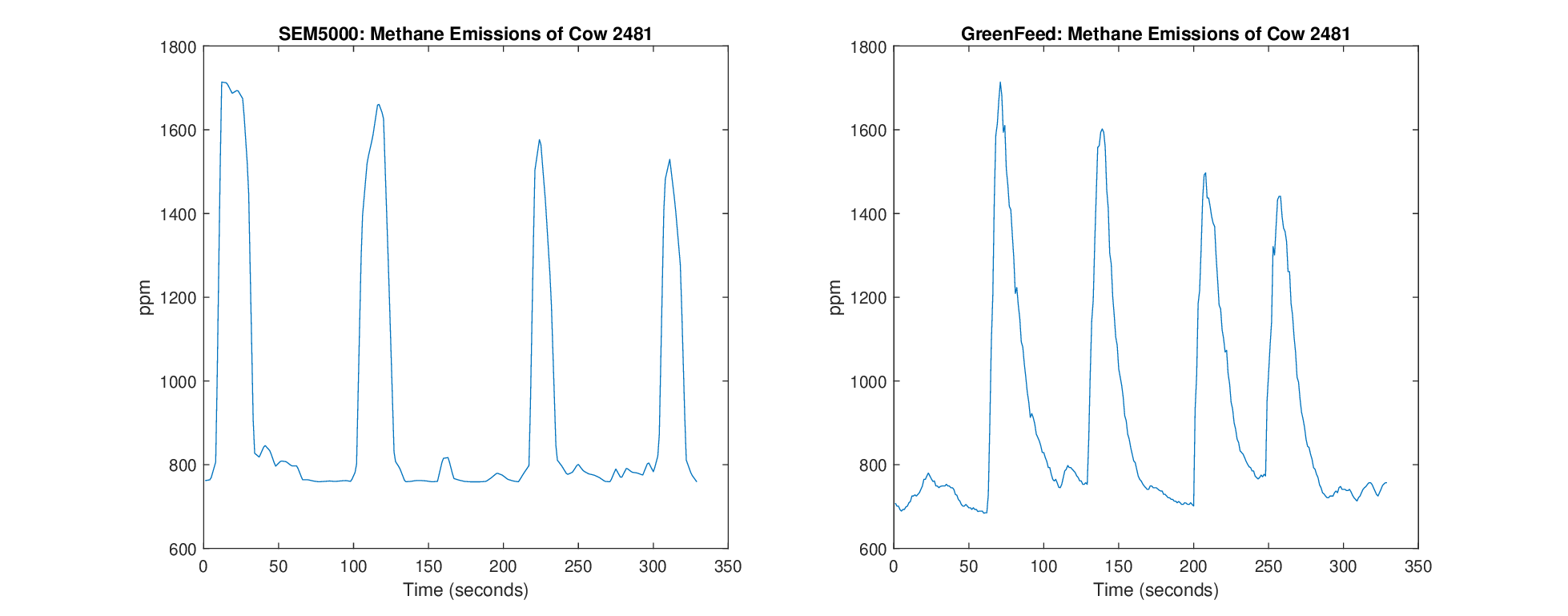}
    \caption{Methane levels (in parts per million) for the same cow, as measured at different times by the SEM5000 sensor and the GreenFeed system. The readings exhibit similar dynamics and values.}
    \label{fig:single_cow_comparison}
\end{figure*}

Table \ref{tbl.comparison1} presents a comparison between measurements taken using the SEM5000 and the Greenfeed sensors for four cows. The results demonstrate that while the SEM5000 sensor exhibits a marginally higher data variance, the key properties, such as average and median emissions levels align closely. The observed differences not only meet the stringent criteria set out by Verra's VM41 protocol \cite{VM41} and the CDM Meth Panel Guidance on Addressing Uncertainty in the Estimation of Emissions Reductions for CDM Project Activities but also fall below the $15\%$ threshold for sensors' compliance defined in the IPCC 2006 Guidelines, Volume 2, Chapter 2, Tables 2.2 to 2.6 \cite{eggleston20062006}. This assertion of compliance is further validated by the Mann-Whitney U Test results \cite{nachar2008mann}, which suggest that (for each cow) both data streams likely derive from the same source. This analysis establishes the credibility of the SEM5000 sensor for measuring enteric methane emissions, affirming its readings to be as dependable as those from the Greenfeed system. The results from the SEM500 and the GreenFeed system, when measuring the same cow, are depicted in Figure \ref{fig:single_cow_comparison}. Both measurements showcase comparable emissions levels and temporal dynamics.

\begin{table*}[htb]
    \centering
    \begin{tabularx}{\textwidth}{lXXXXX}
        \toprule
        Metric & Cow 2071 & Cow 2299 & Cow 2481 & Cow 2849 & Avg. \% Change \\
        \midrule
        Mean $CH_4$ & 159/167 & 123/129 & 909/934 & 127/133 & 4.8\% \\
        Median $CH_4$ & 84/87 & 70/73 & 773/782 & 66/68 & 3.6\% \\
        Mean of top 25\% $CH_4$ readings & 410/418 & 295/326 & 1318/1397 & 335/328 & 5.6\% \\
        STD of $CH_4$ readings & 196/235 & 123/168 & 261/326 & 161/221 & 29.6\% \\
        Mann-Whitney U Test p-value & 0.065245 & 0.022432 & 0.000019 & 0.009071 & N/A \\
        \bottomrule
    \end{tabularx}
    \caption{Comparison between measurements taken using the Greenfeed system and the SEM5000 sensor, for four different cows (values are shown as Greenfeed/SEM5000 for each cow). The close alignment in these metrics between the two sensors underscores their comparable performance. The Mann-Whitney U Test \cite{nachar2008mann}, a non-parametric statistical test, is employed to determine whether two independent samples were drawn from a population with the same distribution. In this context, the test's results suggest a high probability that the measurements from both sensors are from the same distribution. This conclusion further suggests that data captured using the SEM5000 can, with a high degree of confidence, be used as a proxy for results from the Greenfeed system, paving the way for broader and more flexible deployment of these sensors in methane measurement campaigns.}
    \label{tbl.comparison1}
\end{table*}

\begin{figure*}[htbp]
    \centering
    \includegraphics[width=\textwidth]{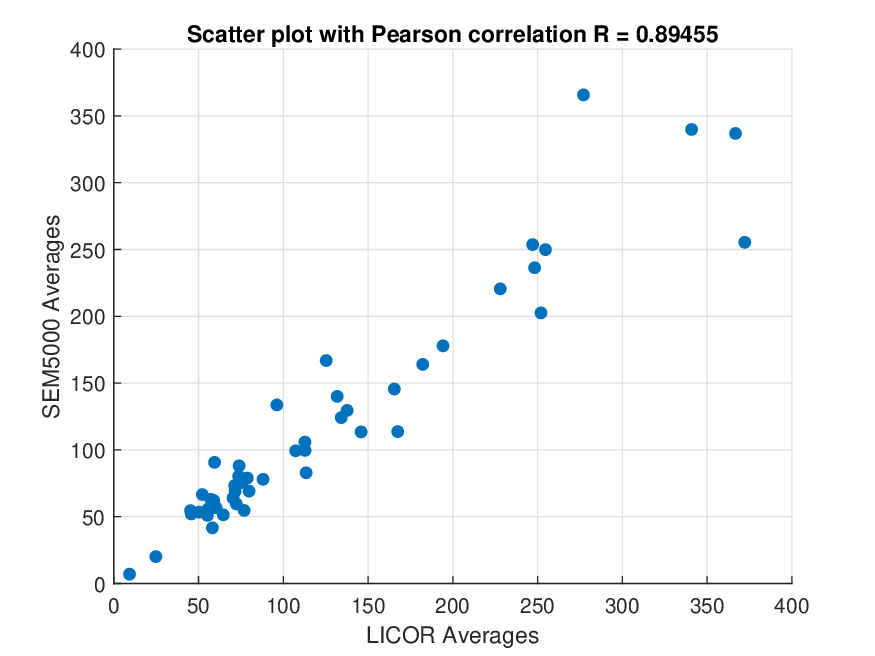}
    \caption{Scatter plot comparing readings from the SEM5000 sensor and the LICOR 7810 laser-based sensor. Each point represents the averaged methane emissions from a unique cow, as measured by both sensors. The Pearson regression line is also depicted, with an accompanying \( R^2 \) value, indicating the strength and direction of the linear relationship between the two sets of measurements. This high correlation implies that the two sensors produce consistent and comparable results, further validating the reliability and accuracy of the SEM5000 in measuring enteric methane emissions. The average discrepancy (noise) between measurements from the two sensors was found to be 14\%, with a median discrepancy of 10\%. Additionally, the Root Mean Square Error (RMSE) between their measurements stood at 28.}
    \label{fig:comparison-scatter}
\end{figure*}

\begin{figure*}[htbp]
    \centering
    \includegraphics[width=\textwidth]{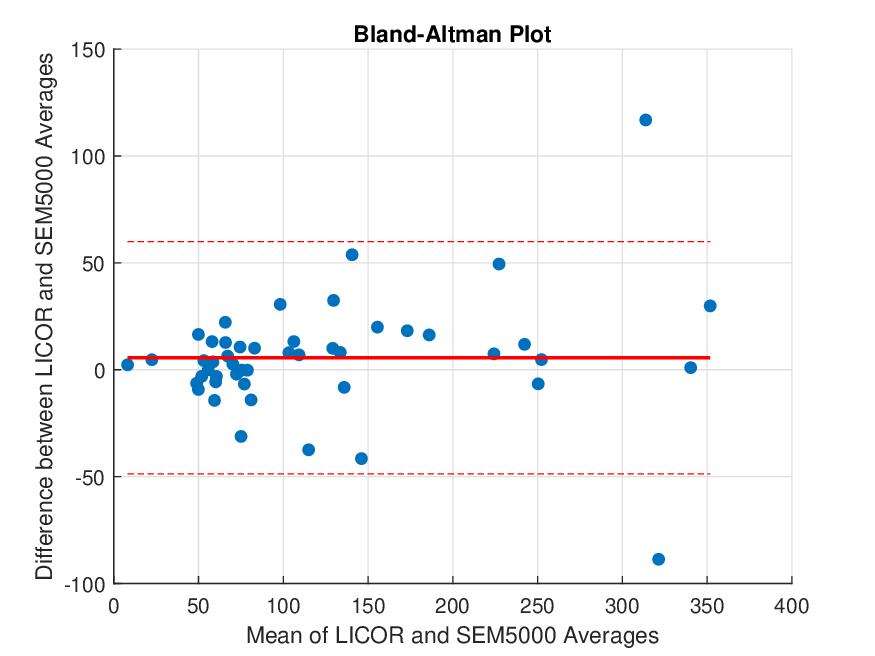}
    \caption{Bland-Altman analysis comparing the SEM5000 and the LICOR 7810 laser-based sensor. The Bland-Altman test\cite{bland1986statistical} is used to assess the agreement between two different instruments measuring the same parameter. In the plot, the difference between the two sensors' readings is plotted against their average. The central line represents the mean difference, while the outer lines depict the limits of agreement, which are calculated as the mean difference $\pm$ 1.96 times the standard deviation of the differences. Apart from two outliers, all data points lie within the confidence limits, indicating that the measurements from the two sensors are largely in agreement and can be used interchangeably for most practical purposes\cite{giavarina2015understanding}.}
    \label{fig:LICOR-comparison-BlandAltman}
\end{figure*}

Figures \ref{fig:comparison-scatter} and \ref{fig:LICOR-comparison-BlandAltman} further bolster the credibility of the SEM5000 sensor for enteric methane measurements. In these figures, methane emissions from 48 distinct cows, measured over a span of 4 weeks by both sensors, are showcased. A discernible strong correlation between the measurements from the two sensors is evident. Coupled with the robust correlation, the Bland-Altman analysis further corroborates these observations. The Bland-Altman method is primarily utilized to assess the agreement between two different measurement techniques, determining the consistency and discrepancy in results. Its affirmation in this context emphasizes the reliability and similarity of readings between the two sensors.

\subsubsection{Periodic Methane Measurement: Duration and Consistency}

One of the core strengths of our research methodology revolves around the repeated measurements of each cow throughout an extended 12-week period post-treatment. This strategy is deeply rooted in the need to validate the persistent efficacy—or potential lack thereof—of the additives. Over time, factors such as changes in feed quality, external environmental conditions, or the cow's inherent physiology might impact methane emission levels. By measuring emissions repeatedly over several weeks, we ensure that the observed effects (or non-effects) of the additive remain consistent.

In addition, to enhance data accuracy and reduce volatility, methane emissions were consistently measured at the same hour of the day for each test session, thereby minimizing the impact of diurnal variations.

The importance of extended measurement periods is not just an internal research principle but is also underscored by the standards set by external bodies. Specifically, the VM41 protocol \cite{VM41} for enteric methane measurement and reduction, established by the Verra agency, mandates that projects measure emissions for at least 8 weeks to be compliant with its guidelines \cite{rey2019comparison}. This timeframe is recommended to ensure a thorough assessment of the additive's performance. In acknowledgment of the significance of these guidelines, and with an aim to enhance the statistical robustness of our results, we opted for an extended 12-week measurement duration for our study.

At each time point, the efficacy of the feed additive was calculated by comparing the current measurements of the cows that were available and measured on that day to their emission levels recorded before the commencement of the trial. This implies that the exact composition of cows measured may differ between consecutive farm visits. However, this does not compromise the accuracy of the efficacy calculation at each point. The reason being, the cows in both the treatment and control groups are always compared to their individual baseline emissions level, which was established prior to the initiation of the trial. This procedure ensures that the efficacy determination is based on individualized comparisons, thereby maintaining the overall reliability of the results.

We also took steps to manage the potential variability in our measurements due to cow evasiveness. The percentage of cows that managed to evade measurement was consistently kept below 20\%, reducing the overall impact of this phenomenon on our data set. Consequently, we believe that the influence of this variability on our overall findings is minimal. This aspect of the study underscores the complexities of field research in livestock environments and the need for adaptable research methodologies.

\subsubsection{Methane Reading Methodology}

When measuring methane emissions from cows, the variability in the data due to diverse observation durations and transient spikes presents significant challenges. Our methodology needed to effectively address these discrepancies to yield reliable and consistent metrics.

\paragraph*{\textbf{Ambient Noise Filtering:}}
Methane measurements can capture ambient readings, especially before and after the actual approach to the cow. To filter out these irrelevant readings and hone in on the cow's emissions, we considered only values above 5 parts per million (ppm). This threshold ensures we primarily focus on the cow's emissions, excluding most ambient interference.

\paragraph*{\textbf{Data Consolidation and Noise Reduction:}}
To condense the varied readings from each visit into a single representative number and simultaneously mitigate the noise (like sudden spikes due to burping), we employed the median value. As a measure of central tendency, the median is inherently robust against outliers, offering a more stable representation of the cow's typical methane emission.

Formally, given a set of methane readings for a specific cow $c$ taken on a sequence of time stamps
$\{ R_{c,t_1}, R_{c,t_2}, \dots, R_{c,t_N} \}$ on a specific day $d$, the consolidated value for that day and cow is computed as:
\[ \hat{R}_{c,d} = \text{median}\left( \{ R_{c,t_i} | R_{c,t_i} > 5 \text{ ppm} \} \right) \]

By implementing this methodology, we ensure a single, consistent methane reading per cow for each visit, establishing a dependable foundation for subsequent comparative analysis across various visits and cows.

\subsection{Feed Additives}

The landscape of feed additives, designed to mitigate methane emissions, is rich and varied, with each product leveraging a unique biological strategy. These formulations are designed to interact with the bovine digestive process in various ways to reduce the production of methane, a major byproduct \cite{adesogan2013mitigation}. The efficacy of these additives is largely determined by the specific biological pathway they target, underlining the need for personalized application based on each farm's specific conditions and requirements \cite{beauchemin2008nutritional}. From methane inhibitors and direct-fed microbials to natural plant extracts and chemical compounds, the range of solutions showcases the vast scope of scientific novelty directed towards curbing this environmental concern.
In this study we have tested the following widely used and commercially available additives.

\subsubsection{Agolin}

Ruminant (Agolin) is a commercially available blend of essential oils (coriander seed oil, eugenol, geranyl acetate, and geraniol) which has been demonstrated to reduce greenhouse gas emissions in dairy cows and improve energy corrected milk and feed efficiency \cite{brambila2022evaluation} at a daily dose of 0.8 to 1 gram per animal.
Agolin increases milk production in cows producing moderate milk yield (~30 kg/d), however, this response depends on duration of feeding (5 to 8 wk min). Some observed consistent and convincing 2-3\% increase in yields of milk or ECM \cite{carrazco2020impact,fouts2022enteric}. Agolin is shown to inhibit ruminal methane production or intensity by 8\% on average while no apparent change in dry matter intake (DMI) nor on milk composition was described. Exact mode of action is yet to be elucidated \cite{fouts2022enteric}.

\subsubsection{Relyon}

Manufactured by Phibro Animal Health, this tannins flavonoid and essential oils-based additive was shown to mitigate ruminal methene emission in by 13\% on average, while no change in milk yield or its composition was observed \cite{phibro2021,honan2021feed}. While more rigorous scientific studies are desirable to substantiate Relyon's promising role in also enhancing feed conversion and stimulating appetite in ruminants, the preliminary results presented to date are encouraging.

\subsubsection{Kexxtone (Elanco)}

Kexxtone is a Monensin containing intraruminal bolus for administration 3-4 weeks pre-calving to help the peri-parturient dairy cow/heifer maintaining an appropriate energy balance and thereby preventing many peri-parturient metabolic based diseases \cite{appuhamy2013anti}. The Kexxtone bolus releases Monensin for a period of 95 days in the rumen \cite{kexxtone13}.

Ionophores such as Monensin improve methane mitigation by enhancing digestive efficiency to favor propionate production over acetate, which reduces $H_2$ for methanogens. This methanogenesis inhibition becomes more pronounced in diets with higher fat content \cite{goodrich1984influence,van1977effect}. Meta-analyses of Monensin conclude an effect on methanogenesis inhibition of up to 10\% reduction on average in dairy \cite{marumo2023enteric,fouts2022enteric}.

\subsubsection{Allimax}

Allimax bolus (Garlic, Allicin) has been developed for the purpose of alternative anti-microbial activity in dairy. The natural extract Allicin, which is the main active ingredient of the sulfur-containing organic compounds in garlic, has anti-inflammatory, anticancer, antioxidant, and antibacterial properties \cite{ma2016effect}. However, the specific mechanism underlying its effect on mastitis in dairy cows needs to be further studied \cite{borlinghaus2014allicin,ahn1effect}. The supplementation of Allicin has been observed to elevate the levels of propionate and butyrate during partial incubation periods, suggesting its potential role in curtailing methane emissions \cite{tao2021understanding}. Even though compelling in vitro evidence demonstrates the ability of Allicin to mitigate methane emissions by up to 38\% \cite{fouts2022enteric,honan2021feed,vranken2019reduction}, in vivo studies confirming these findings remain scarce to date \cite{kekana2021effects,ahn1effect}.

\subsection{From Microbial Data to Additive Efficacy Prediction}
\label{sec.flow.short}

This section offers a concise overview of the methodology used in this study. Detailed discussions and formal mathematical delineations concerning data processing, training, validation, and the employed algorithms can be found in Sections \ref{sec.field.study} and \ref{sec.analytics.detailed}.

\paragraph*{\textbf{Input:}}

This study was carried our in partnership with 34 trial sites. We collected 15 microbiome samples from each site, which were subsequently subjected to deep shotgun metagenomic sequencing. At every site, one or more feed additives were administered to distinct groups, each consisting of 20 cows. Additionally, a control group of 20 cows was established at each site, receiving no treatment. Throughout a period of 3 months, starting from the onset of the trial, we periodically recorded methane emissions from each cow. This consistent monitoring allowed us to determine the normalized efficacy of each additive across the various sites.

\paragraph*{\textbf{Unsupervised Detection of Microbial DNA Patterns:}}

In our research, we directly utilized the raw sequencing data -- strings of 100-150 nucleotides -- without delving into specific identification of microbes or strains. Our distinctive DNA analytics algorithm, with a network-oriented approach, processed the data from all gathered microbiome samples (refer to Section \ref{sec.analytics.detailed}). This analysis yielded a plethora of DNA patterns. Each pattern has been analytically validated to be improbable to emerge spontaneously in random microbial genetic samplings, making them statistically likely to correlate with a phenotypic trait, regardless of its relevance to the study's goals.

The initial phase of our method involves an unsupervised data analysis, serving as an effective dimensionality reduction technique. This process can efficiently handle billions of raw sequences, each 100-150 bases in length. It distills these into a computationally tractable number of clusters containing ``statistically significant'' substrings. Importantly, this approach sidesteps any bias toward predetermined feature spaces, data preprocessing, or the inherent semantics of the problem.

This phase can be updated as new data becomes available, leading to the identification of new DNA patterns that contribute to the system's predictive capabilities for the same or new properties.

\paragraph*{\textbf{Filtering the Microbial DNA Patterns using Semantic Labels:}}

For each DNA pattern (actually a collection of 100-150 long DNA bases), we can filter only those whose frequency (i.e., the number of times they occur in a sample) correlates strongly with the property we aim to predict (i.e., an additive's efficacy, defined by its methane reduction capacity, normalized for the control group on the same farm). This phase is executed once for each group of labels (i.e., once per additive).

\paragraph*{\textbf{Output:}}

The process yields a collection of DNA sequence groups, which are statistically validated to correlate with our target property. These segments, termed ``microbiome markers'', subsequently serve as a reference against which samples from new farms are contrasted. This comparison produces an anticipated efficacy score ranging from 0 (no efficacy) to 1 (maximal efficacy) for the given additive.

\section{Field Study Design}
\label{sec.field.study}

\subsection{Definitions}
\label{sec.definitions}

Below are the definitions of groups and annotations used in the description of the study:

\begin{itemize}
    \item $F$: The set of all farms participating in the study, $F = \{f_1, f_2, \ldots, f_{N}\}$.
    \item $F_A$: The subset of farms selected for testing a specific additive $A$.
    \item $C_{u,i}$: The set of ``Learning Microbiome Cows'' (LMCs) for each farm $f_i \in F$, selected for unsupervised learning of the microbiome.
    \item $P_X$: This represents the collection of ``microbial genetic patterns'' derived from each microbiome sample $X$. Each sample gives rise to a distinct network $G_X$ comprised of a constant set of $M = |V|$ nodes and unique edges, $E_X$. Patterns are extracted both from individual network analysis and from the superposition of networks, enabling exploration of a broad spectrum of combinatorial possibilities based on various criteria.
    \item $P_{S}$: represent the patterns derived from a combination of networks, where $S$ is the set of samples considered for superposition.
    \item $C_{t,i}$ and $C_{v,i}$: The partition of the LMCs into a ``Microbiome Train Group'' and a ``Microbiome Validation Group'' for each farm $f_i \in F_A$.
    \item $C_{m,i}$: The 40 cows selected for methane measurement in each farm $f_i \in F_A$.
    \item $C_{mc,i}$, $C_{mt,i}$, $C_{mv,i}$: The division of $C_{m,i}$ into three groups - ``Control Methane Cows'', ``Train Methane Cows'' and ``Test Methane Cows'' (or ``Validation Methane Cows'').
    \item $\hat{R}_{c,d}$ : the level of methane emission for a cow $c$ for a day $d$ (calculated as the median of methane readings above a certain ``ambiance threshold'').
    \item $M_{\text{pre}}(c)$ and $M_{\text{post}}(c)$: The pre-additive and post-additive methane levels for a cow $c$. Notably, since the cows were measured multiple times over the 12-week period following the introduction of the additive, we obtain multiple $\hat{R}_{c,d}$ values for each cow (each corresponding to a respective day $d$, whereas $D_{\text{pre}}$ and $D_{\text{post}}$ represent the days of pre-additive treatment and post-additive treatment respectively), of which we take the mean:
        \[ M_{\text{post}}(c) = \text{mean} \left( \{\hat{R}_{c,d} | d \in D_{\text{post}} \} \right) \]
        \[ M_{\text{pre}}(c) = \text{mean} \left( \{\hat{R}_{c,d} | d \in D_{\text{pre}} \} \right) \]
        This repeated sampling further strengthens the statistical significance of the measurements.
    \item $T_{pre,i}$, $T_{post,i}$, $C_{pre,i}$, $C_{post,i}$: The mean pre-additive and post-additive methane levels for the treatment and control cows in a farm $f_i$.

    \item $\eta_{A, f_i}$: The methane efficacy for farm $f_i$ and additive $A$, calculated as:
     \[
     \eta_{A, f_i} = \frac{T_{post,i} / T_{pre,i}}{C_{post,i} / C_{pre,i}} =
        \frac{T_{post,i} \cdot C_{pre,i}}{T_{pre,i} \cdot C_{post,i} }
     \]
\end{itemize}

Following are some key notes regarding the groups and their properties:

\begin{itemize}
    \item The Learning Microbiome Cows ($C_{u,i}$) and the cows selected for methane measurement ($C_{m,i}$) in each farm are disjoint, i.e. $C_{u,i} \cap C_{m,i} = \emptyset$ for each farm $f_i \in F_A$.
    \item The ``Microbiome Train Group'' ($C_{t,i}$) and ``Microbiome Validation Group'' ($C_{v,i}$) are also disjoint for each farm $f_i \in F_A$, i.e. $C_{t,i} \cap C_{v,i} = \emptyset$.
    \item Similarly, the ``Control Methane Cows'', ``Train Methane Cows'' and ``Test Methane Cows'' groups are pairwise disjoint for each farm $f_i \in F_A$, i.e. $C_{mc,i} \cap C_{mt,i} = \emptyset$, $C_{mc,i} \cap C_{mv,i} = \emptyset$, and $C_{mt,i} \cap C_{mv,i} = \emptyset$.
    \item The efficacy of an additive $A$ in a farm $f_i$ is a time series measurement, represented as a set $\eta_{A, f_i} = \{e_1, e_2, \ldots, e_n\}$ where each $e_k$ is calculated from the ratio of mean pre-additive and post-additive methane levels for the treatment and control cows.
\item The division of cows into ``control methane cows'' (CMC), ``train methane cows'' (TMC), and ``test methane cows'' (TeMC) has been optimized to minimize bias. This has been achieved by exhaustively examining all possible allocations of cows into the three groups, and selecting the assignment that minimizes the maximum discrepancy among the distributions of age ($AGE_i$), days in lactation ($DIL_i$), and average milk yield ($AMY_i$) across the groups. Let $Gr_{CMC}$, $Gr_{TMC}$, and $Gr_{TeMC}$ represent the groups of cows. The condition for the optimal assignment can be formally expressed as:
    For all $i$, and for any two distinct groups $Gr_x$ and $Gr_y$ from $\{Gr_{CMC}, Gr_{TMC}, Gr_{TeMC}\}$, we choose the assignment that minimizes the following quantity:
    \begin{align*}
        \max \left\{
        \begin{array}{l}
            | P(AGE_i | Gr_x) - P(AGE_i | Gr_y) |,  \\
            | P(DIL_i | Gr_x) - P(DIL_i | Gr_y) |,  \\
            | P(AMY_i | Gr_x) - P(AMY_i | Gr_y) |
        \end{array}
        \right\}
    \end{align*}
    This guarantees that the selected division of cows into groups ensures the least possible bias across all characteristics, given the distributions of age, days in lactation, and average milk yield among the cows.

\end{itemize}

\subsection{Training the Model}

This stage is executed per each feed additive $A$. Note that whereas not all farms are included in this stage (since feed additive $A$ may have been tested by only a subset of the available farms), all of the data patterns extracted in the unsupervised phase is used to train its efficacy prediction model. We denote the subset of farms used for this stage as $F_A \subseteq F$.

For each farm $f_i \in F_A$, we further partition the LMCs into a ``Microbiome Train Group'' $C_{t,i} \subseteq C_{u,i}$ and a ``Microbiome Validation Group'' $C_{v,i} = C_{u,i} \setminus C_{t,i}$.

Also for each farm $f_i$, we introduce a set of 40 cows $C_{m,i}$ for methane measurement. We divide this set into three groups: ``Control Methane Cows'' $C_{mc,i}$, ``Train Methane Cows'' $C_{mt,i}$ and ``Test Methane Cows'' $C_{mv,i}$.

Let $M_{pre}(c)$ and $M_{post}(c)$ denote the pre-additive and post-additive methane levels for a cow $c$, respectively. $M_{pre}(c)$ and $M_{post}(c)$ are calculated as the median of methane levels over 30 to 120 seconds, excluding values smaller than 5 parts per million.

The methane efficacy $\eta_{A, f_i}$ for farm $f_i$ and additive $A$ is calculated as follows:

\begin{itemize}
    \item $T_{pre,i} = \frac{1}{|C_{mt,i}|}\sum_{c \in C_{mt,i}} M_{pre}(c)$, the mean pre-additive methane for the treatment cows in the farm,
    \item $T_{post,i} = \frac{1}{|C_{mt,i}|}\sum_{c \in C_{mt,i}} M_{post}(c)$, the mean post-additive methane for the treatment cows in the farm,
    \item $C_{pre,i} = \frac{1}{|C_{mc,i}|}\sum_{c \in C_{mc,i}} M_{pre}(c)$, the mean pre-additive methane for the control cows in the farm,
    \item $C_{post,i} = \frac{1}{|C_{mc,i}|}\sum_{c \in C_{mc,i}} M_{post}(c)$, the mean post-additive methane for the control cows in the farm.
\end{itemize}

Having the overall efficacy calculated as:
\[
\eta_{A, f_i} = \frac{T_{post,i} \cdot C_{pre,i}}{T_{pre,i} \cdot C_{post,i} }
\]

This efficacy, along with the microbiome samples from $C_{t,i}$, is used for the supervised learning process, serving as the label for all microbiome cows at farm $f_i$. Specifically, the efficacy derived from the training cows $C_{mt,i}$, is utilized as labels for the features generated by the training microbiome cows $C_{t,i}$. Likewise, the efficacy determined for the validation cows, $C_{mv,i}$, is employed as labels for the features from the validation microbiome cows $C_{v,i}$.

\subsection{Flowchart}

\begin{figure*}[htbp]
    \centering
    \includegraphics[width=0.6\textwidth,viewport=160 290 625 530]{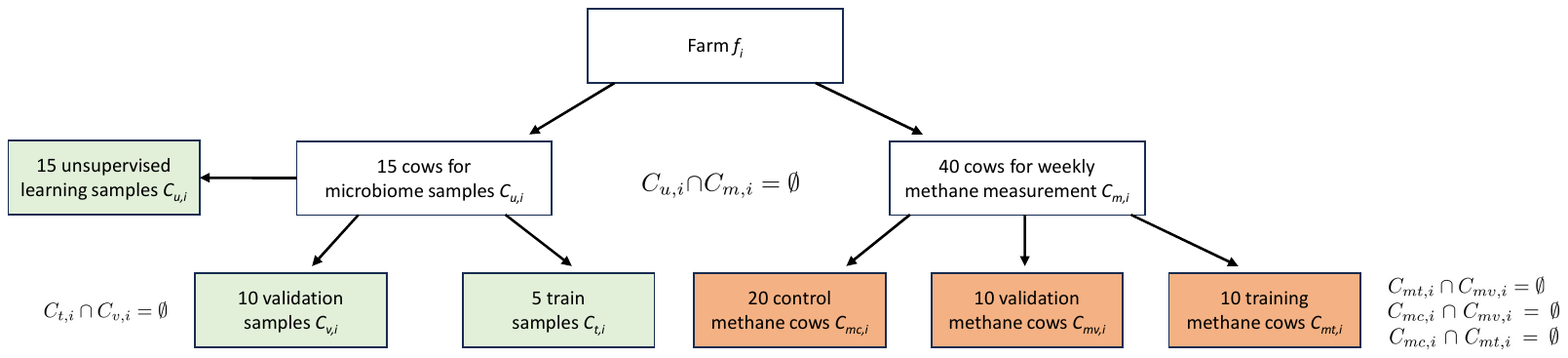}
    \caption{Schematic representation of the field study design tailored for a single feed additive. The initial unsupervised phase identifies genetic patterns from all 15 microbiome samples of a farm and is detailed further in Figure \ref{fig:flowchart2}. Crucially, this phase is conducted once and is applicable across all additives. Of the 15 microbiome samples, 5 are designated for model training while the remaining 10 are earmarked for validation. The microbiome samples form the feature set of the model, with labels being generated based on the average performance of a distinct group of methane cows. One of the pivotal strengths of our design is the assurance that the genetic markers identified by the model are not just emblematic of the microbiome cow group but resonate with the broader farm context. This robustness is reinforced by two layers of separation involving the methane cows: the initial distinction from the microbiome cows and subsequently ensuring the cows used for model training differ from those in the validation, each group being wholly independent. This separation takes place both among the microbiome cows as well as the methane measurement cows.}
    \label{fig:flowchart1}
\end{figure*}

\begin{figure*}[htbp]
    \centering
    \includegraphics[width=0.7\textwidth,viewport=235 100 710 550]{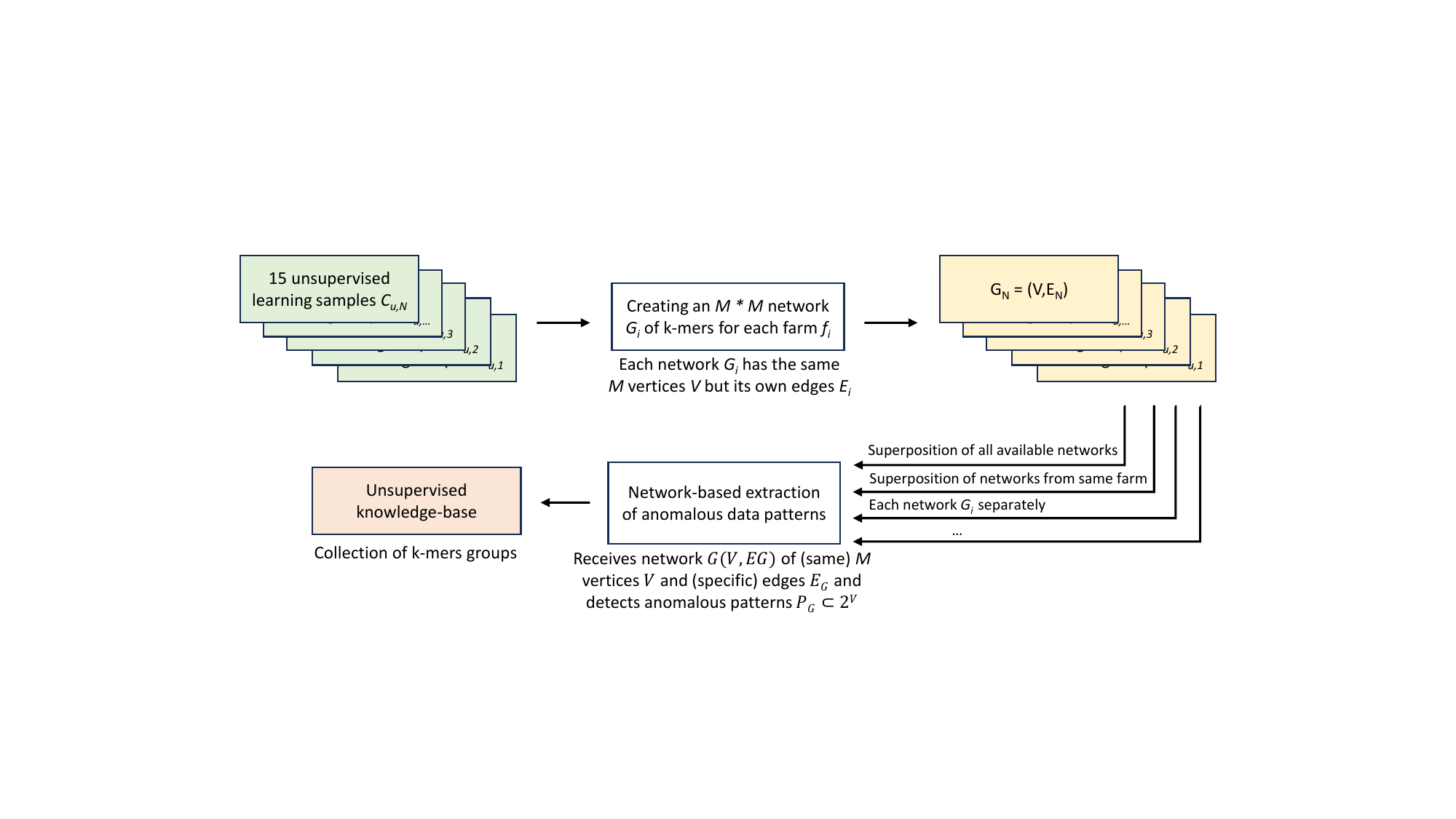}
    \caption{Illustration detailing the unsupervised learning phase. This foundational stage is executed just once, harnessing all accessible microbiome samples. The essence of this phase lies in constructing networks from the genetic material, necessitating varied combinations of these samples. Initially, individual samples stand as solitary data sources to form their respective networks. Subsequently, a comprehensive network is created by pooling samples from an entire farm, resulting in a denser structure that potentially offers a more holistic representation of farm-level features. Multiple other networks can emerge, shaped by diverse criteria like geography, weather conditions, or even an aggregation of all available data. Intriguingly, the formation of these intricate networks is computationally straightforward. By overlaying selected foundational networks, a superposition network is birthed -- akin to executing a simple boolean OR operation on their edges. The establishment of networks specific to each sample paves the way for a nimble and robust integration of fresh data or new samples. It is crucial to emphasize the adaptability of this phase: it is indiscriminate to data sources, allowing for the amalgamation of microbiome samples from diverse entities like cows, sheep, soil, or even humans. Furthermore, its generic nature ensures that identical genetic markers are relevant across varied label groups linked to any targeted biological condition or trait.}
    \label{fig:flowchart2}
\end{figure*}

\begin{figure*}[htbp]
    \centering
    \includegraphics[width=0.6\textwidth,viewport=185 100 765 550]{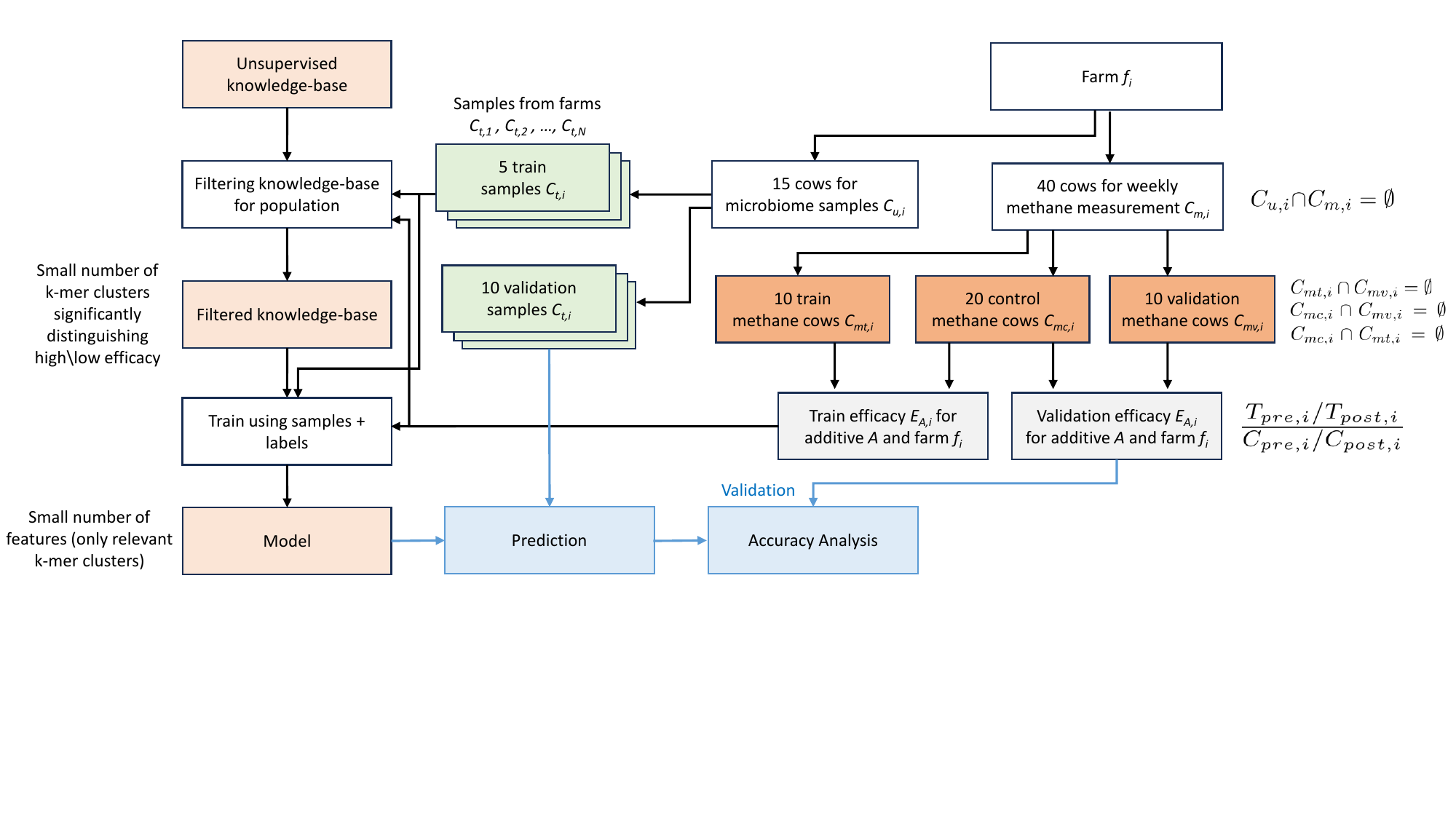}
    \caption{An illustration of the supervised learning phase paired with validation, highlighting the detection of a given additive's expected efficacy. It is important to emphasize the distinction between the microbiome cows, which are used as feature sets, and the methane cows, which contribute to label creation for both training and validation. Additionally, note the deployment of distinct groups of cows for methane measurements during training and validation. The control group is consistently reused to guarantee consistent data normalization. A comprehensive definition of additive efficacy can be found in Section \ref{sec.definitions}.}
    \label{fig:flowchart3}
\end{figure*}

The following figures offer a visual breakdown of our study's flowchart. Figure \ref{fig:flowchart1} outlines the foundational design tailored for one feed additive. This design is reiterated for various additives, with both the control group and the microbiome group being reused. Figure \ref{fig:flowchart2} showcases the unsupervised learning phase. Meanwhile, the supervised learning segment, followed by validation, is illustrated in Figure \ref{fig:flowchart3}.

\subsection{Microbiome Markers used in this Study}
\label{sec.markers}

As highlighted in Section \ref{sec.flow.short} and expounded upon in Section \ref{sec.analytics.detailed}, our proposed AI-driven analytical methodology interprets sequenced microbial data in tandem with corresponding attribute labels, crafting a predictive model applicable for subsequent microbiome samples. Versatile in its design, this approach can formulate a ``microbiome marker'' for any given attribute presented as a label. In the context of this study it is associated with cows exhibiting high efficacy towards a specific feed additive. However, in future works this could equally pertain to other biological attributes such as heightened survival rates against certain diseases and so on. This biomarker comprises two sets of short DNA sequences, their prevalence in microbiome samples serves as a predictor of the target attribute. The first set, termed the ``top list'' (or ``positive list''), features DNA segments indicative of a high likelihood of association with the desired biological condition. Conversely, the ``bottom list'' (or ``negative list'') captures DNA segments that exhibit a low probability of such an association.

The explicit sets of DNA segments identified from the microbiome data and methane measurements used in this study are presented in Tables \ref{tbl.markers.Agolin}, \ref{tbl.markers.Allimax}, \ref{tbl.markers.Kexxtone} and \ref{tbl.markers.Relyon}.
Future studies and commercial projects can leverage these lists to predict the efficacy of the additives evaluated in this study. Given these lists and microbiome samples from cows $c_1, c_2, \ldots$ in a farm $f$ we formally define the prediction score for the efficacy of additive $A$ in $f$ as follows:

\begin{enumerate}
\item For each cow \(c_i\) in farm \(f\) identify the top-1000 most popular k-mers.
\item Compute the score for cow \(c_i\) as:
      \[C_{\text{top}} - C_{\text{bottom}}\]
      where:
    \begin{itemize}
      \item $C_{\text{top}}$ is the ratio of the number of k-mers from the ``top list'' for additive \(A\) present in the top-1000 most popular k-mers for cow \(c_i\) to the total length of the ``top list'' for additive \(A\). This results in values ranging from 0 (no presence in top-1000) to 1 (all k-mers in the top-list are present in top-1000).
      \item $C_{\text{bottom}}$ is defined similarly using the ``bottom list'' for additive $A$.
    \end{itemize}
Consequently, each cow can now have scores between -1 and 1.
\item For farm \(f\) compute the average of its cows' scores. This farm-level score will also range between -1 and 1. To normalize this score, add 1 and then divide by 2, yielding values between 0 (indicative of expected low efficacy) and 1 (indicative of expected high efficacy). Scores around 0.5 suggest insufficient information for prediction.
\end{enumerate}

For the scope of this study it is assumed that both the ``top list'' and ``bottom list'' consist of k-mers of length \(k = 30\). Nonetheless, the analysis detailed in this study can be promptly adapted to encompass k-mers of various lengths. Additionally, k-mers of different lengths that are found to be associated with the efficacy in question can be seamlessly integrated to boost prediction accuracy.

In this study, we opted for a straightforward method to compute both the cow-level and farm-level scores. This decision was made to bolster robustness and minimize the risk of over-fitting. Clearly, the chosen value of $1000$ and the counting method for the k-mers, which disregards their specific rank or absolute popularity, can be substituted with a more refined mechanism. Additionally, this approach could be superseded by advanced machine learning techniques that train models on the presence of identified k-mers, potentially improving predictive accuracy.

\begin{table*}[htbp]
\centering
\begin{tabularx}{\textwidth}{|X|X|}
\toprule
\textbf{Top K-mers} & \textbf{Bottom K-mers}\\
\midrule
ACGTGATCAGTGCATGATCAGTCACGTGAT & AAAGGTACGAAAATTTTAGCTAATCACAAC \\
AGGTGTCGCGCGGCTCAGCTGGCGAGTATC & ACCTTGCAAAGGTACGAAAATTTTAGCTAA \\
AGTATCAGGCAGATGAGCGGGCAGGTGTCG & ACGCGTGGACGCGTGGACGCGTGGACGCGT \\
AGTGCATGATAGCCACGTGATCAGTGCATG & ATAATAATAATAATAATAATAATAATAATA \\
ATAGCCACGTGATCAGTGCATGATCAGTCA & ATGACCTTGCAAAGGTACGAAAATTTTAGC \\
ATCAGTGCATGATAGCCACGTGATCAGTGC & CGTGGACGCGTGGACGCGTGGACGCGTGGA \\
ATCATGCACTGATCACGTGACTGATCATGC & CTTATACACATCTCGAGCCCACGAGACCTA \\
ATCATGCACTGATCACGTGGCTATCATGCA & CTTATACACATCTCGAGCCCACGAGACGCT \\
ATGATAGCCACGTGATCAGTGCATGATCAG & GACGCATGACGCATGACGCATGACGCATGA \\
ATGATCAGTCACGTGATCAGTGCATGATCA & GCCAAGCTGTTCTTGGCGTAAGATGCAATG \\
ATGCACTGATCACGTGGCTATCATGCACTG & GCGTAAGATGCAATGGCTGAGAACTTGACT \\
ATTGGGGATTGGGGATTGGGGATTGGGGAT & GCTGAGAACTTGACTTTCAAGAGTTCTTTT \\
CAGCTGGCGAGTATCAGGCAGATGAGCGGG & GCTGTTCTTGGCGTAAGATGCAATGGCTGA \\
CGCGGCTCAGCTGGCGAGTATCAGGCAGAT & GTTCTTGGCGTAAGATGCAATGGCTGAGAA \\
CGTGATCAGTGCATGATAGCCACGTGATCA & GTTGAGAGTTGAGAGTTGAGAGTTGAGAGT \\
CTCATCTGCCTGATACTCGCCAGCTGAGCC & GTTGATGACCTTGCAAAGGTACGAAAATTT \\
GCATGATCAGCCACGTGATCAGTGCATGAT & TAAGATGCAATGGCTGAGAACTTGACTTTC \\
GCGAGTATCAGGCAGATGAGCGGGCAGGTG & TAGGCCAAGCTGTTCTTGGCGTAAGATGCA \\
GCTCAGCTGGCGAGTATCAGGCAGATGAGC & TCATGCGTCATGCGTCATGCGTCATGCGTC \\
GGCAGGTGTCGCGCGGCTCAGCTGGCGAGT & TCTCTTATACACATCTACGCTGCCGACGAC \\
GGGATTGGGGATTGGGGATTGGGGATTGGG & TCTTATACACATCTCCAGCCCACGAGACTT \\
GTGCATGATCAGTCACGTGATCAGTGCATG & TCTTATACACATCTCGAGCCCACGAGACTT \\
GTGTGTGTGTGTGTGTGTGTGTGTGTGTGT & TCTTATACACATCTTGACGCTGCCGACGAC \\
TCATGCACTGATCACGTGGCTGATCATGCA & TGCAAAGGTACGAAAATTTTAGCTAATCAC \\
TGTCGCGCGGCTCAGCTGGCGAGTATCAGG & TGCAATGGCTGAGAACTTGACTTTCAAGAG \\
 & TGTCAAGCGGCAACCGATCGGTTACGCTGA \\
 & TTATCTCATTGCTTTTCACCTCACACATTT \\
 & TTCAAGAGTTCTTTTCTCTTTCTGATTGCC \\
 & TTCACCTCACACATTTCAGTGTCAAGCGGC \\
 & TTCAGTGTCAAGCGGCAACCGATCGGTTAC \\
 & TTGACTTTCAAGAGTTCTTTTCTCTTTCTG \\
 & TTGCTTTTCACCTCACACATTTCAGTGTCA \\
 & TTGGCGTAAGATGCAATGGCTGAGAACTTG \\
\bottomrule
\end{tabularx}
\caption{Top and Bottom k-mers markers for feed additive Agolin. See more details in Section \ref{sec.markers}}
\label{tbl.markers.Agolin}
\end{table*}

\begin{table*}[htbp]
\centering
\begin{tabularx}{\textwidth}{|X|X|}
\toprule
\textbf{Top K-mers} & \textbf{Bottom K-mers}\\
\midrule
AAACATGGGCAGGCCTATGAAACCCACCGC & AAAATTAGATAAATTTAAAGAAGTTAAAGA \\
AAAGAGAGGTGAGAAACATGGGCAGGCCTA & AACATTATTAGTATTAAAATTAGATAAATT \\
AAATTAATGTTTATATATGTTAAATTAATG & AATAATAATAATAATAATAATAATAATAAT \\
AACGCTGTACAAGAAGCGCCTGAACACCGA & AATCCCCAATCCCCAAAACCCAAAACCCAA \\
ACGCATGACGCATGACGCATGACGCATGAC & AATGGGGATTGGGGATTGGGGATTGGGGAT \\
ATATGTTAAATTAATGTTTATATATGTTAA & AATTGGGGATTGGGGATTGGGGATTGGGGA \\
ATGACGCATGACGCATGACGCATGACGCAT & AGATAAATTTAAAGAAGTTAAAGAAGAACA \\
ATGCGTCATGCGTCATGCGTCATGCGTCAT & AGTATTAAAATTAGATAAATTTAAAGAAGT \\
ATGGGCAGGCCTATGAAACCCACCGCAGTC & ATTAAAATTAGATAAATTTAAAGAAGTTAA \\
CAGGCCTATGAAACCCACCGCAGTCAAGAA & ATTAGATAAATTTAAAGAAGTTAAAGAAGA \\
CCAGACCCTCAGCGACATCGGAACGACCGC & ATTAGTATTAAAATTAGATAAATTTAAAGA \\
CGTCATGCGTCATGCGTCATGCGTCATGCG & ATTATTAGTATTAAAATTAGATAAATTTAA \\
CTCTGCTCTGCTCTGCTCTGCTCTGCTCTG & ATTATTATTATTATTATTATTATTATTATT \\
CTCTTATACACATCTCGAGCCCACGAGACA & ATTGGGCCCAATCCCCAATCCCCAAACCCC \\
GAGAAACATGGGCAGGCCTATGAAACCCAC & ATTGGGGATTGGGGATTGGGGATTGGGCCC \\
GGTGAGAAACATGGGCAGGCCTATGAAACC & CCAATCCCCAAAACCCAAAACCCCAAACCC \\
GTTAAATTAATGTTTATATATGTTAAATTA & CCAATCCCCAATCCCCAATACCCAAAACCC \\
TCTTTTTCTTTTTCTTTTTCTTTTTCTTTT & CCAATCCCCAATCCCCAATCCCCAATCCCC \\
TTTTCTTTTTCTTTTTCTTTTTCTTTTTCT & CCCAATCCCCAATCCCCAAAACCCAAAACC \\
 & GATTGGGGATTGGGGATTGGGGATTGGGGG \\
 & GGGATTGGGGAGTGGGGATTGGGGATTGGG \\
 & GGGGATTGGGGATTGGGGATTGGGGATTGG \\
 & TAAATTTAAAGAAGTTAAAGAAGAACAATT \\
 & TCCCCAATCCCCAATCCCCAATCCCCATTA \\
 & TCTTATACACATCTCGAGCCCACGAGACGA \\
 & TGGGGATTGGGGATTGGGGAGTGGGGATTG \\
 & TTGGGGATTGGGGAGTGGGGATTGGGGATT \\
 & TTGGGGATTGGGGATTGGGGATTGGGGCCA \\
\bottomrule
\end{tabularx}
\caption{Top and Bottom k-mers markers for feed additive Allimax. See more details in Section \ref{sec.markers}}
\label{tbl.markers.Allimax}
\end{table*}

\begin{table*}[htbp]
\centering
\begin{tabularx}{\textwidth}{|X|X|}
\toprule
\textbf{Top K-mers} & \textbf{Bottom K-mers}\\
\midrule
AAACACCATATATATTGAGAAAGAGAGGTG & AAACGCCTCAGGAGGCTTGACTCCCTTGAG \\
AAACATGGGCAGGCCTATGAAACCCACCGC & AAGAAGAAGAAGAAGAAGAAGAAGAAGAAG \\
AAATTAATGTTTATATATGTTAAATTAATG & AGGTACGACGGCGAGGTCAGTGAGCCTCTC \\
AAATTTAAAGAAGTTAAAGAAGAACAATTA & AGTGCATGATAGCCACGTGATCAGTGCATG \\
AAATTTATCTAATTTTAATACTAATAATGT & ATCAGTGCATGATAGCCACGTGATCAGTGC \\
AACTTCTTTAAATTTATCTAATTTTAATAC & ATCATGCACTGATCACGTGACTGATCATGC \\
AAGAAGAAGAAGAAGAAGAAGTTGAACATG & ATCTCGCGACCTCTCTCCAAACGCCTCAGG \\
AAGAAGAAGAAGAAGAAGTTGAACATGAAG & ATGCACTGATCACGTGGCTGATCATGCACT \\
AATTTTAATACTAATAATGTTAATAATATG & CACACACACACACACACACACACACACACA \\
AATTTTAATACTAATAATGTTACTGATATG & CAGGAGGCTTGACTCCCTTGAGTCCACCCA \\
ACACTAAACACCATATATATTGAGAAAGAG & CATGATAGCCACGTGATCAGTGCATGATCA \\
ACCATATATATTGAGAAAGAGAGGTGAGAA & CCTCAGGAGGCTTGACTCCCTTGAGTCCAC \\
AGATAAATTTAAAGAAGTTAAAGAAGAACA & CCTCTCTCCAAACGCCTCAGGAGGCTTGAC \\
AGGCCTATGAAACCCACCGCAGTCAAGAAG & CGGCTCGGCTCGGCTCGGCTCGGCTCGGCT \\
ATAAATGGGGATTGGGGATTGGGGATTGGG & CGTGATCAGTGCATGATAGCCACGTGATCA \\
ATCTAATTTTAATACTAATAATGTTAATAA & CTCCAAACGCCTCAGGAGGCTTGACTCCCT \\
ATGCGTCATGCGTCATGCGTCATGCGTCAT & CTGATCATGCACTGATCACGTGGCTATCAT \\
ATGGGGATTGGGGATTGGGGATTGGGGATT & GCGACCTCTCTCCAAACGCCTCAGGAGGCT \\
ATTATTATTATTATTATTATTATTATTATT & GTCCACCCAGTGAGCTCCAAGAGATACCCG \\
CCAATCCCCAATCCCCAATCCCCAATCCCC & TCTTATACACATCTCGAGCCCACGAGACTC \\
CGTCATGCGTCATGCGTCATGCGTCATGCG &  \\
CTCTGCTCTGCTCTGCTCTGCTCTGCTCTG &  \\
CTCTTATACACATCTCGAGCCCACGAGACG &  \\
CTTATACACATCTCGAGCCCACGAGACAAC &  \\
CTTATACACATCTCGAGCCCACGAGACTGT &  \\
CTTTAAATTTATCTAATTTTAATACTAATA &  \\
CTTTAACTTCTTTAAATTTATCTAATTTTA &  \\
GATTGGGGATTGGGGATTGGGGATTGGGGA &  \\
TTTATCTAATTTTAATACTAATAATGTTAA &  \\
\bottomrule
\end{tabularx}
\caption{Top and Bottom k-mers markers for feed additive Kexxtone. See more details in Section \ref{sec.markers}}
\label{tbl.markers.Kexxtone}
\end{table*}

\begin{table*}[htbp]
\centering
\begin{tabularx}{\textwidth}{|X|X|}
\toprule
\textbf{Top K-mers} & \textbf{Bottom K-mers}\\
\midrule
AATCATGCTGCTCAGCTGGCAATAATCAAG & AATACCCAAAACCCAAAACCCAAAACCCAA \\
AATCTTCCATTCGAGTTGCGAAGGAAAGCT & AATCCCCAATCCCCAAAACCCAAAACCCCA \\
ACACACACACACACACACACACACACACAC & AATTTTAATACTAATAATGTAACTAATATG \\
ACCTGCCCGCTCATCTGCCTGATACTCGCC & AGAGCAGAGCAGAGCAGAGCAGAGCAGAGC \\
ACGTGATCAGTGCATGATCAGTCACGTGAT & ATAATAATAATAATAATAATAATAATAATA \\
ACTGACCTCGCCGTCGTACCTCGTGAGAAA & ATATTAGTTACATTATTAGTATTAAAATTA \\
AGGTGTCGCGCGGCTCAGCTGGCGAGTATC & ATTATTATTATTATTATTATTATTATTATT \\
AGTATCAGGCAGATGAGCGGGCAGGTGTCG & ATTGGGCCCAATCCCCAATCCCCAAACCCC \\
AGTGCATGATAGCCACGTGATCAGTGCATG & ATTGGGGATTGGGGATTGGGGAGTGGGGAT \\
ATAGCCACGTGATCAGTGCATGATCAGTCA & CAATCCCCAAAACCCAAAACCCCAAACCCC \\
ATCACGTGACTGATCATGCACTGATCACGT & CACTGACTGCAGTGATAACACTGACTGCAG \\
ATCAGGCAGATGAGCGGGCAGGTGTCGCGC & CCAATCCCCAATCCCCAAACCCCAATCCCC \\
ATCAGTGCATGATAGCCACGTGATCAGTGC & CCAATCCCCAATCCCCAAACCCCCAAACCC \\
ATCATGCACTGATCACGTGACTGATCATGC & CCAATCCCCAATCCCCAATACCCAAAACCC \\
ATCATGCACTGATCACGTGGCTATCATGCA & CCCAATCCCCAATCCCCAATCCCCAATACC \\
ATCATGCACTGATCACGTGGCTGATCATAC & CCCCAATCCCCAAAACCCAAAACCCCAAAC \\
ATGATAGCCACGTGATCAGTGCATGATCAG & CCCCAATCCCCAATCCCCAAAACCCAAAAC \\
ATGATCAGTCACGTGATCAGTGCATGATCA & CTTATACACATCTCGAGCCCACGAGACACT \\
ATGCACTGATCACGTGGCTATCATGCACTG & CTTATACACATCTCGAGCCCACGAGACCTA \\
ATGCACTGATCACGTGGCTGATCATACACT & CTTATACACATCTCGAGCCCACGAGACGCT \\
CAGCTGGCGAGTATCAGGCAGATGAGCGGG & GACTGCAGTGATAACACTGACTGCAGTGAT \\
CATGATCAGTCACGTGATCTGTGCATGATC & GATAACACTGACTGCAGTGATAACACTGAC \\
CGCGGCTCAGCTGGCGAGTATCAGGCAGAT & GATTGGGGATTGGGGAGTGGGGATTGGGGA \\
CGTGATCAGTGCATGATAGCCACGTGATCA & GGGATTGGGGATTGGGGATTGGGGAGTGGG \\
CTCATCTGCCTGATACTCGCCAGCTGAGCC & GGGGATTGGGGATTGGGGAGTGGGGATTGG \\
CTGATCATGCACTGATCACGTGGCTATCAT & TATTGGGGATTGGGGATTGGGGATTGGGGA \\
CTTCCATTCGAGTTGCGAAGGAAAGCTGGG & TGCTTGCTTGCTTGCTTGCTTGCTTGCTTG \\
GCATGATCAGCCACGTGATCAGTGCATGAT & TTGGGGAGTGGGGATTGGGGATTGGGGATT \\
GCGAGTATCAGGCAGATGAGCGGGCAGGTG &  \\
GCTCAGCTGGCGAGTATCAGGCAGATGAGC &  \\
GGCAGGTGTCGCGCGGCTCAGCTGGCGAGT &  \\
GTGCATGATCAGTCACGTGATCTGTGCATG &  \\
GTGTGTGTGTGTGTGTGTGTGTGTGTGTGT &  \\
TCATGCACTGATCACGTGGCTGATCATGCA &  \\
TGCATGATCAGTCACGTGATCAGTGCATGA &  \\
TGTCGCGCGGCTCAGCTGGCGAGTATCAGG &  \\
\bottomrule
\end{tabularx}
\caption{Top and Bottom k-mers markers for feed additive Relyon. See more details in Section \ref{sec.markers}}
\label{tbl.markers.Relyon}
\end{table*}

\section{Results}

For each feed additive $A$ and for each farm $f_i \in F_A$, the microbiome samples from $C_{v,i}$ were used to predict the additive's efficacy. This predicted efficacy was then contrasted with the actual efficacy determined through the analysis of methane emissions from $C_{mc,i}$ and $C_{mv,i}$.

This design allows for general applicability to different additives and use cases, with potential for synergistic improvement as more data is added to the unsupervised learning stage. Importantly, the division of cows into various groups is done in a way that reduces bias for factors like age of cows, their days in lactation, and average milk yield.

\subsection{Additive Efficacy Measurements}

The following Tables \ref{tbl.rawAdditiveAgolin}, \ref{tbl.rawAdditiveAllimax}, \ref{tbl.rawAdditiveKexxtone}, and \ref{tbl.rawAdditiveRelyon} provide a comprehensive summary of the results obtained from measuring methane emissions across various farms and for different feed additives, as measured for the test group of cows $C_{mv,i}$, and normalized by the control group $C_{mc,i}$. Specifically, each table represents one unique additive. The columns in the table are as follows:

\begin{itemize}
  \item \textbf{Farm}: Identifier for each farm where measurements were taken.
  \item \textbf{$CH_4$ Change Treatment}: This column displays the mean percentage change in methane emissions for cows that received the feed additive. For example, a value of 0 indicates no change compared to the values taken prior to the initiation of the trial, -10 indicates a 10\% reduction, and 20 indicates a 20\% increase in emissions.
  \item \textbf{$CH_4$ Change Control}: This column shows the mean percentage change in methane emissions for cows that did not receive the feed additive, serving as the control group.
  \item \textbf{Normalized Efficacy $\eta_{A, f_i}$}: Reflecting the normalized mean percentage change in methane emissions, accounting for variations in control and treatment groups (see formal definition in Section \ref{sec.definitions}).
  \item \textbf{Effect Size}: This is a statistical measure that quantifies the size of the difference between the two groups, widely used to assess medical and nutritional treatment efficacy \cite{hess2004robust}.
  \item \textbf{Cohen's D}: A Statistical measure of effect size, indicating the standardized difference between the means in units of standard deviation \cite{rosenthal1994parametric}.
\end{itemize}

\begin{table*}[htbp]
\centering
\begin{tabularx}{\textwidth}{lXXXXX}
\toprule
Farm & $CH_4$ Change (Treatment) & $CH_4$ Change (Control) & Normalized Efficacy $\eta_{A, f_i}$ & Effect Size & Cohen's D \\ \midrule
AB1 & -46.8\% & -41.2\% & -9.5\% & -8.3 & -0.10 \\
BT1 & 4.5\% & 20.7\% & -13.4\% & -20.6 & -0.18 \\
FG1 & -56.6\% & -49.1\% & -14.8\% & -11.6 & -0.23 \\
GR1 & 121.2\% & 155.5\% & -13.4\% & -47.5 & -0.11 \\
LV1 & 18.9\% & 22.3\% & -2.8\% & -9.5 & -0.04 \\
MP1 & -58.6\% & -48.3\% & -19.9\% & -20.5 & -0.23 \\
RZ1 & -27.4\% & -28.2\% & 1.1\% & 0.8 & 0.14 \\
SI1 & -38.7\% & -25.5\% & -17.7\% & -13.7 & -0.30 \\
YE1 & -26.0\% & -24.3\% & -2.3\% & -2.0 & -0.03 \\
JN2 & 102.9\% & 100.7\% & 1.1\% & -2.3 & -0.01 \\
ST2 & 59.6\% & 73.0\% & -7.7\% & -15.8 & -0.08 \\
TS2 & 32.3\% & 42.1\% & -6.9\% & -12.8 & -0.08 \\
YK2 & -19.5\% & -19.4\% & -0.1\% & -0.5 & -0.00 \\
\bottomrule
\end{tabularx}
\caption{Summary of methane emission changes across farms for the additive Agolin. See a detailed explanation above for full interpretation of the columns.}
\label{tbl.rawAdditiveAgolin}
\end{table*}

\begin{table*}[htbp]
\centering
\begin{tabularx}{\textwidth}{lXXXXX}
\toprule
Farm & $CH_4$ Change (Treatment) & $CH_4$ Change (Control) & Normalized Efficacy $\eta_{A, f_i}$ & Effect Size & Cohen's D \\ \midrule
LV1 & -27.9\% & 22.3\% & -41.1\% & -224.4 & -0.76 \\
YE1 & -41.6\% & -24.3\% & -22.9\% & -21.2 & -0.28 \\
BT2 & -24.4\% & -14.7\% & -11.3\% & -12.2 & -0.16 \\
FG2 & 78.7\% & 155.6\% & -30.1\% & -84.4 & -0.33 \\
SI2 & 161.0\% & 163.0\% & -0.8\% & -0.7 & -0.01 \\
AB3 & -52.1\% & -36.3\% & -24.7\% & -24.9 & -0.32 \\
JN3 & -10.4\% & 16.5\% & -23.1\% & -27.5 & -0.75 \\
KS3 & 0.3\% & 34.6\% & -25.5\% & -67.3 & -0.26 \\
LV3 & 2.2\% & 47.8\% & -30.9\% & -60.9 & -0.52 \\
MP3 & -61.0\% & -63.2\% & 6.0\% & 2.4 & 0.11 \\
VL3 & -26.8\% & 19.6\% & -38.8\% & -71.0 & -0.55 \\
\bottomrule
\end{tabularx}
\caption{Summary of methane emission changes across farms for the additive Allimax. See a detailed explanation above for full interpretation of the columns.}
\label{tbl.rawAdditiveAllimax}
\end{table*}

\begin{table*}[htbp]
\centering
\begin{tabularx}{\textwidth}{lXXXXX}
\toprule
Farm & $CH_4$ Change (Treatment) & $CH_4$ Change (Control) & Normalized Efficacy $\eta_{A, f_i}$ & Effect Size & Cohen's D \\ \midrule
AB2 & -46.2\% & -11.0\% & -39.6\% & -45.4 & -0.62 \\
JN2 & 30.1\% & 100.7\% & -35.2\% & -73.0 & -0.46 \\
KS2 & 106.3\% & 94.8\% & 5.9\% & 21.2 & 0.05 \\
LV2 & 21.2\% & 66.6\% & -27.2\% & -47.0 & -0.66 \\
MP2 & -50.6\% & -39.7\% & -18.2\% & -21.0 & -0.18 \\
RZ2 & -62.2\% & -56.2\% & -13.8\% & -6.9 & -0.14 \\
ST2 & 24.9\% & 73.0\% & -27.8\% & -57.7 & -0.29 \\
TS2 & 3.7\% & 42.1\% & -27.0\% & -48.4 & -0.30 \\
BT3 & -20.7\% & -7.1\% & -14.7\% & -19.5 & -0.15 \\
FG3 & -44.4\% & -43.7\% & -1.3\% & -0.9 & -0.04 \\
SI3 & -43.9\% & -3.3\% & -42.0\% & -90.5 & -0.58 \\
SR3 & -47.7\% & -48.1\% & 0.7\% & 0.2 & 0.01 \\
VL3 & -5.1\% & 19.6\% & -20.6\% & -34.7 & -0.33 \\
YE3 & -10.2\% & 2.9\% & -12.7\% & -14.2 & -0.27 \\
\bottomrule
\end{tabularx}
\caption{Summary of methane emission changes across farms for the additive Kexxtone. See a detailed explanation above for full interpretation of the columns.}
\label{tbl.rawAdditiveKexxtone}
\end{table*}

\begin{table*}[htbp]
\centering
\begin{tabularx}{\textwidth}{lXXXXX}
\toprule
Farm & $CH_4$ Change (Treatment) & $CH_4$ Change (Control) & Normalized Efficacy $\eta_{A, f_i}$ & Effect Size & Cohen's D \\ \midrule
AB1 & -47.3\% & -41.2\% & -10.4\% & -8.5 & -0.11 \\
GR1 & 108.7\% & 155.5\% & -18.3\% & -64.9 & -0.17 \\
JN1 & -38.3\% & -36.5\% & -2.8\% & -2.6 & -0.04 \\
KS1 & 86.8\% & 120.1\% & -15.1\% & -45.9 & -0.11 \\
LV1 & 17.4\% & 22.3\% & -4.0\% & -12.6 & -0.04 \\
MP1 & -67.8\% & -48.3\% & -37.8\% & -40.2 & -0.56 \\
YE1 & -25.1\% & -24.3\% & -1.0\% & -0.9 & -0.01 \\
KS2 & 71.8\% & 94.8\% & -11.8\% & -45.6 & -0.14 \\
AB3 & -41.5\% & -36.3\% & -8.0\% & -7.6 & -0.05 \\
JN3 & 8.7\% & 16.5\% & -6.7\% & -8.7 & -0.25 \\
LV3 & 12.0\% & 47.8\% & -24.2\% & -46.3 & -0.43 \\
YE3 & -13.6\% & 2.9\% & -16.1\% & -17.6 & -0.35 \\
\bottomrule
\end{tabularx}
\caption{Summary of methane emission changes across farms for the additive Relyon. See a detailed explanation above for full interpretation of the columns.}
\label{tbl.rawAdditiveRelyon}
\end{table*}

\begin{figure*}[htbp]
  \centering
  \includegraphics[width=\textwidth]{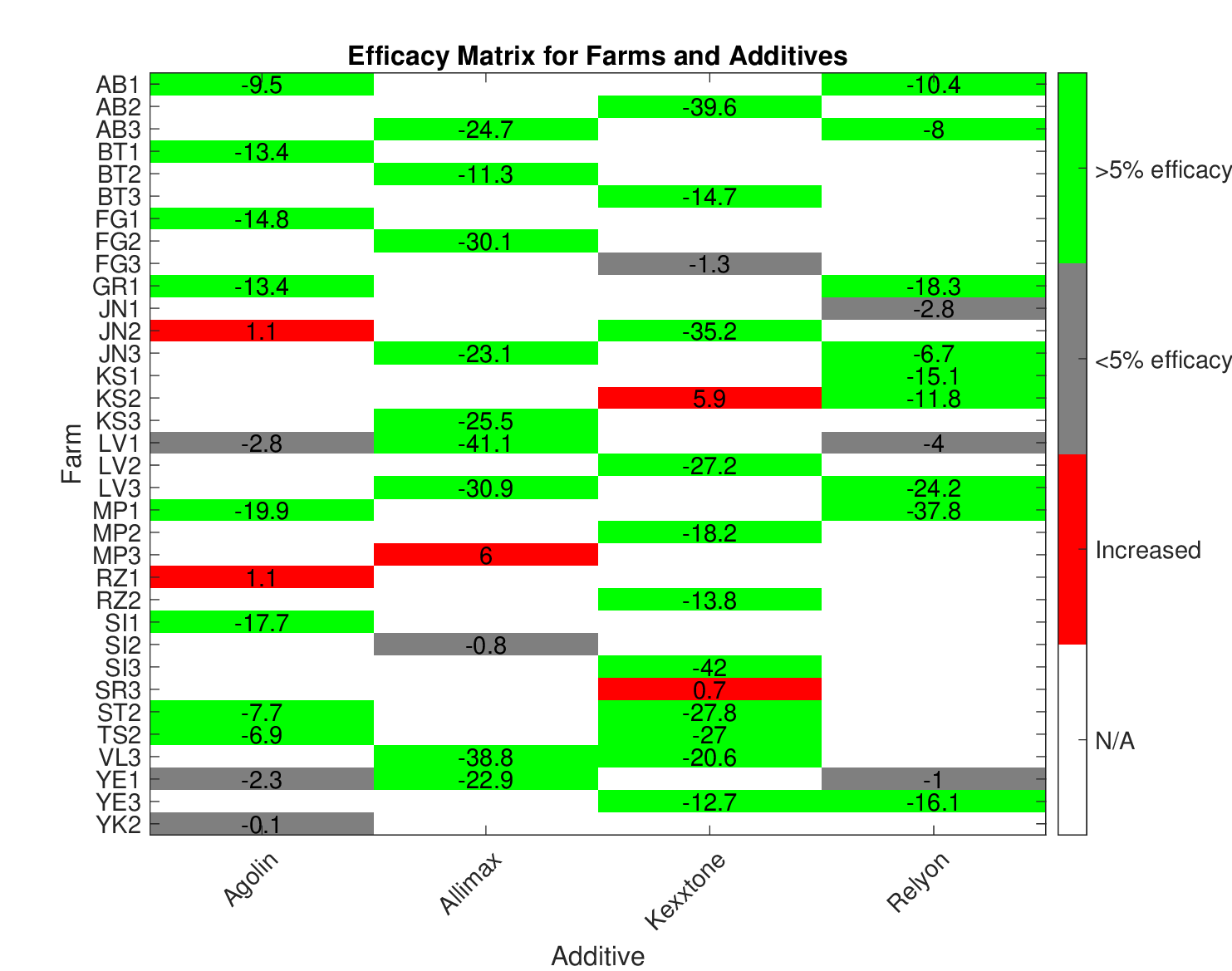}
  \caption{Efficacy matrix of various feed additives across multiple farms, illustrating how different feed additives affect methane emissions in a variety of farms. Each cell in the matrix represents the mean change in methane emissions, in percentage terms, for a particular farm-additive combination. The numbers are compared to the emissions level measured before the trial began, and are normalized by the respected change of the control group of the same farm. For instance, a value of '0' indicates no change in emmissions (or the same change that the control cows underwent), '-10' indicates a 10\% normalized reduction, and '20' signifies a 20\% increase. The color coding further aids interpretation: green cells indicate significant reductions in emissions (below -5\%), red cells highlight increases, and gray cells show negligible change (between -5\% and 0). Notably, each feed additive has a varying level of efficacy across different farms. There are instances where a single additive either fails to lower emissions or even increases them in a subset of farms. This variation underscores the necessity for a nuanced approach in methane reduction strategies, as a one-size-fits-all solution may be ineffective or counterproductive.}
  \label{fig.efficacyMatrix}
\end{figure*}

In evaluating the effectiveness of various feed additives for reducing methane emissions across multiple farms, we observe a distinct variability in efficacy. As illustrated in an efficacy matrix (see Figure \ref{fig.efficacyMatrix}), each additive performs differently depending on the farm where it is applied. Notably, for every additive, there are at least a few farms where it either fails to reduce emissions or even exacerbates them. Similarly, the effectiveness of additives varies within individual farms, underscoring the complexity of methane reduction strategies and suggesting that a 'one-size-fits-all' approach may not be viable. This variability also highlights the economic and business challenges associated with the adoption of additives. Negative or non-existent efficacy, even if relatively rare, may discourage farmers from incorporating additives into their practices.

\begin{figure*}[htbp]
  \centering
  \includegraphics[width=\textwidth]{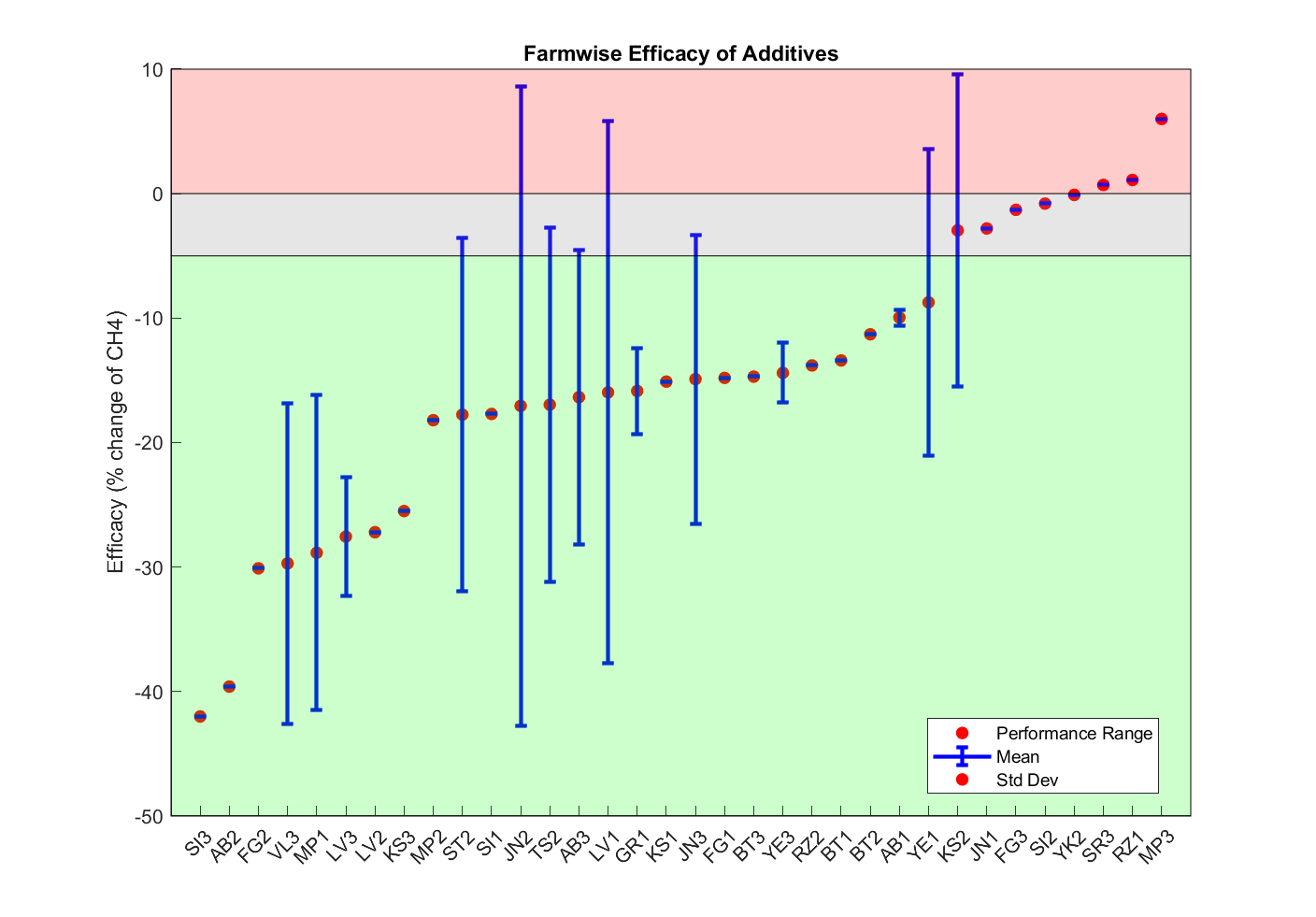}
  \caption{Efficacy bar chart of the farms across multiple feed additives.
    Each ``stick'' represents a single farm, extending from its lowest to highest additive performance. The mean efficacy is marked by a red dot, and the range of standard deviation is depicted by blue error bars. The chart is color-coded to facilitate interpretation: the green zone (below -5\%) represents significant reductions in methane emissions, the gray zone (between -5\% and 0\%) indicates negligible changes, and the red zone (above 0\%) highlights increases in emissions. While a cursory look at the data might suggest that most of the additive-farm combinations result in positive efficacy (as indicated by the prevalence of data in the green zone), a closer examination reveals a complex picture. Although the bulk of the data points indicate reduced methane emissions, the presence of occasional inefficacies (or even increases in emissions) significantly skews the overall performance. This is particularly evident when considering the upper range of efficacy (upper end of the error bars), which often falls within the gray or even red zones for many farms. This volatile efficacy performance underscores the risks associated with a non-strategic or arbitrary choice of additives, potentially explaining the hesitancy among farmers to adopt them.}
  \label{fig.efficacyBar}
\end{figure*}

Figure \ref{fig.efficacyBar} illustrates the variability in the efficacy of different additives across multiple farms. While a majority of the additives generally demonstrate positive efficacy—reducing methane emissions by at least 5\% -- the data also reveals cases where the additives either have a negligible impact or paradoxically even increase emissions. This high volatility in efficacy at the farm level suggests that farmers who lack a rigorous selection methodology for additives are at a greater risk of experiencing poor outcomes. This variability can deter farmers from adopting additives, as a small number of poor matches can significantly undermine overall performance and satisfaction.

\subsection{Optimized Additive Deployment}

The following figures present the improvements in feed additive efficacy achieved using our proposed microbiome-based, AI-assisted predictive model. These improvements have significant potential economic implications by enabling more targeted and efficient use of additives. Such targeted approach not only maximizes methane emission reduction but also optimizes resource allocation, thereby offering a compelling value proposition that could accelerate the widespread adoption of sustainable farming practices.

\begin{figure*}[htbp]
    \centering
    \subfloat[Relyon]{\includegraphics[width=0.5\textwidth]{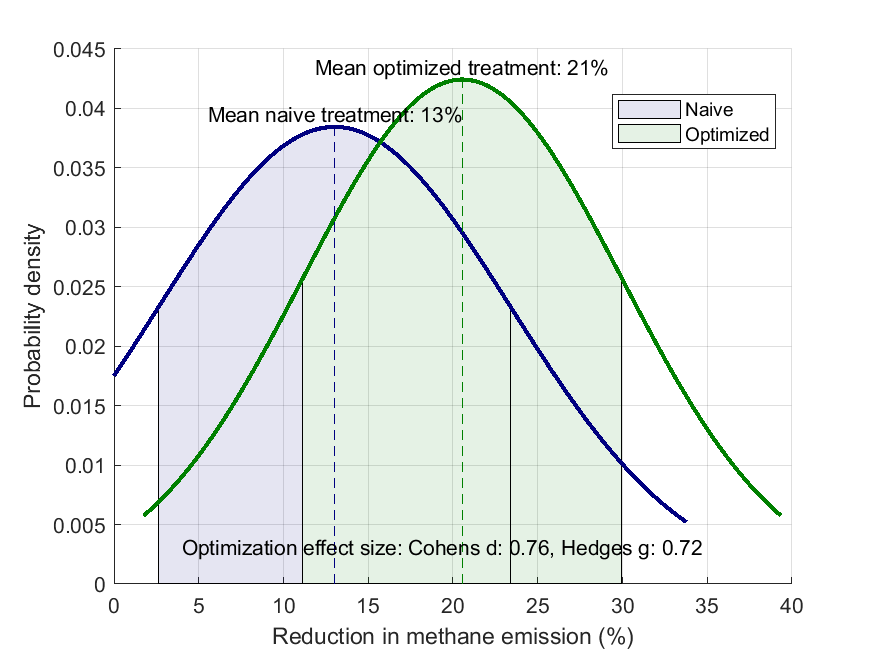}\label{fig:dist_relyon}}
    \hfill
    \subfloat[Agolin]{\includegraphics[width=0.5\textwidth]{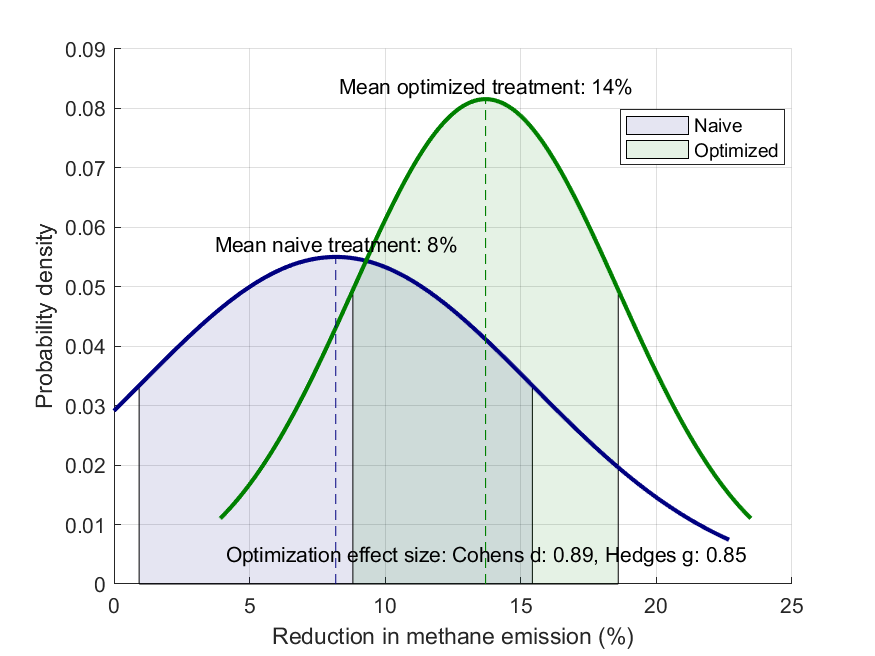}\label{fig:dist_agolin}}
    \\
    \subfloat[Allimax]{\includegraphics[width=0.5\textwidth]{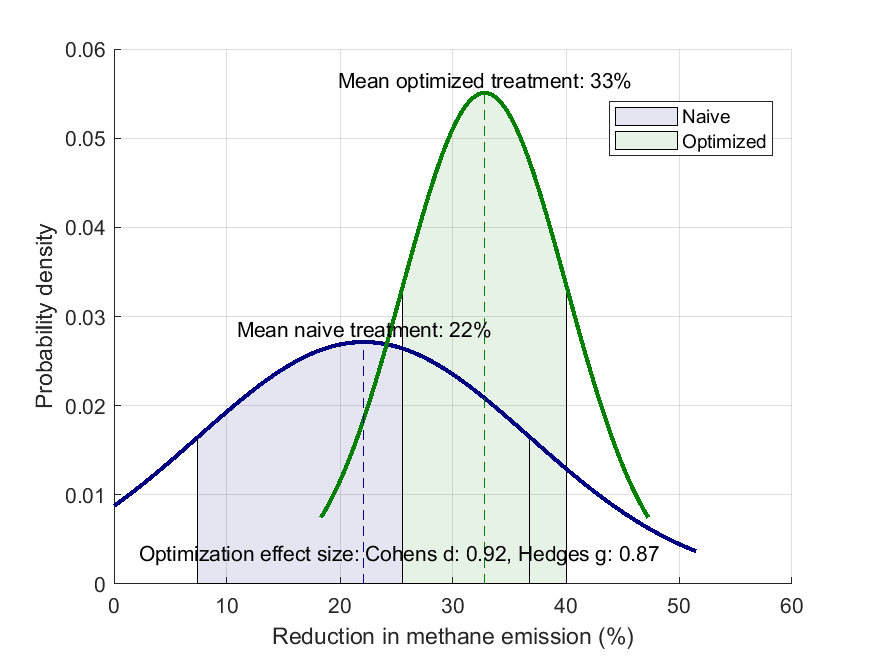}\label{fig:dist_allimax}}
    \hfill
    \subfloat[Kexxtone]{\includegraphics[width=0.5\textwidth]{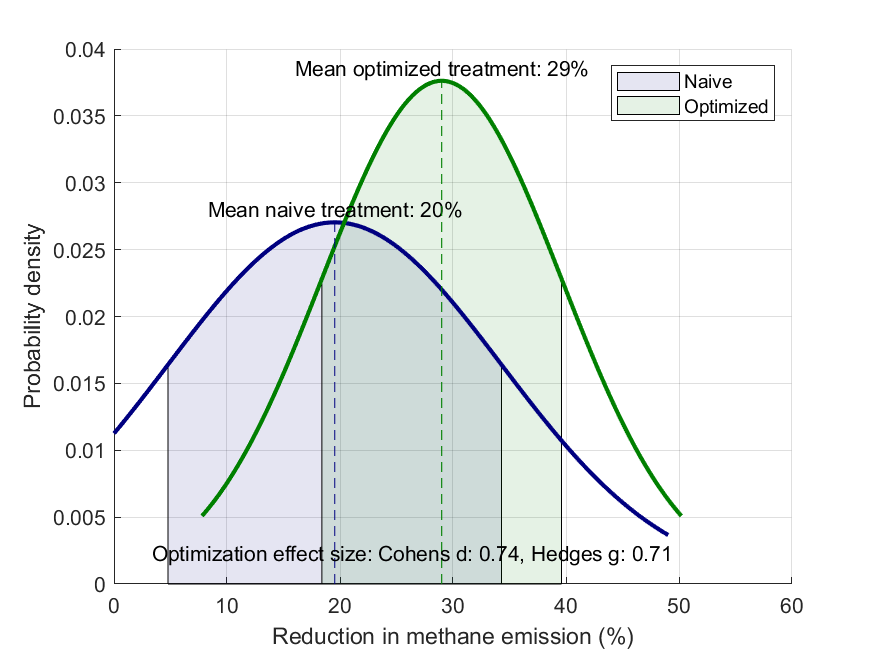}\label{fig:dist_kexxtone}}
    \caption{An illustration of the normalized efficacy distribution for the four feed additives measured in this study: Relyon, Agolin, Allimax, and Kexxtone. The data is based on Tables \ref{tbl.rawAdditiveAgolin}, \ref{tbl.rawAdditiveAllimax}, \ref{tbl.rawAdditiveKexxtone}, and \ref{tbl.rawAdditiveRelyon} and has been regressed to fit a Normal distribution. Each subfigure presents two distributions: one depicting the raw data from all farms (Dark Blue, denoted as `Naive'), and the other (Green, denoted as `Optimized') showing efficacy across the top 50\% of farms as predicted by our microbiome-based model. Included in each chart are the Cohen's \(d\) \cite{hess2004robust} and Hedge's \(g\) \cite{hedges1981distribution} metrics, indicating strong statistical significance of the  observed effects under optimized conditions.}
    \label{fig.efficacy.distributions}
\end{figure*}

Figure \ref{fig.efficacy.distributions} presents the efficacy analysis of the four additives examined in this study, as derived from the data in Tables \ref{tbl.rawAdditiveAgolin}, \ref{tbl.rawAdditiveAllimax}, \ref{tbl.rawAdditiveKexxtone}, and \ref{tbl.rawAdditiveRelyon}. The efficacy is initially represented as a Normal distribution under naive deployment conditions, without farm selection (denoted as ``Naive Deployment''). This is contrasted with an optimized deployment strategy where each additive is applied to only 50\% of the farms, specifically selected based on our microbiome-based predictive model (denoted as ``Optimized Deployment''). The comparison reveals that the optimized approach substantially improves additive efficacy by targeting farms where the highest impact is expected. This leads to an approximate 60\% increase in the effectiveness of the additives in reducing methane emissions.

\begin{figure*}[htbp]
  \centering
  \includegraphics[width=\textwidth]{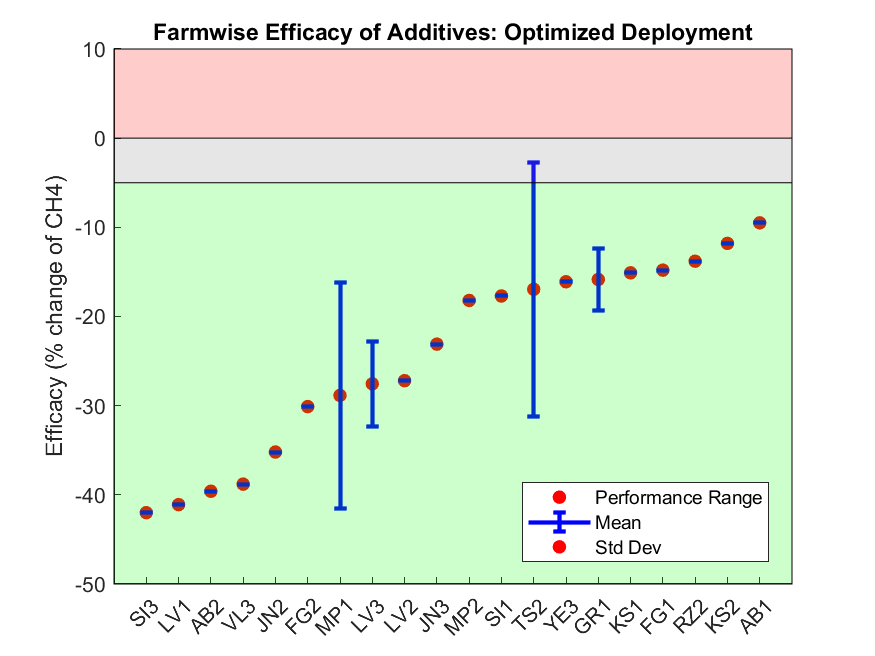}
  \caption{Efficacy bar chart of the farms across multiple feed additives, emphasizing the advantages of our microbiome-based, AI-assisted efficacy prediction. Each `stick' represents a farm's performance range with different additives, extending from the lowest to the highest efficacy. The mean efficacy is highlighted by a red dot, and the standard deviation range is shown as blue error bars. The background color-coding aids in interpretation: the green zone (below -5\%) signifies substantial reductions in methane emissions, the gray zone (-5\% to 0\%) suggests negligible impact, and the red zone (above 0\%) indicates increases in emissions. Contrasted with the findings in Figure \ref{fig.efficacyBar}, the advantages of targeted optimization are clear. With the exception of a single farm, all data points are predominantly located within the green zone, indicating that farmers who employ this optimized strategy are likely to experience consistently positive results.}
  \label{fig.efficacyBar_optimized}
\end{figure*}

Figure \ref{fig.efficacyBar_optimized} provides a complementary analysis to Figure \ref{fig.efficacyBar}, incorporating the optimization phase based on our microbiome-based predictive model. Observing this Figure it can be seen that not only does the optimized approach enhance the average additive performance by approximately 60\%, but it also fundamentally alters the experience for farmers by shifting from a pattern of mixed successes and failures to a consistently positive performance profile. In other words, the targeted deployment avoids instances where additives could yield poor or even detrimental outcomes. This transformative impact is likely to be a significant driver in increasing farmers' willingness to adopt feed additives, as it removes the unpredictability that has been a barrier to widespread adoption.

\begin{figure*}[htbp]
  \centering
  \includegraphics[width=\textwidth]{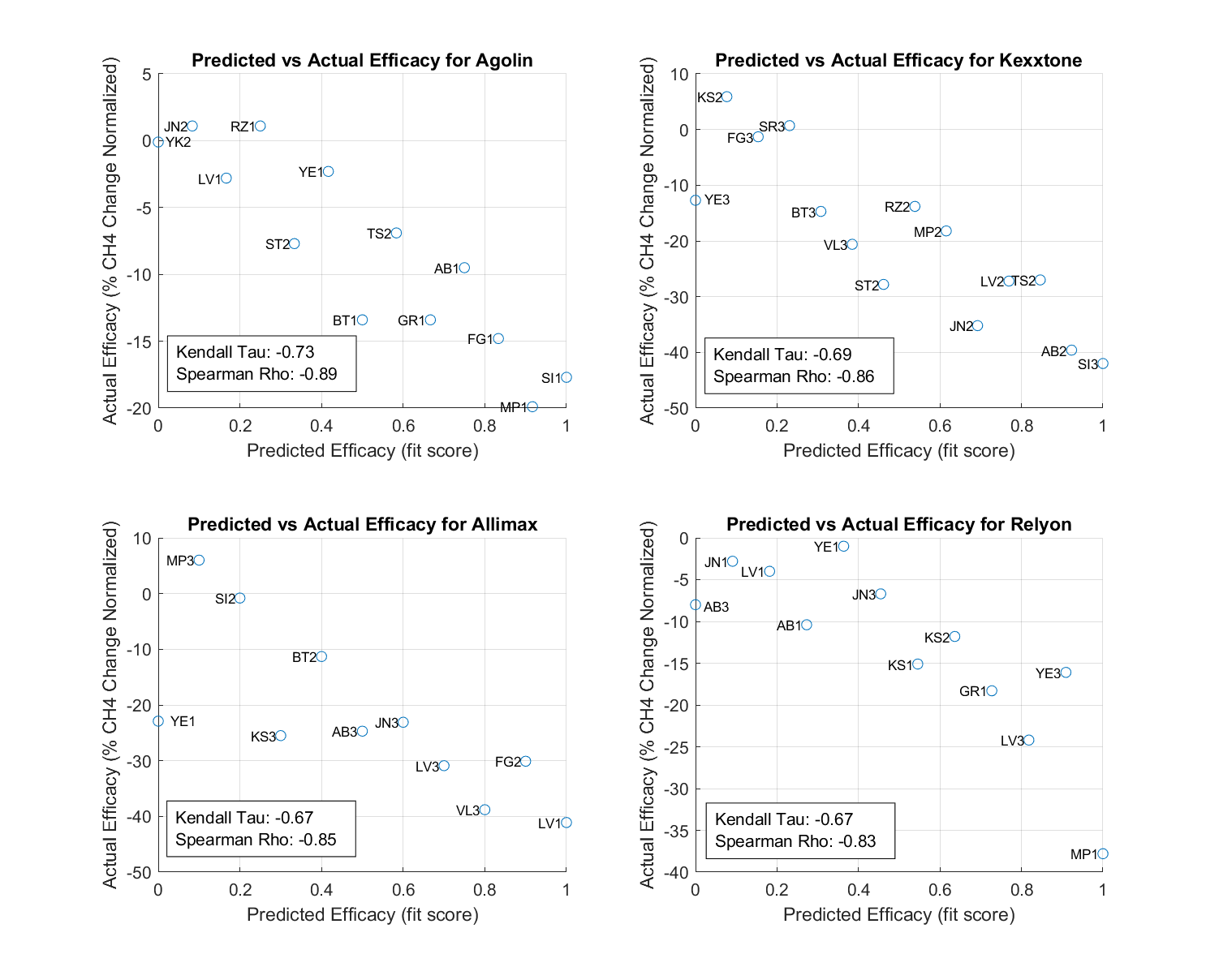}
  \caption{Evaluation of predictive model accuracy across four additives. Each subplot corresponds to a specific feed additive (Agolin, Kexxtone, Allimax, and Relyon) and presents a scatter plot of farms, ranked by their predicted efficacy (x-axis) against their actual, measured efficacy (y-axis). A more accurate model would manifest as a monotonically decreasing line, given that negative values indicate reduced methane emissions. Spearman's \(\rho\) and Kendall's \(\tau\) are also displayed in each subplot, serving as statistical measures of correlation between the predicted and actual efficacies. These measures provide insights into the model's ability to correctly rank farms based on the anticipated benefits of each additive.}
  \label{fig.prediction.quality}
\end{figure*}

Figure \ref{fig.prediction.quality} showcases the proficiency of our predictive model in accurately identifying the farms that are most likely to benefit from each specific additive. The primary objective is to rank farms based on the anticipated efficacy of these additives, as estimated by the prediction model. For each additive, the scatter plot displays farms sorted by their predicted efficacy (x-axis) against their actual, post-factum measured efficacy (y-axis). Ideally, an accurate model would yield a scatter plot that approximates a monotonically decreasing line, since negative values indicate a reduction in methane emissions. Additionally, each subplot provides two statistical measures: Spearman's $\rho$ and Kendall's $\tau$. Spearman's $\rho$ quantifies the strength and direction of the association between the predicted and actual efficacies. Kendall's $\tau$ serves as a non-parametric measure to evaluate the strength of the correlation, focusing on the similarity in the ordering of data when both sets of quantities are ranked.

In Table \ref{tbl.impact} we present an extensive analysis of individual farm performances when applying our microbiome-based, AI-assisted predictive model for additive selection. Each farm is evaluated based on the average efficacy of feed additives for which the farm ranks in the top 33\% or top 50\% in terms of predicted efficacy, among the overall participating farms. These percentages represent the fraction of farms for which the model anticipates the highest potential for methane emission reduction through the use of a particular additive. In other words, we choose to deploy additives in farms only for these farms that are predicted to benefit them the most, and if a farm is predicted to benefit from more than a single additive, we arbitrarily choose between them (taking the mean efficacy).
A value of 'N/A' for a given farm implies that the farm does not fall within the top portion of predicted efficacy for any of the additives examined, and hence would not be administered any additive according to this targeted approach.

It is crucial to understand that although our strategy may leave some farms without additives, the optimization is primarily geared towards enhancing the overall reduction of methane emissions and increasing yield. These are key metrics not only for environmental regulators but also from a return on investment standpoint. This selective model is designed to optimize the use of resources dedicated to methane mitigation, thereby maximizing both environmental impact and profitability for farmers. The results of implementing this strategy are compelling: adopting the top 50\% strategy results in additive deployment at 62\% of farms and achieves an average emissions reduction efficacy of approximately 24\%. Conversely, the more rigorous top 33\% strategy is applicable to 44\% of farms but delivers a higher efficacy, exceeding 27\% in emissions reduction. Importantly, this performance surpasses the individual efficacy of each additive and closely aligns with the ambitious 30\% reduction target set by major dairy stakeholders. Moreover, this tailored approach is likely to be more cost-effective than a naive deployment of the best—and potentially most expensive—additives, as it matches each farm with the most suitable, and often more economical, additive options.

Additionally, the scalability of our proposed model lends itself to easy integration with new additives. As we expand our catalog of additives, we anticipate improvements in two key areas: firstly, our ability to cater to a larger proportion of farms, and secondly, an increase in the overall average efficacy of the treatments.

\begin{table*}[htbp]
\centering
\begin{tabularx}{\textwidth}{lXX}
\toprule
Farm & Deployment for Top 33\% & Deployment for Top 50\% \\ \midrule
AB1 & -9.50\% & -9.50\% \\
AB2 & -39.60\% & -39.60\% \\
AB3 & N/A & N/A \\
BT1 & N/A & -13.40\% \\
BT2 & N/A & N/A \\
BT3 & N/A & N/A \\
FG1 & -14.80\% & -14.80\% \\
FG2 & -30.10\% & -30.10\% \\
FG3 & N/A & N/A \\
GR1 & -15.85\% & -15.85\% \\
JN1 & N/A & N/A \\
JN2 & -35.20\% & -35.20\% \\
JN3 & N/A & -23.10\% \\
KS1 & N/A & -15.10\% \\
KS2 & N/A & -11.80\% \\
KS3 & N/A & N/A \\
LV1 & -41.10\% & -41.10\% \\
LV2 & -27.20\% & -27.20\% \\
LV3 & -27.55\% & -27.55\% \\
MP1 & -28.85\% & -28.85\% \\
MP2 & N/A & -18.20\% \\
MP3 & N/A & N/A \\
RZ1 & N/A & N/A \\
RZ2 & N/A & -13.80\% \\
SI1 & -17.70\% & -17.70\% \\
SI2 & N/A & N/A \\
SI3 & -42.00\% & -42.00\% \\
SR3 & N/A & N/A \\
ST2 & N/A & N/A \\
TS2 & -27.00\% & -16.95\% \\
VL3 & -38.80\% & -38.80\% \\
YE1 & N/A & N/A \\
YE3 & -16.10\% & -16.10\% \\
YK2 & N/A & N/A \\ \midrule
\textbf{Average efficacy} & -27.42\% & -23.65\% \\
\textbf{Farms treated} & 15 out of 34 (44\%) & 21 out of 34 (62\%) \\
\bottomrule
\end{tabularx}
\caption{Individual farm efficacy based on targeted additive allocation. Each row represents a farm and shows the average efficacy of feed additives for which the farm is ranked in the top 33\% or 50\% in terms of predicted additive efficacy. Values are expressed in percentages. The label 'N/A' indicates that the farm does not rank in the top 33\% or 50\% for any of the examined additives. This targeted approach is designed to optimize the aggregate reduction of methane emissions while maximizing economic returns. While some farms do not receive any additive under this model, the overall methane reduction efficacy is notably increased, aligning with both environmental conservation and economic objectives. In terms of scope and efficacy, following the top 50\% strategy results in additive deployment at 62\% of the farms, achieving an average efficacy of approximately 24\% in emissions reduction. On the other hand, the more stringent top 33\% strategy covers 44\% of the farms but results in a higher average efficacy, exceeding 27\% in emissions reduction.}
\label{tbl.impact}
\end{table*}

\subsection{Additional Benefits of Additive Optimization}

Although this paper primarily targets methane emissions reduction, we have also observed a consequential increase in yield, which took place when additives' efficacy was at its peak. This yield enhancement is not merely a fortuitous result, but it is inextricably linked to our methane reduction efforts. This relationship can be attributed to the metabolic energy redirection within the organism. As less energy is channeled towards methane production, more becomes accessible for other essential biological processes, such as milk production or body mass increase in cattle.

The correlation between methane emissions and yield has been well-documented in the literature. Studies like \cite{boland2020feed,melgar2021enteric}, provide robust evidence substantiating this association. Our prediction model, initially designed to effectively facilitate methane emissions reduction, also shows substantial promise in the sphere of yield maximization. This exciting potential demonstrates the dual environmental and economic benefits of our approach.

Although we will not delve into a comprehensive exploration of yield maximization in this paper, it's worth highlighting its relevancy and the utility of our predictive model in this context. We plan to detail the role of our model in maximizing yield alongside minimizing emissions in a forthcoming paper, thereby contributing to the ongoing effort for sustainable farming practices.

\section{Microbial Data Analytics}
\label{sec.analytics.detailed}

\subsection{Motivation and Overview}

This research is predicated on the analysis of numerous microbiome samples collected from bovine subjects across diverse farm settings (ranging from different geographical locations, environmental conditions, herd sizes, and management practices). A subset of these subjects have been administered a feed additive, and subsequent methane emissions were measured, creating an experimental group, while others remained as a control group. Each sample encapsulates a plethora of `reads', each representing sequences of 100 to 150 nucleotides.
Our objective lies in the identification of significant microbial genetic patterns pertinent to the trait of interest, which, in this case, is the high efficacy of the feed additive.

A ``k-mer'' is a contiguous subsequence of length $k$ derived from a longer string of nucleotides. In the context of genomics, a k-mer typically refers to a sequence of $k$ nucleotides within a larger DNA or RNA sequence.

In a more formal mathematical context, if we denote the original longer sequence of nucleotides as the string $S$ and its length as $n$, then a k-mer is a substring of $S$ of length $k$. Given $S[i:j]$ represents the substring of $S$ starting at position $i$ and ending at position $j$ (inclusive), a k-mer of $S$ starting at position $i$ would be represented as $S[i:i+k-1]$ (for $1 \leq i \leq n-k+1$). Consequently, the total number of distinct k-mers that can be extracted from a sequence $S$ of length $n$ is $n-k+1$.

Furthermore, considering the biological context where each position in the string can be one of four nucleotides (A, T, C, or G), the total number of possible k-mers of length $k$, without considering any specific longer sequence, is $4^k$.

Our analytical approach is characterized by an unbiased exploration of large k-mers, specifically those with $k=30$, though not confined to this value. Previous studies have illustrated the optimal expressivity of k-mers of length 30 or longer for predictive applications \cite{chikhi2014informed}. However, their use is typically constrained to cases of extreme data sampling or pre-set filtering criteria, both of which can introduce bias. Conversely, models that leverage k-mers as features in machine learning typically limit $k$ to values of 6 or less, driven by concerns of data scarcity and potential model overfitting. Traditionally, an unbiased analysis of longer k-mers would be considered computationally impractical due to the vast number of possible combinations, approximately $2^{60}$. Additionally, it would necessitate significant amounts of data to circumvent overfitting.

However, we leverage the understanding that the distribution of these 30-mers within DNA does not follow a uniform pattern but instead conforms to a power-law. This inherent property allows us to implement efficient analytic techniques and extract a significant number of k-mer groups automatically. Each of these groups is assuredly associated with a particular epigenetic trait. However, the relevancy of such traits to our current interest may vary.

The innovation presented within this study manifests in a dual capacity. Firstly, we extend our analysis beyond merely long k-mers, thereby enhancing their expressivity, to encompass groups of k-mers, which, in turn, fortifies their role as potent predictive features. Secondly, we address data paucity by utilizing a technique that capitalizes on our power-law distributed data, as opposed to a brute force examination of ``all k-mers'' or ``all groups''.

This technique facilitates the efficient detection of ``correlated anomalies'' -- localized groups giving rise to network structures which do not naturally arise in power-law networks. Analytically, the presence of such groups is indicative of an underlying causality within the data, signifying an association with a specific property relevant to the group of genetic information.

Additionally, the nature of our approach, predicated on the holistic examination of microbiome samples, allows for each group of k-mers to potentially comprise DNA fragments derived from heterogeneous sources. This denotes that functionalities emanating from diverse microbes may concurrently contribute to the observed behavior of interest.

\subsection{Architecture and Key Strengths}

Outlined below is a systematic overview of our method to analyze microbial data. Each constituent step is elucidated in subsequent sections for deeper understanding:

\begin{enumerate}
  \item \textbf{Representation of Sequenced Data:} Each sequenced microbial dataset is encapsulated as a network. While each network maintains a constant node count denoted by $M$, their edge configurations are susceptible to variability.
  \item \textbf{Formation of Superposition Networks:} To achieve a more nuanced analysis, we formulate superposition networks. This is accomplished by layering individual networks based on specific criteria—be it samples originating from identical farms, those sourced from a particular geographic locale, share some biological attribute, or samples acquired under similar meteorological conditions.
  \item \textbf{Unsupervised Analysis:} Both the original one-sample networks and the constructed superposition networks undergo an in-depth unsupervised exploration. The procedure for each network can be distilled into the following sub-steps:
  \begin{enumerate}
        \item For each unique degree $d$ present in the network, nodes bearing the degree $d$ are analyzed.
        \item We then empirically assess the ``internal connectivity'' of this node cluster. This entails calculating the ratio of edges that both originate and culminate within this cluster to the edges that begin within this cluster but terminate externally.
        \item Should this ratio surpass the benchmark delineated by Theorem \ref{thm.beta}, we infer analytically that the node cluster (e.g. k-mers) probably bears semantic relevance. The degree of this statistical confidence is symbolized by $\epsilon$ which we can calibrate as per our requirements.
    \end{enumerate}
  \item \textbf{Storage of k-mer Clusters:} All discerned k-mer clusters are cataloged for subsequent utilization during the supervised phase. While every cluster possesses its unique statistical confidence metric conducive for granular analysis, they collectively uphold a robustness standard that exceeds our initial $\epsilon$ setting.
  \item \textbf{Retrospective Genetic Investigation:} The utility of superposition networks extends to retrospective analysis as well, offering invaluable insights when certain shared properties among cows are identified post-data collection. For instance, if disparate cows nationwide are later found to possess a common susceptibility to a specific disease, a dedicated superposition network can be easily crafted to encompass these particular cows. This allows for the extraction of unique genetic patterns potentially correlated with the identified trait. Consequently, superposition networks facilitate not only a dynamic reanalysis of the existing data but also enable researchers to unveil subtle, yet crucial, genetic markers tied to various biological characteristics discovered in hindsight. These markers then serve as pivotal reference points for future genetic investigations and strategies aimed at addressing the specific traits or susceptibilities uncovered.
\end{enumerate}

The unsupervised learning phase, central to our approach, is initiated by utilizing all available microbiome samples. The essence of this phase lies in constructing networks derived from genetic material, necessitating the amalgamation of various sample combinations. Initially, individual samples stand autonomously, each constituting a unique source to form distinct networks. Following this, an aggregated network is derived by pooling samples from a complete farm. This composite structure, being inherently denser, serves as a robust representation of farm-level features.

Furthermore, our methodology permits the formation of several nuanced networks, shaped by criteria as diverse as geography, specific weather conditions, or a collation of all accessible data. One might assume that the creation of such intricate networks would be computationally arduous. However, by overlaying selected foundational networks, a superposition network is efficiently created, by executing a boolean OR operation on their edges. This design, which earmarks a dedicated network to each sample, streamlines the integration of new data or samples.

A pivotal aspect of this phase is its adaptability and inclusivity. Irrespective of the data source -- whether being microbiome samples from cows, sheep, soil, or even humans -- this phase seamlessly integrates diverse data. Moreover, its design ensures that the genetic markers identified are universally relevant across various label groups, all linked to a targeted biological condition or trait.

One of the salient features of our design is its capacity to harness the entirety of available raw data, including the long-tail genetic information. Long-tail data, denoting genetic information from microbes present in minimal quantities within a sample, often impedes mainstream metagenomic analytics, curtailing the diversity and precision of predictions obtainable from microbiome data. By leveraging our approach, we transcend this limitation, moving beyond traditional DNA read analysis and focusing on the intricate dynamics of networks instead.

Additionally, it is imperative to emphasize the unbias nature of our approach. Contrary to being contingent on existing reference databases in scholarly literature, our method stands apart as inherently data-driven, refraining from zeroing in on selective information subsets.

\subsection{Data Representation}

As previously defined, from each farm $f_i \in F$ a random set of cows $C_{u,i}$ are selected, and their microbiome sampled and sequenced.

Each microbiome sample, denoted as $X$, contributes to the generation of a unique network, $G_X$, comprised of $M$ nodes and $E_X$ edges. The nodes' consistency across all networks originates from the initial data processing: during an automated preliminary analysis of the data, we exclude infrequently appearing k-mers (i.e., only k-mers that appear at least twice in at least two samples are retained\footnote{This filtering criterion can, of course, be made stricter, further reducing the number of k-mers.}). We are thus left with a manageable quantity $M$ that is conducive to our network analytics. This provision is facilitated by the power-law distribution of k-mer repetitions across the data, which guarantees that the majority of k-mers will indeed be unique and subsequently filtered out, while predicting that there will still be a substantial number of k-mers recurring multiple times.

Specifically, we work with 30-mers\footnote{The number $k=30$ was arbitrarily selected for this study. It's large enough to allow for sufficient expressivity of the k-mers but low enough to allow different k-mers to appear in the same read, subsequently manifested as an edge in the k-mers network. Note that changing the value of $k$ trades the maximum number of possible k-mers for the maximum number of possible connections between them. We believe various values from $k=20$ to $k=80$ could be used with our proposed method, potentially producing DNA patterns that would further enhance the model's predictive capabilities.}. From a theoretical search space of $4^{30}$ possible 30-mers, we end up using approximately one million 30-mers, translating into networks of one million nodes, a scale that is computationally feasible to handle.

An enhanced model that repeats the flow described here for various values of $k$, each resulting in a different set of k-mers, can still be managed computationally efficiently. Even a model with k-mers for 60 different values of $k$ would result in networks of around 100 million nodes. Although such networks may initially seem too large, due to the nature of our network analytics, they can still be analyzed with practical computational resources.

\subsection{New Statistical Dimensions of Microbial Genetic Data}

The realm of microbiome data has rapidly emerged as a thrilling frontier in the biological sciences. Rich in its complexity, it holds the potential to unlock myriad mysteries about agriculture efficiency, environmental challenges, as well as human health and disease. However, much of the current research in this space has been predominantly rooted in modeling. Modeling offers a structured way to simulate, predict, and infer mechanisms that underlie microbial dynamics. Yet, while it can be invaluable, modeling often tends to abstract away some of the intricate details and nuances that the raw data encapsulates. This orientation towards modeling, rather than a more exploratory, data-driven approach, might inadvertently overlook or overshadow key insights waiting to be discovered.

A significant limitation of model-based analysis in microbiome research is its inherent dependency on reference databases. Such databases, though comprehensive, are not exhaustive. Relying heavily on them can introduce biases and potentially skew interpretations of microbial communities. Conversely, adopting a purely data-driven approach allows for an unbiased lens, grounding the analysis in the very essence of the data.

In our study, we found that microbiome genetic data exhibits distinct power-law dynamics. This pattern is evident in the frequency of raw reads, the prevalence of k-mers within these reads, and the co-occurrence of these k-mers in the same samples. When we model k-mers as vertices in a network, with edges representing their co-occurrence in the same sample, the power-law dynamics are further underscored.

To the best of our knowledge, this observation stands as both unique and unprecedented in the realm of microbiome research. The recognition of such patterns may bear profound implications, potentially revolutionizing our understanding of microbial interactions and dynamics, serving as robust indicators or predictors of specific microbiome behaviors or states. Recognizing the potential of this discovery, in the following section, we delve into how these observed power-law dynamics can be harnessed for prediction purposes, potentially offering a novel toolset for researchers and practitioners in the field.

\subsection{The Universal Footprint of Emergent Power Law Dynamics Across Varied Domains}

The ubiquity of power law dynamics across a spectrum of systems and domains attests to its fundamental nature in characterizing complex phenomena. From the vast expanse of celestial bodies to the intricacies of minute cellular structures, power law distributions emerge as an underlying pattern that ties disparate realms into a coherent tapestry of universal behaviors.

\paragraph*{\textbf{Social Networks:}}
The topology of social networks often follows a power law distribution. A minority of nodes (individuals) possess a disproportionately large number of connections, while the majority have relatively few. This phenomenon, often referred to as the ``rich get richer'' mechanism or preferential attachment, exemplifies the emergence of 'hubs' in networks, and it has profound implications for the spread of information, behaviors, and even epidemics \cite{CSS-Lazer-Science-2009}.

\paragraph*{\textbf{Transportation:}}
In transportation networks, whether examining the distribution of passenger flights among airports or the traffic flow across urban intersections, a power law dynamic becomes evident \cite{gonzalez:2008nature}. This pattern extends beyond traditional modes of transportation to include recent advancements in mobility services, such as ride-sharing systems \cite{shmueli2015ride}. In these systems, similar power law characteristics are observable, where a small number of hubs or nodes (major cities or busy urban centers, for instance) account for a disproportionately large volume of ride-sharing activities. This pattern reflects the concentration of demand and supply in these systems, further underscoring the ubiquity of power law dynamics across various transportation models, both conventional and modern \cite{altshuler2019modeling}.

\paragraph*{\textbf{Financial Activity:}}
Stock market fluctuations, company sizes, and trading volumes frequently exhibit power law characteristics, a pattern that is emblematic of many economic systems \cite{somin2020network}. This is particularly evident in the distribution of wealth, where a small fraction of entities control a disproportionately large share of resources, exemplifying classical power law dynamics in economics. Such distributions are not random but follow predictable patterns that can be quantitatively analyzed. Understanding these underlying patterns is not merely an academic exercise; it is instrumental in developing various predictive mechanisms \cite{somin2022beyond}. These mechanisms are crucial for forecasting market trends, assessing economic risks, and devising strategies for financial management and policy-making. By leveraging the insights provided by power law dynamics, economists and financial analysts can better anticipate market behaviors, contributing to more robust and resilient economic systems \cite{somin2022remaining}.

\paragraph*{\textbf{Cyber Crime and Terrorism:}}
Power law dynamics manifest significantly in the realms of cyber crime and counter-terrorism, providing a unique lens through which these complex and clandestine activities can be understood and addressed \cite{SR-IIS}. In the context of cyber crime, particularly with the rise of sophisticated botnets, power law distributions can often be observed in the behavior and spread of these networks \cite{zwang2018detecting}.
Similarly, in the domain of blockchain and cryptocurrency, power law dynamics can be instrumental in uncovering illicit activities. This insight is particularly valuable for identifying hidden groups of illicit wallets which might be involved in money laundering or funding terrorist activities. By analyzing transaction patterns and identifying outliers that deviate from expected power law distributions, it becomes possible to flag suspicious activities and detect networks of wallets that may otherwise remain concealed within the vast expanse of blockchain transactions \cite{socialphysicsandcybercrime2018}.

\paragraph*{\textbf{Large Language Models:}}
The distribution of word popularity, as well as the context in which words appear, also exhibits a power law dynamic, a phenomenon that is particularly evident in the field of natural language processing (NLP). This observation is grounded in the principle that a small subset of words is used exceedingly frequently, while the vast majority are used much less often. Such a distribution can be observed in various languages and texts, ranging from everyday communication to specialized literature. This characteristic is also integral to the principles underpinning Large Language Models (LLMs) such as the Generative Pre-trained Transformer (GPT) series \cite{wolfram2023chatgpt}.

These aforementioned examples provide just a glimpse into the myriad of systems governed by power law dynamics. Recognizing and understanding these patterns is not just an academic exercise; it is crucial for optimizing and regulating these systems and predicting their behavior under various conditions. As this paper will later delve into, the prevalence of power law dynamics is not limited to these realms; it extends intriguingly into the domain of genetic data, offering profound insights into biological processes and evolutionary mechanics.

\subsubsection{The Uncharted Territory of Power Laws in Raw DNA Sequencing}

In our quest for deeper understanding, we direct our focus towards the statistical macroscopic patterns that naturally emerge from microbiome data. By analyzing the popularity and prevalence of various reads and k-mers, and studying the intricate web of interactions between them, we aim to capture the larger, holistic picture of microbial dynamics. A cornerstone of our investigation is the surprising observation of power-law properties manifesting within raw microbiome sequencing data. Power laws, characterized by phenomena where small occurrences are extremely common and large occurrences extremely rare, might hold the key to understanding overarching trends and rules governing microbial populations.

Historical precedents in biological research have illuminated the emergence of power-law dynamics in diverse systems. For instance, interactions between proteins \cite{ito2000toward}, as well as those between metabolites \cite{jeong2000large,huss2007currency} or cellular activities \cite{barabasi2004network}, have been shown to obey power-law distributions, hinting at universal principles that might be governing these interactions. However, when it comes to raw DNA -- and especially in the context of microbiome DNA -- such observations are virtually non-existent. Despite recent work that call for the use of network theory in the study of microbiome data \cite{layeghifard2017disentangling}, and while power-law properties have been foundational in understanding various biological systems, their presence in sequenced data remains a largely unexplored domain. To the best of our knowledge, we tread new ground, aiming to uncover these elusive patterns and decipher what they signify for the world of microbiomes.

\subsubsection{Macroscopic Insights from Microbial DNA Sequencing}

\begin{figure*}[htbp]
    \centering
    \includegraphics[width=\textwidth]{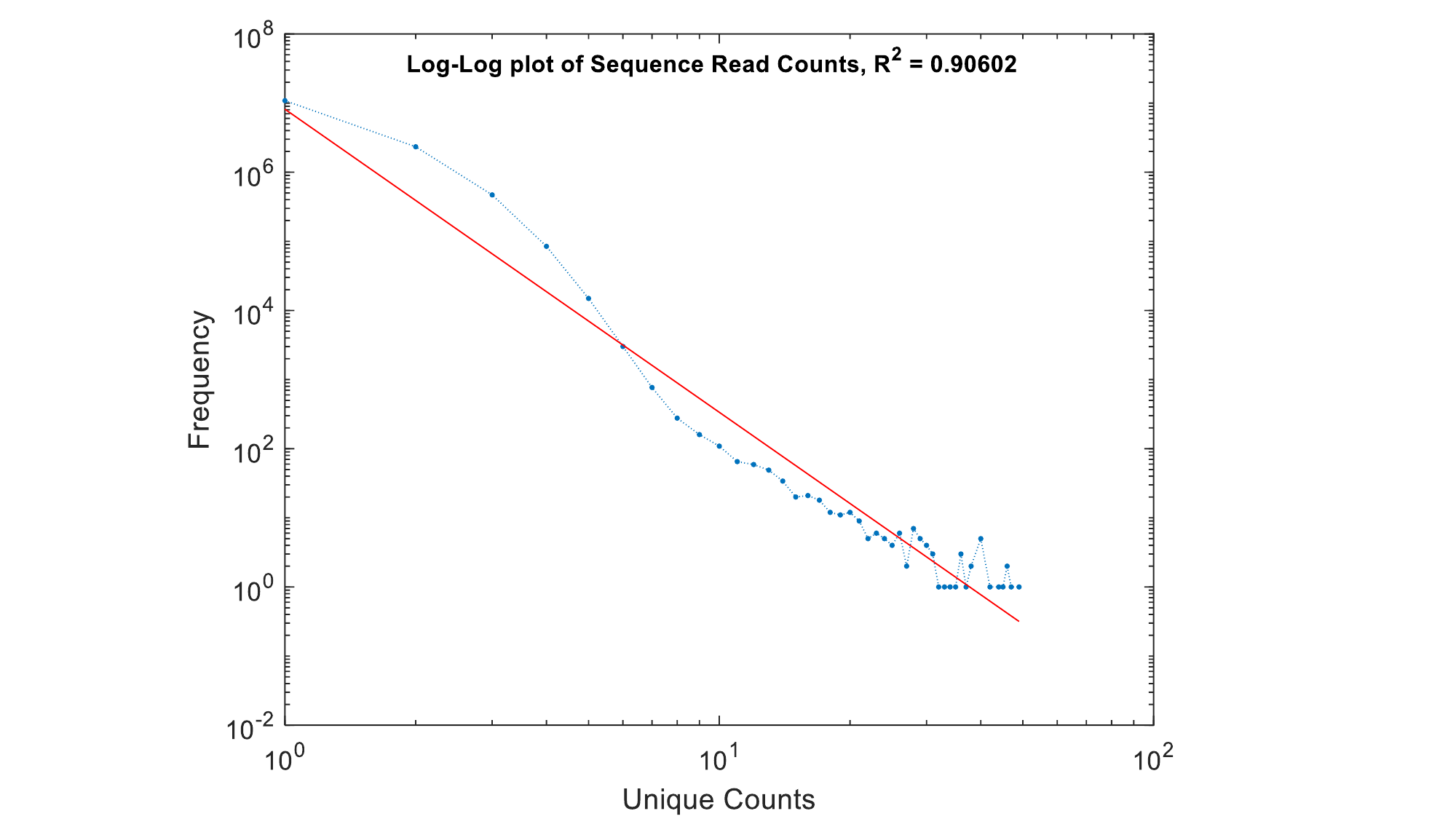}
    \caption{Log-log scale distribution of DNA read popularity from a sequenced microbiome sample of a representative cow. The Y-axis indicates the number of unique reads in the sample that appeared a given number of times, as specified by the X-axis. Notably, over half of the unique reads occur only once. Yet, approximately 0.01\% of them surface 5 times or more. About 100 unique reads manifest 10 times, with a subset surfacing multiple dozens of times. The distribution's dynamics align closely with the power-law model, showcasing a compelling regression fit with $R^2 > 0.9$, and validated using the method suggested in \cite{clauset2009power}.}
    \label{fig:powerlaw1}
\end{figure*}

Our initial analysis began with a singular microbiome sample from an arbitrary cow. We counted the occurrences of each unique sequence, 150 nucleotides in length. Given the sample's roughly 16 million reads, a randomly generated DNA sequence would predict an infrequent recurrence of unique reads. Contrary to this expectation, while over half of the unique segments appeared only once, hundreds repeated more than 10 times. A few segments even surfaced several dozen times. The distribution of these occurrences exhibited power law dynamics, with an $R^2$ value exceeding 0.9 when fit to a power law model. These results are detailed in Figure \ref{fig:powerlaw1}.

\begin{figure*}[htbp]
    \centering
    \includegraphics[width=\textwidth]{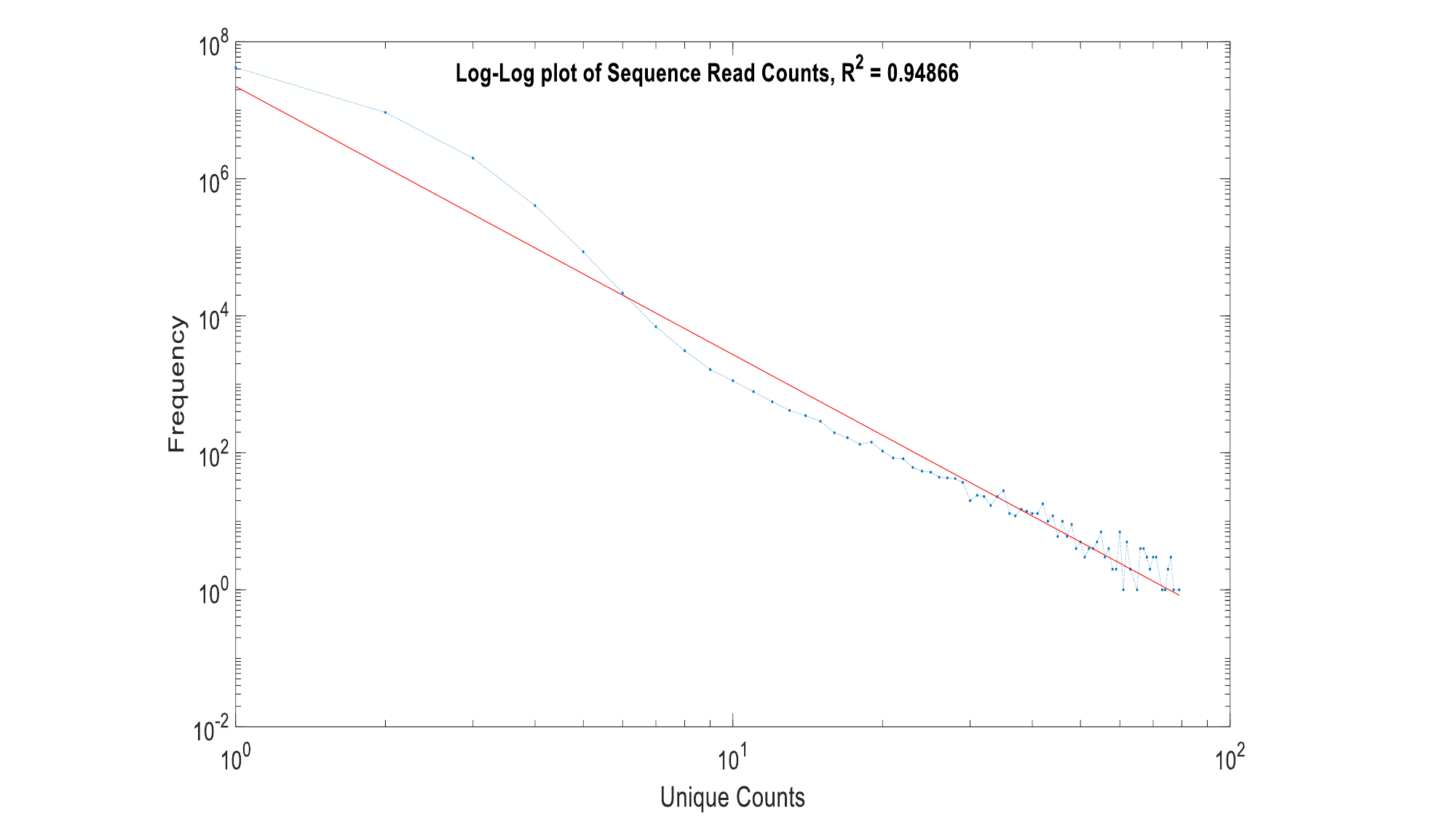}
    \caption{Log-log scale distribution of DNA read popularity from a combined microbiome sample, aggregating data from five cows, each from a distinct herd. The Y-axis represents the number of unique reads in the sample that appear a given number of times, denoted by the X-axis. This aggregated sample showcases dynamics strikingly similar to the individual cow sample depicted in Figure \ref{fig:powerlaw1}. The distribution adheres closely to the power-law model, evidenced by a robust regression fit with $R^2 > 0.94$, and validated using the method suggested in \cite{clauset2009power}.}
    \label{fig:powerlaw2}
\end{figure*}

Expanding our scope, we amalgamated samples from five distinct cows, each from a different herd. The inherent microbiome diversity among herds amplified the number of unique reads. Nonetheless, this aggregated data mirrored the previously observed power law dynamics, with the $R^2$ of the regression fit nearing 0.95, as shown in Figure \ref{fig:powerlaw2}.

\begin{figure*}[htbp]
    \centering
    \includegraphics[width=0.6\textwidth,viewport=185 170 765 550]{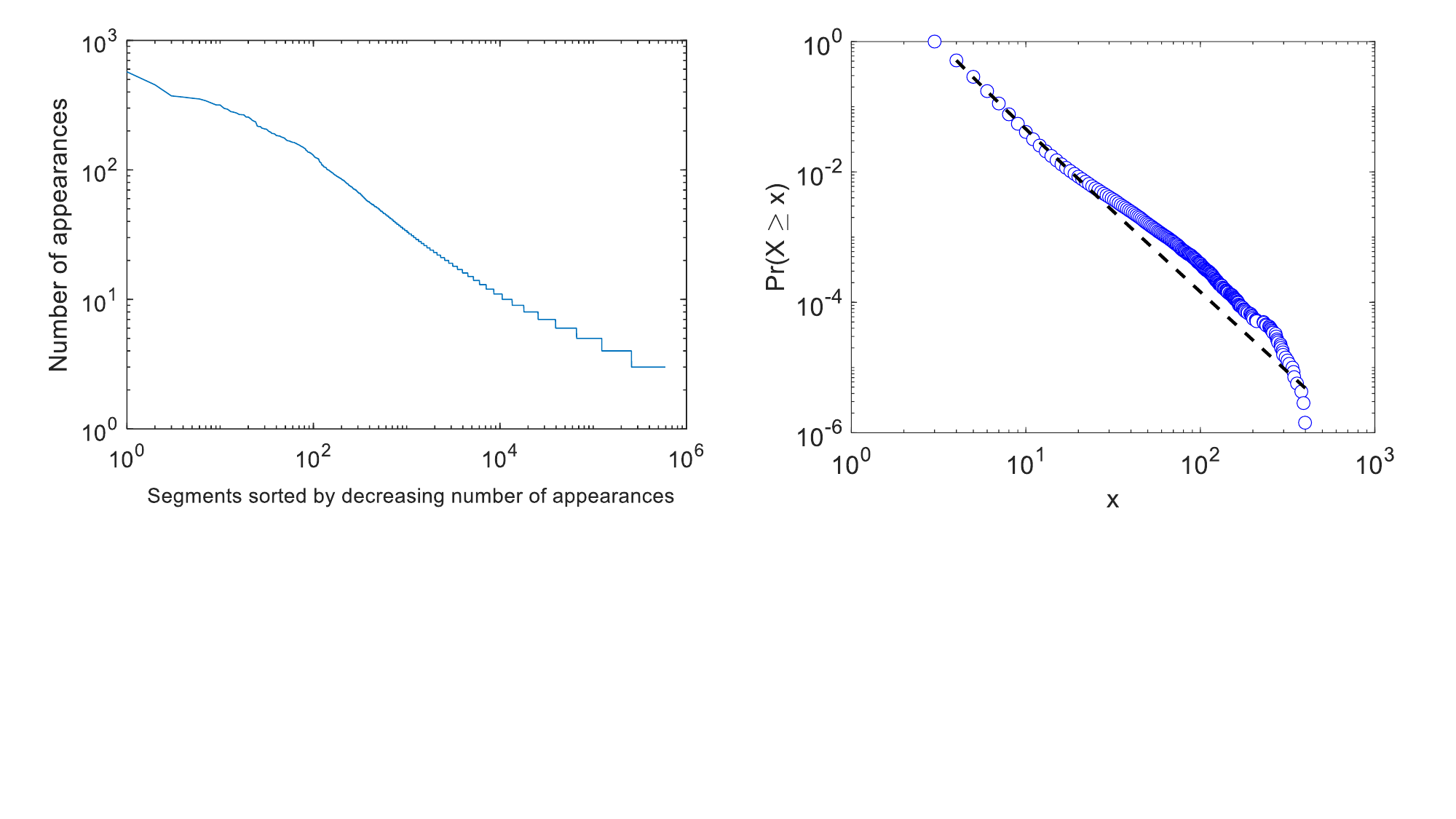}
    \caption{Log-log scale visualization of the popularity distribution for human microbiome data, sourced from a representative sample in the Human Microbiome Project \cite{human2012framework}. The chart on the left illustrates the frequency of all unique segments, organized in descending order based on their popularity. The right-side chart depicts the likelihood of a segment from the sample to manifest at least $x$ times, for varying values of $x$. A power-law fit is superimposed, and its validity is corroborated using the methodology proposed in \cite{clauset2009power}.}
    \label{fig:powerlaw3}
\end{figure*}

\begin{figure*}[htbp]
    \centering
    \includegraphics[width=0.6\textwidth,viewport=185 100 765 550]{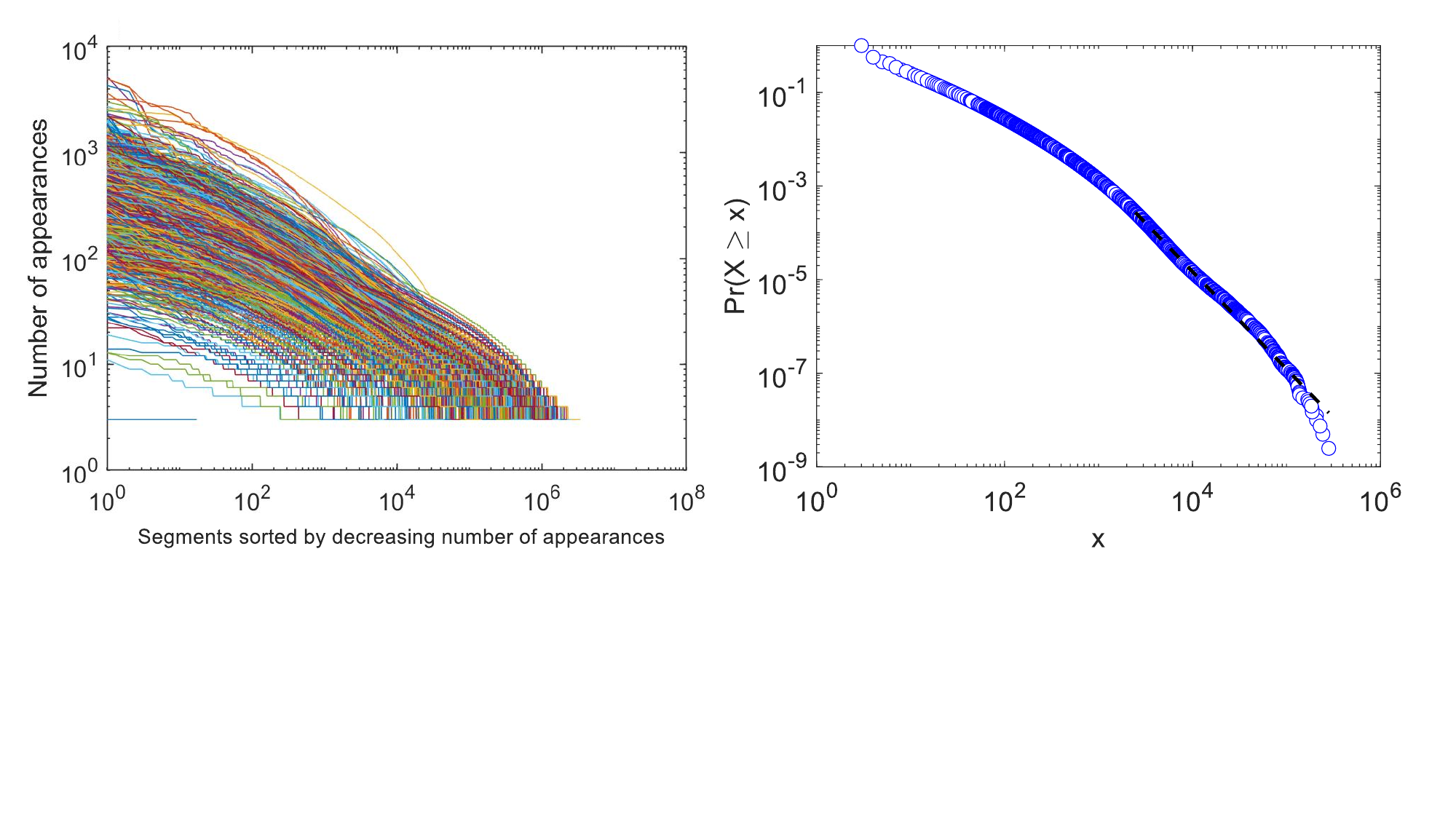}
    \caption{Log-log scale depiction of the popularity distribution derived from 1,300 microbiome samples in the Human Microbiome Project \cite{human2012framework}. The chart on the left showcases each sample as an individual line, with a median regression fit to power-law standing at \(R^2 = 0.88\). On the right, data from all samples is aggregated into a singular dataset, illustrating the probability of a segment occurring at least \(x\) times for various \(x\) values.}
    \label{fig:powerlaw4}
\end{figure*}

To mitigate potential biases from our rumen sampling or the sequencing protocol, we turned to the Human Microbiome Project \cite{human2012structure,human2012framework}, curated by the National Institute of Health. This dataset, boasting over 1,300 analyzable microbiome samples, upheld our earlier observations. The power law dynamics persisted whether we treated the entire dataset as a singular entity or delved into individual samples (Figures \ref{fig:powerlaw3} and \ref{fig:powerlaw4}).

\begin{figure*}[htbp]
    \centering
    \includegraphics[width=0.65\textwidth,viewport=185 160 765 550]{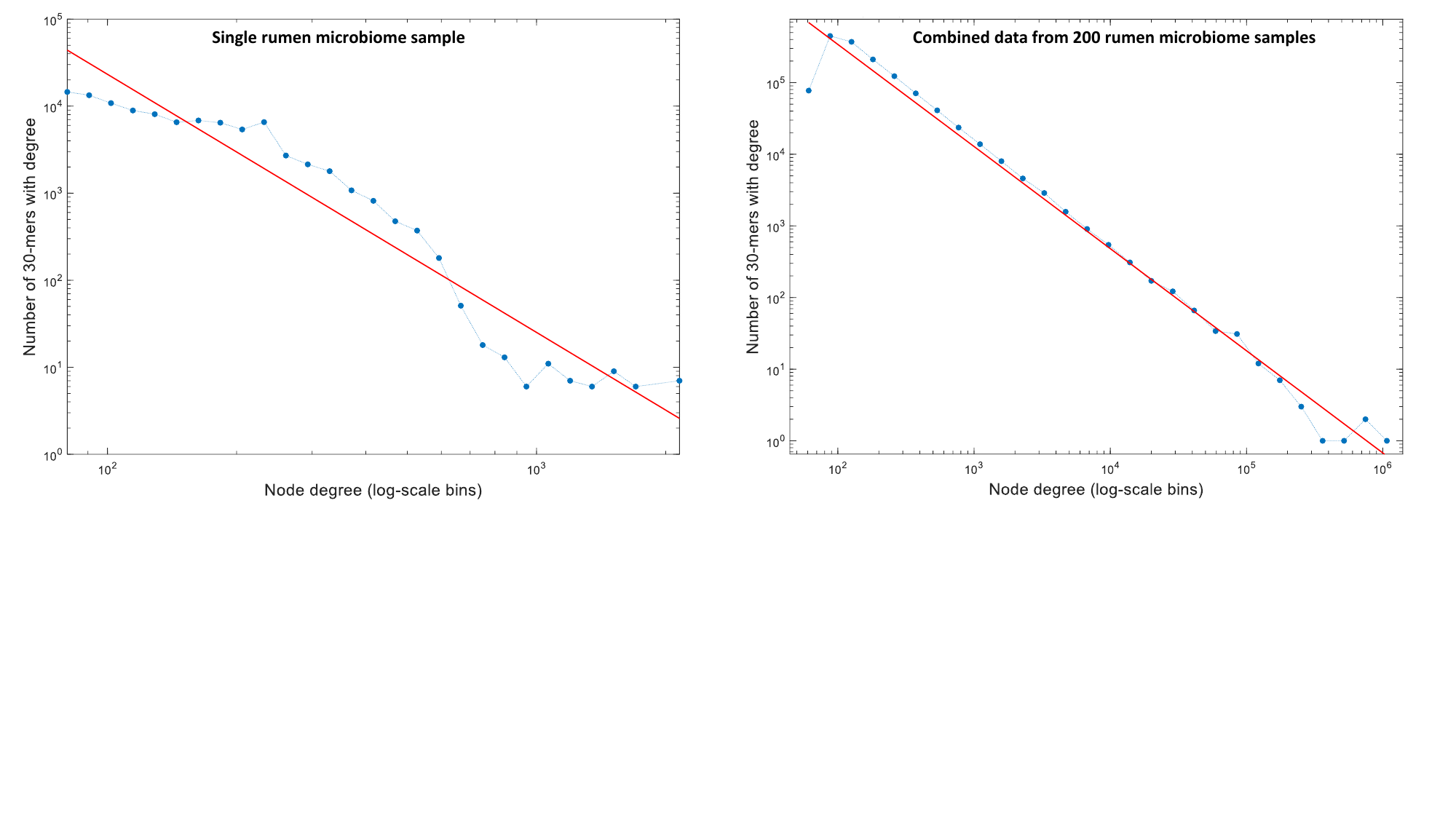}
    \caption{Degree distribution of networks constructed from 30-mer sequences: In this visualization, each node represents a unique 30-mer, and an edge is drawn between two 30-mers if they are found within the same read of a given sample. The left-hand chart displays the distribution from a singular cow's microbiome data, while the right amalgamates data from 200 distinct cows. Both charts exhibit a power-law dynamic in the node degree distribution, with a strong regression fit of \(R^2 > 0.92\). Specifically, the Y-axis signifies the count of nodes with a specific degree \(x\), as represented on the X-axis. The observed power-law trend, consistent across both scales, underscores the inherent structure and shared features in microbiome datasets.}
    \label{fig:powerlaw5}
\end{figure*}

In a subsequent phase, we construed networks from the samples taken from cows. Using k-mers as vertices, and establishing edges when two k-mers co-exist within a read, the resultant unweighted, undirected networks -- after pruning low-occurrence k-mers to manage data volume -- revealed the familiar power law dynamics. This held true for individual samples as well as amalgamated ones, as depicted in Figure \ref{fig:powerlaw5}. These findings further strengthen the argument that the observed dynamics stem from inherent microbial genetic properties rather than being external artifacts.

Lastly, for all instances above, the power law's validity received further affirmation using the methodology proposed by Clauset in \cite{clauset2009power}.

\subsection{Network Generation from k-mers}

The generation of each network, denoted as $G_X$, involves a combination of genomic sequencing and preliminary data analysis. In the initial stage, the microbial DNA is fragmented into a set of k-mers, specifically, 30-mers, each of which serves as a potential node in the network we are going to produce in the next phase. This process results in a theoretical search space of $4^{30}$ k-mers.

To ensure the computational feasibility of our network analysis, we employ a preliminary filtering step. For each microbial sample, only those k-mers that appear at least twice in the sample are retained, significantly reducing the search space. The use of this filtering is made possible thanks to the fact that the repetition of k-mers across the data follows a power-law distribution, allowing us to significantly reduce the size of our search space (as most of the k-mers would indeed have a unique appearance throughout each sample), while still statistically guaranteeing the existence of a large enough selection of k-mers that appear multiple times in each sample. This method results in approximately one million unique 30-mers that form the consistent nodes $V$ used across all networks.

When new samples are received, they can be easily processed such that only k-mers from $V$ are used for their analysis. From time to time, the nodes $V$ can be refreshed by re-running this process, expanding the set of supported k-mers and updating the internal ordering of the networks, allowing for new k-mers originating from new data to be included in the overall analysis.
This process, however, is not mandatory, ensuring flexibility in our analytical approach.

Once the initial construction of the set of supported k-mers is completed, each microbiome sample $X$ can be analyzed and its corresponding network $G_X = G(V, E_X)$ can be constructed as follows:

\begin{enumerate}
\item The sample $X$ contains a large number of reads (usually between 1 and 20 millions), each a string of nucleotides, 100 to 150 bases long.
\item For each read, extract all k-mers that are contained in the read and filter this list of k-mers per the list of supported k-mers $V$.
\item For each pair $v_1$ and $v_2$ of supported k-mers that are contained in this read, add the edge $(v_1, v_2)$ to $E_X$.
\end{enumerate}

Note that whereas each network $G_X$ has its unique set of edges $E_X$, the nodes in the network are the same group $V$, established during the initial filtering phase, using the multitude of available microbiome samples. In addition, we assume that it is enough that the adjacency matrix of the network $G_X$ represents only the existence of connections without storing their actual strengths or quantities. In other words, a Boolean representation can be used for $E_X$, further reducing time and space complexities.

Once the networks $G_X$ that are constructed from single microbial samples are available, further networks $G_S$ can be constructed on demand, for various groups $S$, by easily applying a logical OR operator on the adjacency matrices corresponding to the networks $G_X$ for every $X \in S$. Each of these networks can then be used as input for the network anomalies detection phase, allowing for the discovery of new data patterns, observable through the data-perspective provided by the semantics of the group $S$.

A significant revelation from our exploration of the constructed networks \(G_X\) and \(G_S\) lies in the distribution of their node degrees. An intensive examination of these networks highlighted a consistent pattern: the degree distribution follows a power-law distribution, a finding that stands true for both individual microbial samples as well as composite groups.
Furthermore, the consistency in observing this distribution across both individual and composite samples hints at an inherent and universal structure within microbial genetic data. A strong regression fit of \(R^2 > 0.92\) further solidifies this observation.
This is illustrated in Figure \ref{fig:powerlaw5}.

\subsection{Detection of Local Network Anomalies in the k-mer Networks}

Given a network $G_X (V,E_X)$ (or $G_S (V,E_S)$) we know that the degrees if its nodes closely follows a power-law distribution: $P(k) \sim k^{-\alpha}$.
Literature shows that in many real-world networks that follows a power-law degree distribution this pattern applies degrees larger than some minimal degree $d_{min}$. Namely, for some normalization constant $c$ the probability of a node $v$ to have degree $d \geq d_{min}$ equals:
\[
\forall v \in V \ , \ \forall d \geq d_{min} \ , \ P[deg(v) = d] = c \cdot d^{-\alpha}
\]

Let us first calculate the expected value for the normalization constant $c$ using values easily obtainable from the data:

\begin{definition}
\label{def.vestar}
    Let $V_{\star}$ denote all nodes of degree at least $d_{min}$ and let $E_{\star}$ denote the edges that have at least one side in $V_{\star}$:
    \[
    V_{\star} = \left\{ (v \in V) | (deg(v) \geq d_{min}) \right\}
    \]
    \[
    E_{\star} = \left\{ ((v,u) \in E) | (v \in V_{\star}) \vee (u \in V_{\star})  \right\}
    \]
\end{definition}

\begin{lemma}
\label{lemma.c}

    If:
    \[
    \forall v \in V \ , \ \forall d \geq d_{min} \ , \ P[deg(v) = d] = c \cdot d^{-\alpha}
    \]
    then:

    \[
    c = \frac{2|E_{\star}|}{|V_{\star}| \cdot \sum_{d = d_{min}}^{d_{max}} d^{1-\alpha}}
    \]

\end{lemma}
\begin{proof}
    The sum of a graph's $G(V_{\star},E_{\star})$ degrees equals twice the number of its edges:

    \[
    \sum_{v \in V_{\star}} deg(v) = 2 |E_{\star}|
    \]

    Taking into account the power-law distribution of the graph's nodes, and summing the expected number of nodes with degree $d$ for every degree $d$ from $d_{min}$ to the maximal degree $d_{max}$ we can see that:
    \[
    2 |E_{\star}| = \sum_{v \in V_{\star}} deg(v) =
    \]
    \[
    E\left[  \sum_{d = d_{min}}^{d_{max}} d \cdot \left| (v \in V_{\star}) \wedge (deg(v) = d) \right|  \right] =
    \]
    \[
    \sum_{d = d_{min}}^{d_{max}} d \cdot |V_{\star}| \cdot P\left( v \in V_{\star} , deg(v) = d \right)  =
    \]
    \[
    \sum_{d = d_{min}}^{d_{max}} d \cdot |V_{\star}| \cdot c \cdot d^{-\alpha} = 2 |E_{\star}|
    \]

    Implying:
    \[
    c = \frac{2|E_{\star}|}{|V_{\star}| \cdot \sum_{d = d_{min}}^{d_{max}} d^{1-\alpha}}
    \]
\end{proof}

From this we can easily calculate the expected number of nodes with some degree $\hat{d}$ in a network $G$ with a power-law degree distribution:
\begin{definition}
\label{def.vdhat}
    Let $V_{\hat{d}} \subseteq V_{\star}$ denote all nodes of degree $\hat{d} \geq d_{min}$:
    \[
    V_{\hat{d}} \triangleq \left| (v \in V_{\star}) \wedge (deg(v) = \hat{d}) \right|
    \]
\end{definition}

\begin{lemma}
\label{lemma.num.of.nodes.with.deg.d}
    If:
    \[
    \forall v \in V \ , \ \forall d \geq d_{min} \ , \ P[deg(v) = d] = c \cdot d^{-\alpha}
    \]
    then, for every degree $d_{min} \leq \hat{d} \leq d_{max}$:
    \[
    \left| V_{\hat{d}} \right| \approx \frac{2|E_{\star}| \cdot \hat{d}^{-\alpha}}{\frac{1}{2-\alpha}(d_{\text{max}}^{2-\alpha} - d_{\text{min}}^{2-\alpha})}
    \]

\end{lemma}
\begin{proof}

    The number of nodes with degree $\hat{d}$ is~:
    \[
    \left| V_{\hat{d}} \right| = c \cdot |V_{\star}| \cdot \hat{d}^{-\alpha}
    \]

    which using Lemma \ref{lemma.c} equals:
    \[
    \left| V_{\hat{d}} \right| = |V_{\star}| \cdot \hat{d}^{-\alpha} \cdot \frac{2|E_{\star}|}{|V_{\star}| \cdot \sum_{d = d_{min}}^{d_{max}} d^{1-\alpha}}
    \]

    Namely:
    \[
    \left| V_{\hat{d}} \right| = \frac{2|E_{\star}| \cdot \hat{d}^{-\alpha}}{\sum_{d = d_{min}}^{d_{max}} d^{1-\alpha}}
    \]

    The denominator is a sum that goes from $d_{min}$ to $d_{\text{max}}$. Since $d_{\text{max}}$ is large, we can approximate this sum by the integral:
    \[
    \sum_{d = d_{min}}^{d_{\text{max}}} d^{1-\alpha} \approx \int_{d_{min}}^{d_{\text{max}}} x^{1-\alpha} \, dx
    \]

    The integral can be evaluated by using the power rule for integration:
    \[
    \int_{d_{min}}^{d_{\text{max}}} x^{1-\alpha} \, dx = \left. \frac{x^{2-\alpha}}{2-\alpha} \right|_{d_{min}}^{d_{\text{max}}} =
    \]
    \[
    = \frac{d_{\text{max}}^{2-\alpha}-1}{2-\alpha} - \frac{d_{min}^{2-\alpha}-1}{2-\alpha} =
    \]
    \[
    = \frac{d_{\text{max}}^{2-\alpha} - d_{\text{min}}^{2-\alpha}}{2-\alpha}
    \]

    Substituting this back into the expression for $\left| V_{\hat{d}} \right|$ we obtain:
    \[
    \left| V_{\hat{d}} \right| \approx \frac{2|E_{\star}| \cdot \hat{d}^{-\alpha}}{\frac{1}{2-\alpha}(d_{\text{max}}^{2-\alpha} - d_{min}^{2-\alpha})}
    \]

    This approximation holds for $\alpha > 2$, where the integral is convergent.

\end{proof}

\begin{definition}
\label{def.lambda}
    Let $\lambda$ denote the power-law normalizing constant for the graph $G(V_{\star}, E_{\star})$:
    \[
    \lambda = \frac{2 (2-\alpha) |E_{\star}|}{(d_{\text{max}}^{2-\alpha} - d_{min}^{2-\alpha})}
    \]
    such that Lemma \ref{lemma.num.of.nodes.with.deg.d} can be written as:
    \[
    |V_{\hat{d}}| \approx \lambda \hat{d}^{-\alpha}
    \]
\end{definition}

\begin{definition}
\label{def.edhat}
    Let $E_{\hat{d}} \subseteq E$ denote all the edges that touch at least one node in $V_{\hat{d}}$. We divide these edges to two complementing and mutually exclusive groups ``internal edges'' $E_{\hat{d}}^{in}$ (edges starting and ending in $V_{\hat{d}}$) and ``external edges'' $E_{\hat{d}}^{out}$ (edges that have one side in $V_{\hat{d}}$ and one side in $V \setminus V_{\hat{d}}$):

    \[
    E_{\hat{d}} \triangleq \left\{ (v_1, v_2)  |  (v_1 \in V_{\hat{d}}) \vee (v_2 \in V_{\hat{d}})    \right\}
    \]

    \[
    E_{\hat{d}}^{in} \triangleq \left\{ (v_1, v_2)  |  (v_1 \in V_{\hat{d}}) \wedge (v_2 \in V_{\hat{d}})    \right\}
    \]

    \[
    E_{\hat{d}}^{out} \triangleq E_{\hat{d}} \setminus E_{\hat{d}}^{in}
    \]

\end{definition}
Note that for every value of $\hat{d}$ the values of $|E_{\hat{d}}^{in}|$ and $|E_{\hat{d}}^{out}|$ can trivially be acquired from the data. We can therefore calculate the ratio between the number of ``internal edges'' and the number of overall edges for $\hat{d}$, which we denote as the internal connectivity for degree $\hat{d}$.

\begin{definition}
\label{def.betta}
    Let $\beta_{\hat{d}}$ denote the internal connectivity of the network $G$ for the nodes of degree $\hat{d}$.

    \[
    \beta_{\hat{d}} \triangleq \frac{\left| E_{\hat{d}}^{in} \right|}{\left| E_{\hat{d}}\right|}
    \]

    Since the number of edges in $E_{\hat{d}}$ equals the sum of degrees of the nodes in $V_{\hat{d}}$ minus the edges that has their two sides in $V_{\hat{d}}$ (as these are counted twice), and since the degrees of the nodes in $V_{\hat{d}}$ is exactly $\hat{d}$ we get:
    \[
    \beta_{\hat{d}} = \frac{\left| E_{\hat{d}}^{in}\right|}{|V_{\hat{d}}| \cdot \hat{d} - \left| E_{\hat{d}}^{in} \right|}
    \]

\end{definition}

Note that the value of $\beta_{\hat{d}}$ can be easily calculated from the data for every value of $\hat{d}$. Some groups of nodes of the same degree may have high internal connectivity and some may have a lower one. It is therefore interesting to ask: what is the expected value of the internal connectivity of various degrees, and above which value a given internal connectivity would be considered ``too strong'' (i.e. representing a local network structure whose probability to spontaneously emerge in a network with power-law distribution of the degree is extremely low).

Theorem \ref{thm.beta} provides a criterion that for every degree $\hat{d}$ greater than $d_{min}$ can positively guarantee that the set of nodes of degree $\hat{d}$ are ``too anomalous'' with probability $(1 - \epsilon)$ for every small threshold $\epsilon$. This criterion applies for the internal connectivity of this group of nodes (that is, the ratio between the number of edges connecting nodes of the same degree, and edges connecting nodes of different degrees), and depends only on the degree $\hat{d}$ and the network structural properties $\alpha$, $|E_{\star}|$, $d_{min}$ and $d_{max}$.

\begin{theorem}
\label{thm.beta}
    If:
    \[
    \forall v \in V \ , \ \forall d \geq d_{min} \ , \ P[deg(v) = d] = c \cdot d^{-\alpha}
    \]
    then for any small threshold $\epsilon$ and any degree $\hat{d} > d_{min}$ the group of nodes of degree $\hat{d}$ with internal connectivity $\beta_{\hat{d}}$ is considered ``unlikely to spontaneously emerge'' with respect to the threshold $\epsilon$ if the following is satisfied by $\beta_{\hat{d}}$:

    \[
    \beta_{\hat{d}} \geq
    \]
    \[\max\left\{
    \frac{B_{1} \left( \sqrt{  B_{2}^2  - 4 B_{3} }   -    B_{2} \right)}{\lambda \hat{d}^{1-\alpha} - B_{1} \left( \sqrt{  B_{2}^2  - 4 B_{3} }   -    B_{2} \right)} , B_{4}  \right\}
    \]
    where:
    \[
    B_{1} = \frac{1}{2 + \frac{\lambda \hat{d}^{1-\alpha}}{|E_{\star}|} }
    \]
    \[
    B_{2} = 1 - \frac{\lambda^2 \hat{d}^{2-2\alpha}}{|E_{\star}|} + \left( \Phi^{-1}(1 - \epsilon) \right)^2   \frac{\lambda \hat{d}^{1-\alpha}}{2|E_{\star}|}
    \]
    \[
    B_{3} = \left( 0.5 - \frac{\lambda^2 \hat{d}^{2-2\alpha}}{2|E_{\star}|}  \right)^2   -
    \]
    \[
    \left( \Phi^{-1}(1 - \epsilon) \right)^2   \frac{\lambda^2 \hat{d}^{2-2\alpha}}{2|E_{\star}|} \left(1 - \frac{\lambda \hat{d}^{1-\alpha}}{2|E_{\star}|}\right)
    \]
    \[
    B_{4} = \frac{\lambda^2 \hat{d}^{2 - 2\alpha} |E_{\star}|^{-1} - 1}{2 \lambda \hat{d}^{1 - \alpha} + 1}
    \]
\end{theorem}
\begin{proof}

    For $\hat{d} \geq d_{min}$ and for an edge $e \in E_{\hat{d}}$ we know by definition that at least one of its sides is in $V_{\hat{d}}$. The probability that the second side of this edge also ends in $V_{\hat{d}}$ as well is affected by two factors: first, the more nodes there are in $V_{\hat{d}}$ the greater the chance that they would be selected for the edge $e$. Second, as the network has a power-law degree distribution we assume it also follows the ``preferential attachment'' principle (meaning that nodes with higher degree are more likely to acquire new connections, leading to a skew in the distribution). Under this principle, the probability of a node being chosen for a new connection is proportional to its degree, and the probability for the edge $e$ to have both sides in $V_{\hat{d}}$ is therefore a Bernoulli trial with success probability $p = \frac{\hat{d} \cdot |V_{\hat{d}}|}{ \sum_{d=d_{min}}^{d_{max}}(d \cdot |V_{d}|) } = \frac{\hat{d} \cdot |V_{\hat{d}}|}{2|E_{\star}|}$:

    \[
        E_{\hat{d}}^{in} \sim Binomial\left(\left|E_{\hat{d}}\right|, \frac{\hat{d} \cdot |V_{\hat{d}}|}{2|E_{\star}|}\right)
    \]

    A value of $\beta_{\hat{d}}$ that is ``too high'' (i.e. unlikely to occur with a probability less than some threshold $\epsilon$) means that the number of successes from $E_{\hat{d}}$ trials is greater than some threshold $x_{\text{thres}}$.

    To calculate an upper threshold for the number of successes in our Bernoulli trials that is considered statistically significant at level $\epsilon$, and since our number of trials $N$ is large, and our success probability $p$ is not too close to 0 or 1, the Binomial distribution can be approximated by a Normal distribution. If $X \sim Binomial(N, p)$, it can be approximated as $X \approx Y$ where $Y \sim Normal(\mu, \sigma^2)$ with $\mu = Np$ and $\sigma^2 = Np(1-p)$:
    \[
    X \sim Normal\left(Np, Np(1-p)\right)
    \]

    This approximation allows us to express the cumulative distribution function (CDF) of $X$ in terms of the CDF of the Normal distribution $\Phi$ as follows:

    \[ P(X \leq x) \approx \Phi\left(\frac{x + 0.5 - Np}{\sqrt{Np(1-p)}}\right) \]


    Having $\Phi(x)$ formally defined as:
    \[
    \Phi(x) = \frac{1}{\sqrt{2 \pi}} \int_{-\infty}^{x} e^{-\frac{1}{2}t^{2}} dt
    \]

    To find a threshold $x_{\text{thres}}$ such that the probability of observing more than $x_{\text{thres}}$ successes is less than some small $\epsilon$, we can use the Normal approximation to write:

    \[ 1 - \Phi\left(\frac{x_{\text{thres}} + 0.5 - Np}{\sqrt{Np(1-p)}}\right) < \epsilon \]

    Or equivalently,

    \[ \Phi^{-1}(1 - \epsilon) = \frac{x_{\text{thres}} + 0.5 - Np}{\sqrt{Np(1-p)}} \]

    Solving for $x_{\text{thres}}$ gives:

    \[ x_{\text{thres}} = \Phi^{-1}(1 - \epsilon) \sqrt{Np(1-p)} + Np - 0.5 \]

    This provides the threshold value of the number of successes for which the probability of achieving that number or more is less than $\epsilon$.

    Using the original expressions for $N$ and $p$ this means that the number of internal edges above which a group of nodes with a given degree is considered ``unlikely'' is:

    \begin{align*}
    x_{\text{thres}} = & \Phi^{-1}(1 - \epsilon) \sqrt{\left|E_{\hat{d}}\right| \frac{\hat{d} |V_{\hat{d}}|}{2|E_{\star}|} \left(1 - \frac{\hat{d} |V_{\hat{d}}|}{2|E_{\star}|}    \right)} \\
    & + \left|E_{\hat{d}}\right| \frac{\hat{d} |V_{\hat{d}}|}{2|E_{\star}|} - 0.5
    \end{align*}

    Since we already know that:
    \[
    \left|E_{\hat{d}}\right| = {|V_{\hat{d}}| \cdot \hat{d} - \left| E_{\hat{d}}^{in} \right|}
    \]
    This means that:
    \[
    x_{\text{thres}} = \Phi^{-1}(1 - \epsilon)
    \]
    \[
    \times \sqrt{ \left(|V_{\hat{d}}| \hat{d} - x_{\text{thres}}\right) \frac{\hat{d} |V_{\hat{d}}|}{2|E_{\star}|} \left(1 - \frac{\hat{d} |V_{\hat{d}}|}{2|E_{\star}|} \right)}
    \]
    \[
    + \left(|V_{\hat{d}}| \hat{d} - x_{\text{thres}}\right) \frac{\hat{d} |V_{\hat{d}}|}{2|E_{\star}|} - 0.5
    \]

    Rearranging:
    \[
     x_{\text{thres}} \left(1 +  \frac{\hat{d} |V_{\hat{d}}|}{2|E_{\star}|}  \right)   + 0.5 - \frac{\hat{d}^2 |V_{\hat{d}}^2|}{2|E_{\star}|} =
    \]
    \begin{equation}
    \Phi^{-1}(1 - \epsilon)  \sqrt{ \left(|V_{\hat{d}}| \hat{d} - x_{\text{thres}}\right) \frac{\hat{d} |V_{\hat{d}}|}{2|E_{\star}|} \left(1 - \frac{\hat{d} |V_{\hat{d}}|}{2|E_{\star}|} \right)}
    \label{eq1}
    \end{equation}

    Squaring both sides:
    \[
    A_{1} x_{\text{thres}}^2 + A_{2} x_{\text{thres}} + A_{3} = 0
    \]
    where:
    \[
    A_{1} = \left(1 + \frac{\hat{d} |V_{\hat{d}}|}{2|E_{\star}|}  \right)^2
    \]
    \[
    A_{2} = \left(1 + \frac{\hat{d} |V_{\hat{d}}|}{2|E_{\star}|}  \right) \times
    \]
    \[
    \left( 1 - \frac{\hat{d}^2 |V_{\hat{d}}|^2}{|E_{\star}|} + \left( \Phi^{-1}(1 - \epsilon) \right)^2   \frac{\hat{d} |V_{\hat{d}}|}{2|E_{\star}|}  \right)    \]
    \[
    A_{3} = \left( 0.5 - \frac{\hat{d}^2 |V_{\hat{d}}|^2}{2|E_{\star}|}  \right)^2   -
    \]
    \[
    \left( \Phi^{-1}(1 - \epsilon) \right)^2   \frac{\hat{d}^2 |V_{\hat{d}}|^2}{2|E_{\star}|} \left(1 - \frac{\hat{d} |V_{\hat{d}}|}{2|E_{\star}|}\right)
    \]

    Solving this quadratic equation, and noting that $A_{1} > 0$ , yields:
    \[
    x_{\text{thres}} =     B_{1} \left( \sqrt{  B_{2}^2  - 4 B_{3} }   -    B_{2} \right)
    \]
    \[
    B_{1} = \frac{1}{2 + \frac{\hat{d} |V_{\hat{d}}|}{|E_{\star}|} }
    \]
    \[
    B_{2} = 1 - \frac{\hat{d}^2 |V_{\hat{d}}|^2}{|E_{\star}|} + \left( \Phi^{-1}(1 - \epsilon) \right)^2   \frac{\hat{d} |V_{\hat{d}}|}{2|E_{\star}|}
    \]
    \[
    B_{3} = \left( 0.5 - \frac{\hat{d}^2 |V_{\hat{d}}|^2}{2|E_{\star}|}  \right)^2   -
    \]
    \[
    \left( \Phi^{-1}(1 - \epsilon) \right)^2   \frac{\hat{d}^2 |V_{\hat{d}}|^2}{2|E_{\star}|} \left(1 - \frac{\hat{d} |V_{\hat{d}}|}{2|E_{\star}|}\right)
    \]

    Using Lemma \ref{lemma.num.of.nodes.with.deg.d} this can be written as:
    \[
    x_{\text{thres}} =     B_{1} \left( \sqrt{  B_{2}^2  - 4 B_{3} }   -    B_{2} \right)
    \]
    \[
    B_{1} = \frac{1}{2 + \frac{\lambda \hat{d}^{1-\alpha}}{|E_{\star}|} }
    \]
    \[
    B_{2} = 1 - \frac{\lambda^2 \hat{d}^{2-2\alpha}}{|E_{\star}|} + \left( \Phi^{-1}(1 - \epsilon) \right)^2   \frac{\lambda \hat{d}^{1-\alpha}}{2|E_{\star}|}
    \]
    \[
    B_{3} = \left( 0.5 - \frac{\lambda^2 \hat{d}^{2-2\alpha}}{2|E_{\star}|}  \right)^2   -
    \]
    \[
    \left( \Phi^{-1}(1 - \epsilon) \right)^2   \frac{\lambda^2 \hat{d}^{2-2\alpha}}{2|E_{\star}|} \left(1 - \frac{\lambda \hat{d}^{1-\alpha}}{2|E_{\star}|}\right)
    \]

    Recalling Definition \ref{def.betta} this means that for a given degree $\hat{d}$ a group of nodes would be considered ``unlikely'' for an internal connectivity greater than:
    \[
    \beta_{\hat{d}} \geq \frac{x_{\text{thres}}}{|V_{\hat{d}}| \cdot \hat{d} - x_{\text{thres}}}
    \]

    In addition, recalling Equation \ref{eq1} and noting that both $\Phi^{-1}(1-\epsilon)$ and the square root are always positive, we can see that~:
    \[
     x_{\text{thres}} \left(1 +  \frac{\hat{d} |V_{\hat{d}}|}{2|E_{\star}|}  \right)   + 0.5 - \frac{\hat{d}^2 |V_{\hat{d}}^2|}{2|E_{\star}|} \geq 0
    \]
    meaning that~:
    \[
     x_{\text{thres}} \geq \frac{\hat{d}^2 |V_{\hat{d}}^2| - |E_{\star}|}{\hat{d} |V_{\hat{d}}| + 2|E_{\star}|}
    \]
    and using Lemma \ref{lemma.num.of.nodes.with.deg.d} written as:
    \[
     x_{\text{thres}} \geq \frac{\lambda^2 \hat{d}^{2 - 2\alpha} - |E_{\star}|}{\lambda \hat{d}^{1 - \alpha} + 2|E_{\star}|}
    \]

    Recalling Definition \ref{def.betta} we can see that~:
    \[
    \beta_{\hat{d}} \geq \frac{\frac{\lambda^2 \hat{d}^{2 - 2\alpha} - |E_{\star}|}{\lambda \hat{d}^{1 - \alpha} + 2|E_{\star}|}} {\lambda \hat{d}^{1 - \alpha} -    \frac{\lambda^2 \hat{d}^{2 - 2\alpha} - |E_{\star}|}{\lambda \hat{d}^{1 - \alpha} + 2|E_{\star}|}} \geq
    \]
    \[
    \frac{\lambda^2 \hat{d}^{2 - 2\alpha} |E_{\star}|^{-1} - 1}{2 \lambda \hat{d}^{1 - \alpha} + 1}
    \]

\end{proof}

Note that the memory and time complexity of this approach is remarkably low due to the aforementioned techniques, allowing for regular refreshes as new data is obtained. This low complexity, coupled with the utilization of large k for k-mers, enhances our prediction accuracy significantly. The expressivity of large k-mers, which refers to their ability to encapsulate functionality, contributes to the robustness and precision of our analysis.

\begin{figure*}[htbp]
    \centering
    \includegraphics[width=0.7\textwidth]{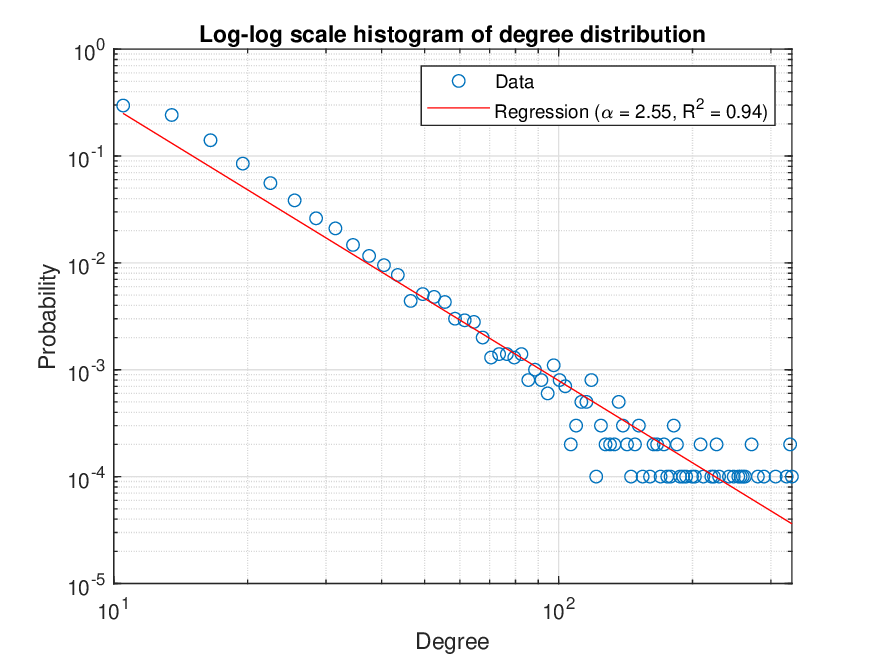}
    \includegraphics[width=\textwidth]{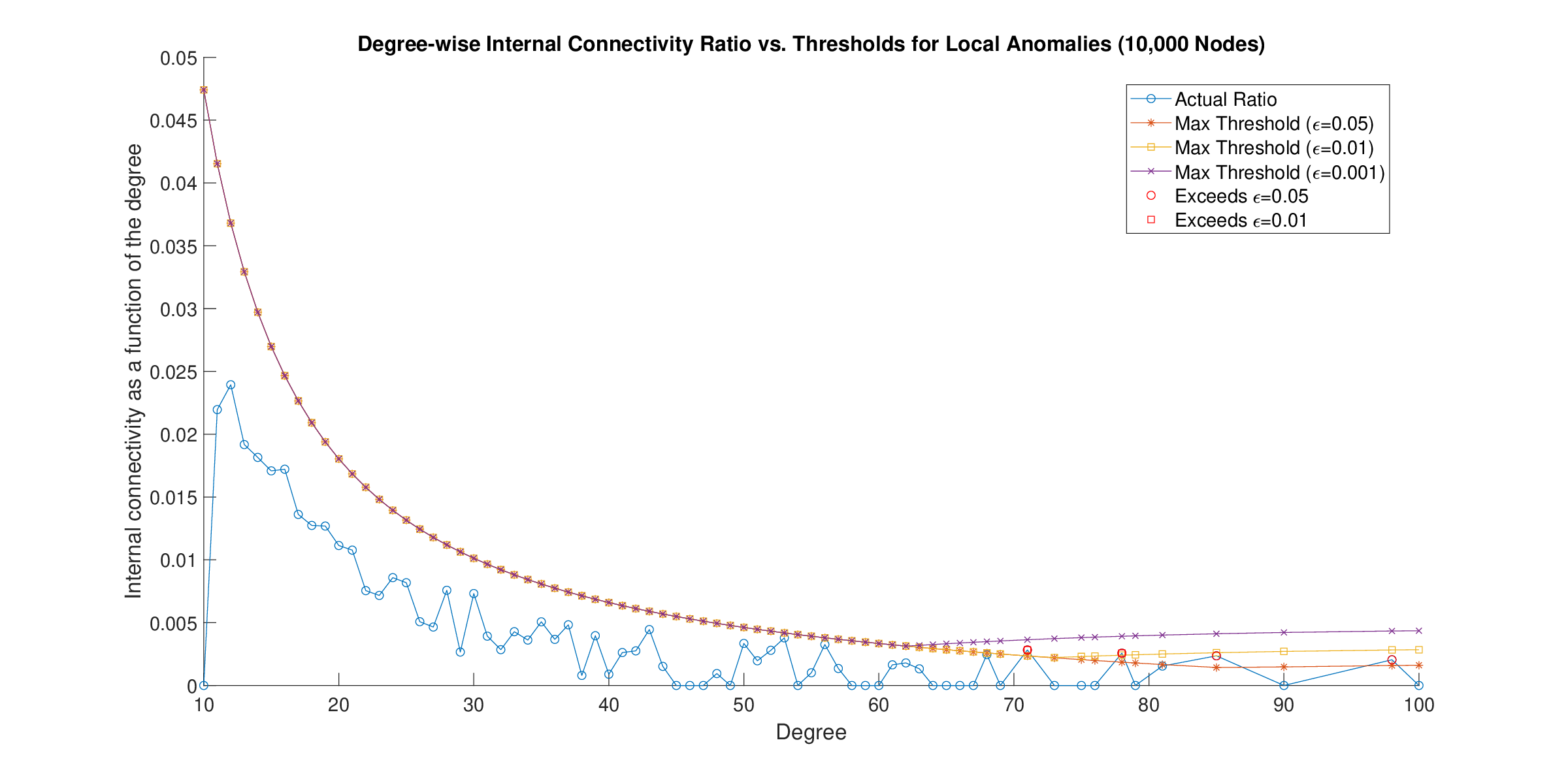}
    \caption{Illustration of Theorem \ref{thm.beta} contrasting the probabilistic limit on the emergence of clusters of nodes with the same degree that exhibit high internal connectivity, against the internal connectivity observed in a randomly generated graph \( G \) using the Preferential Attachment method \cite{newman2001clustering} for 10,000 nodes. The upper chart shows the graph's degree distribution, affirming its adherence to power-law dynamics. The chart below illustrates the theoretical constraints on the internal connectivity of nodes of degree \( d \) across all observed degrees in graph \( G \), for varying statistical confidence levels \( \epsilon = 0.05, 0.01, 0.001 \). As anticipated, a more stringent \( \epsilon \) results in fewer degrees that can form node groups with internal connectivity exceeding the predicted bounds.}
    \label{fig:thm5_10000}
\end{figure*}

\begin{figure*}[htbp]
    \centering
    \includegraphics[width=0.7\textwidth]{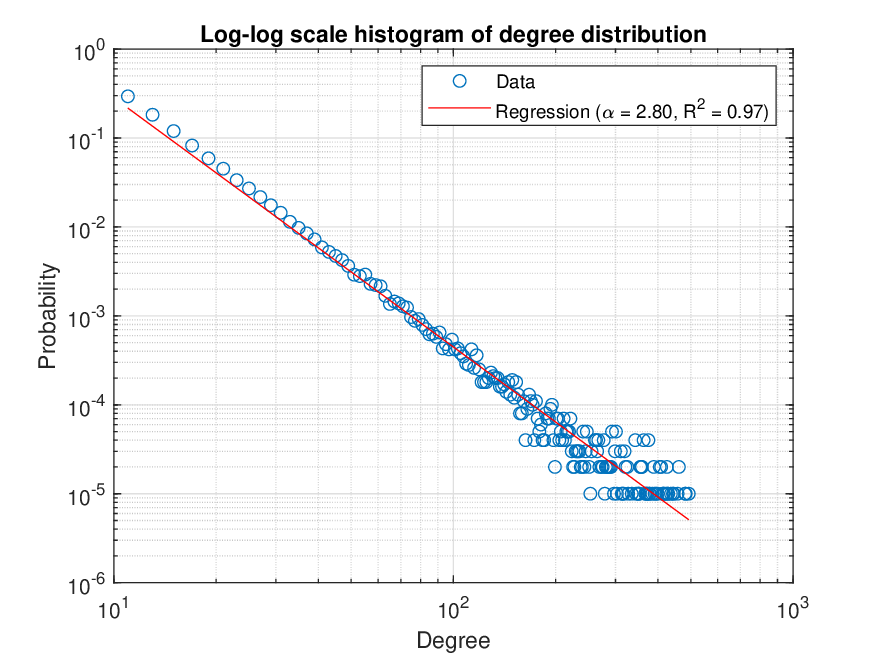}
    \includegraphics[width=\textwidth]{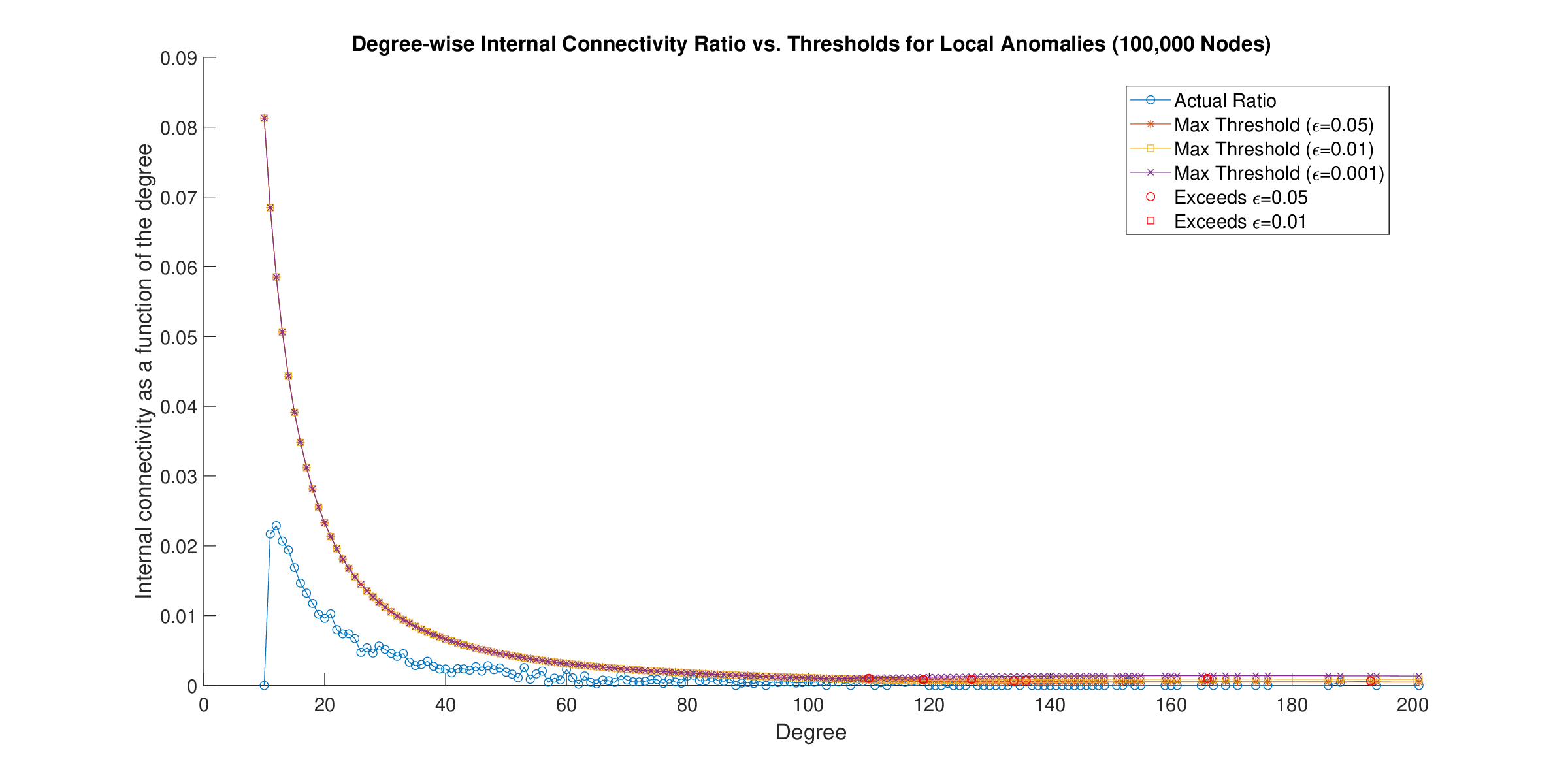}
    \includegraphics[width=\textwidth]{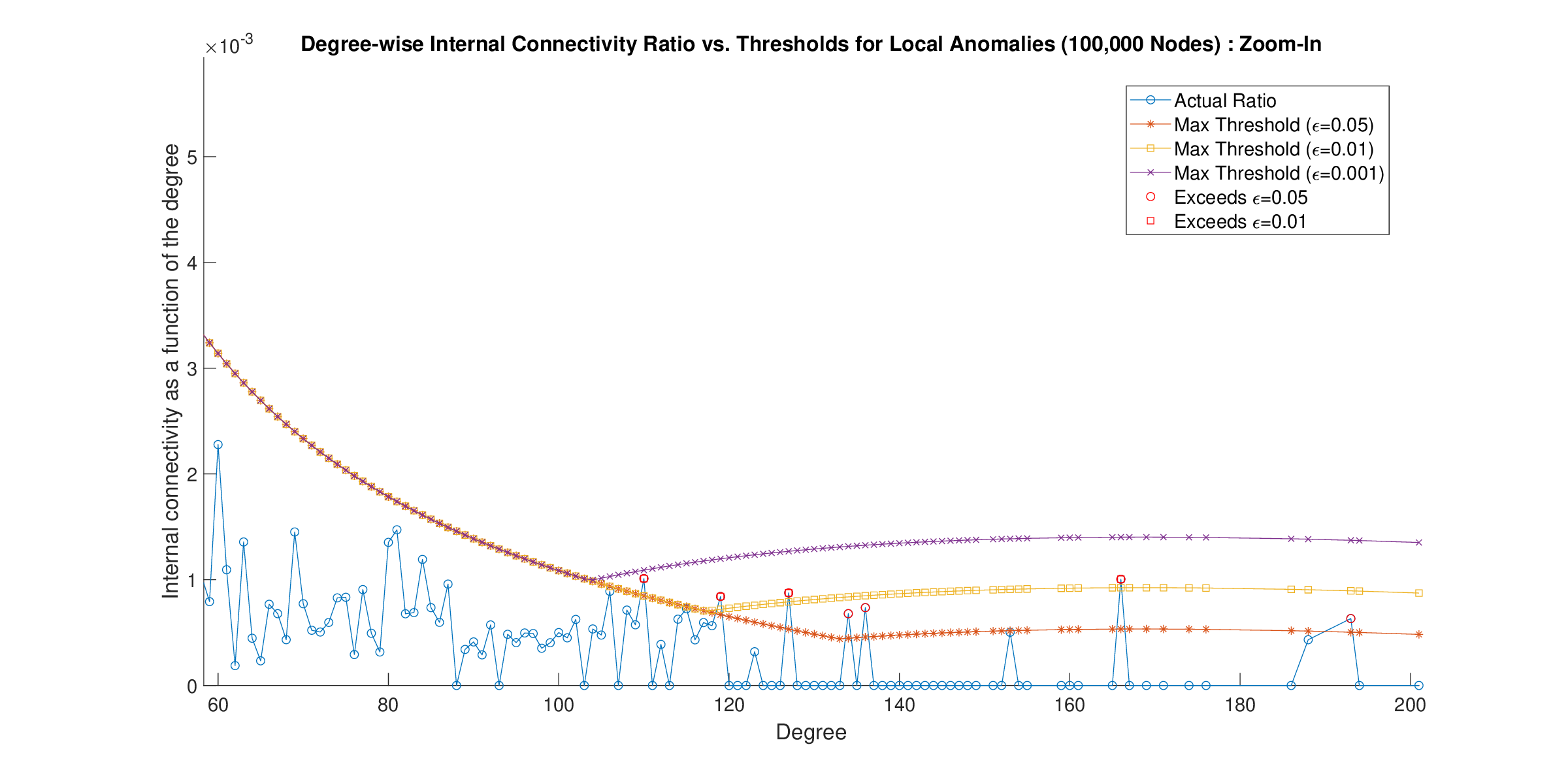}
    \caption{An illustration of Theorem \ref{thm.beta} contraststing the probabilistic boundary regarding the appearance of node clusters of the same degree with elevated internal connectivity, with the observed connectivity in a graph \( G \) of 100,000 nodes, generated via the Preferential Attachment method \cite{newman2001clustering}. The top chart displays the graph's degree distribution, reinforcing its alignment with power-law behavior. The middle chart delineates the theoretical limits on the internal connectivity for nodes of degree \( d \) throughout all present degrees in \( G \), set against distinct statistical confidence levels \( \epsilon = 0.05, 0.01, 0.001 \). As expected, a tighter \( \epsilon \) narrows the scope of degrees where node clusters exceed the anticipated connectivity values. The bottom chart offers a magnified perspective on the data from the middle chart.}
    \label{fig:thm5_100000}
\end{figure*}

\begin{figure*}[htbp]
    \centering
    \includegraphics[width=0.7\textwidth]{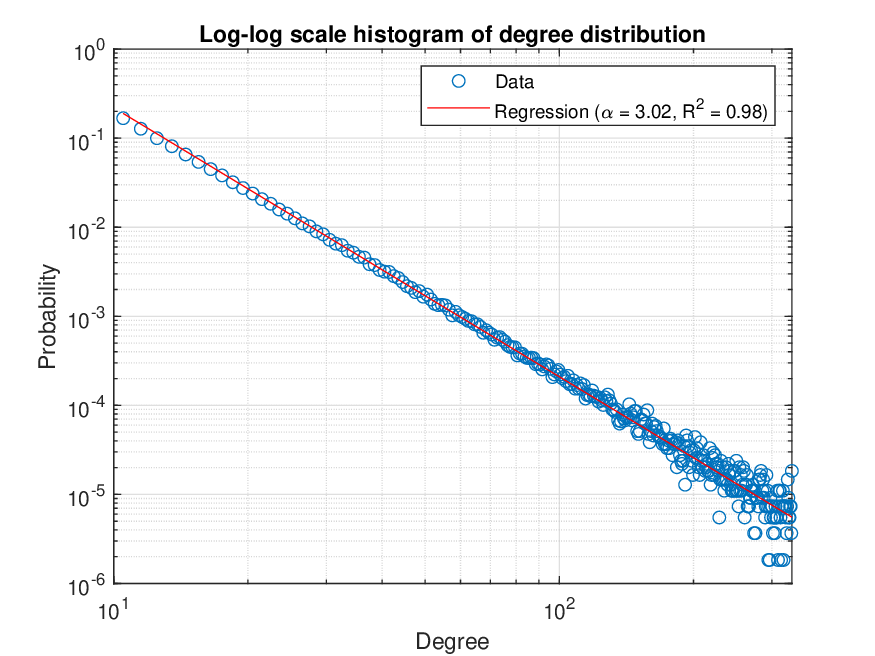}
    \includegraphics[width=\textwidth]{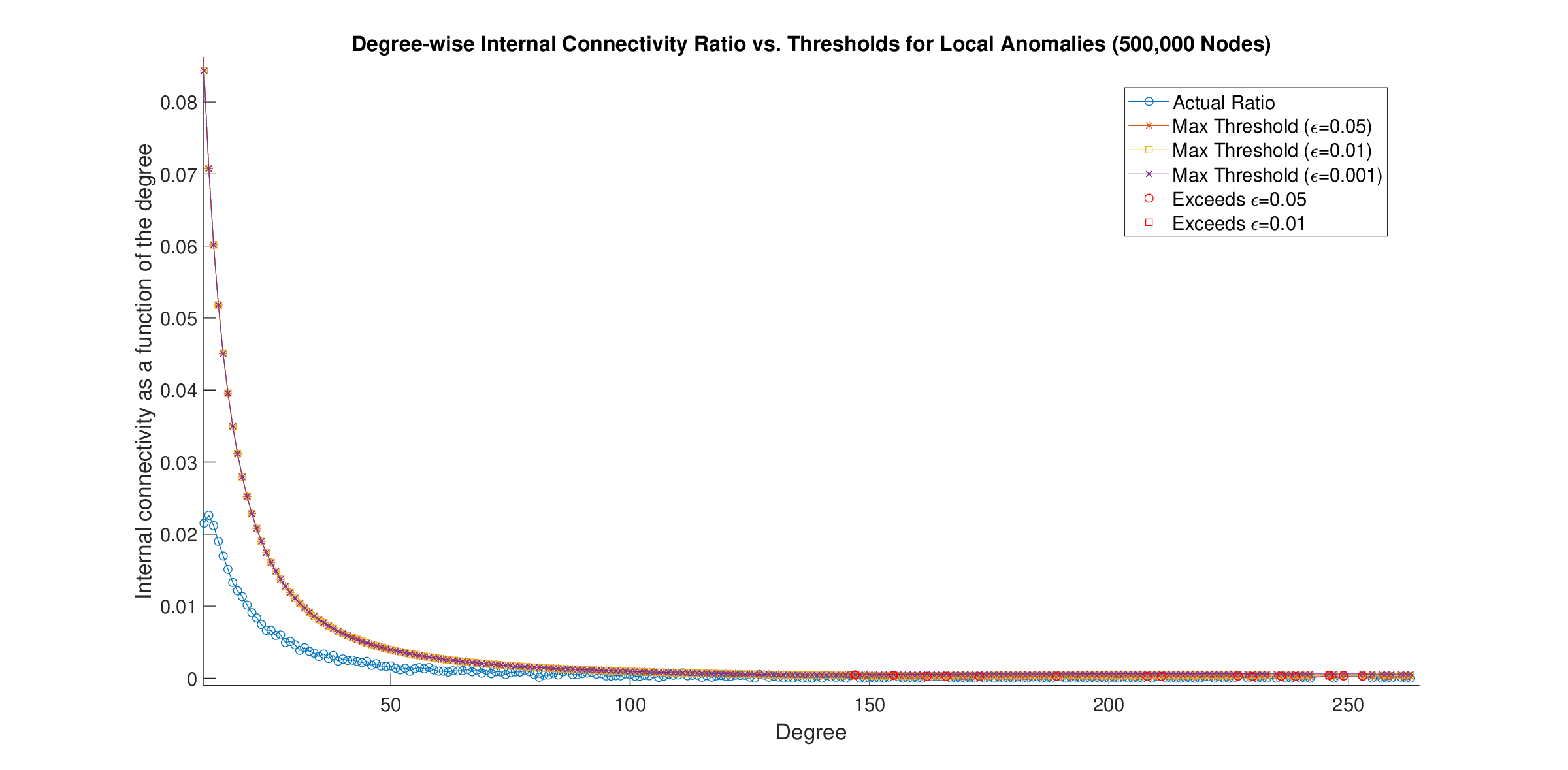}
    \includegraphics[width=\textwidth]{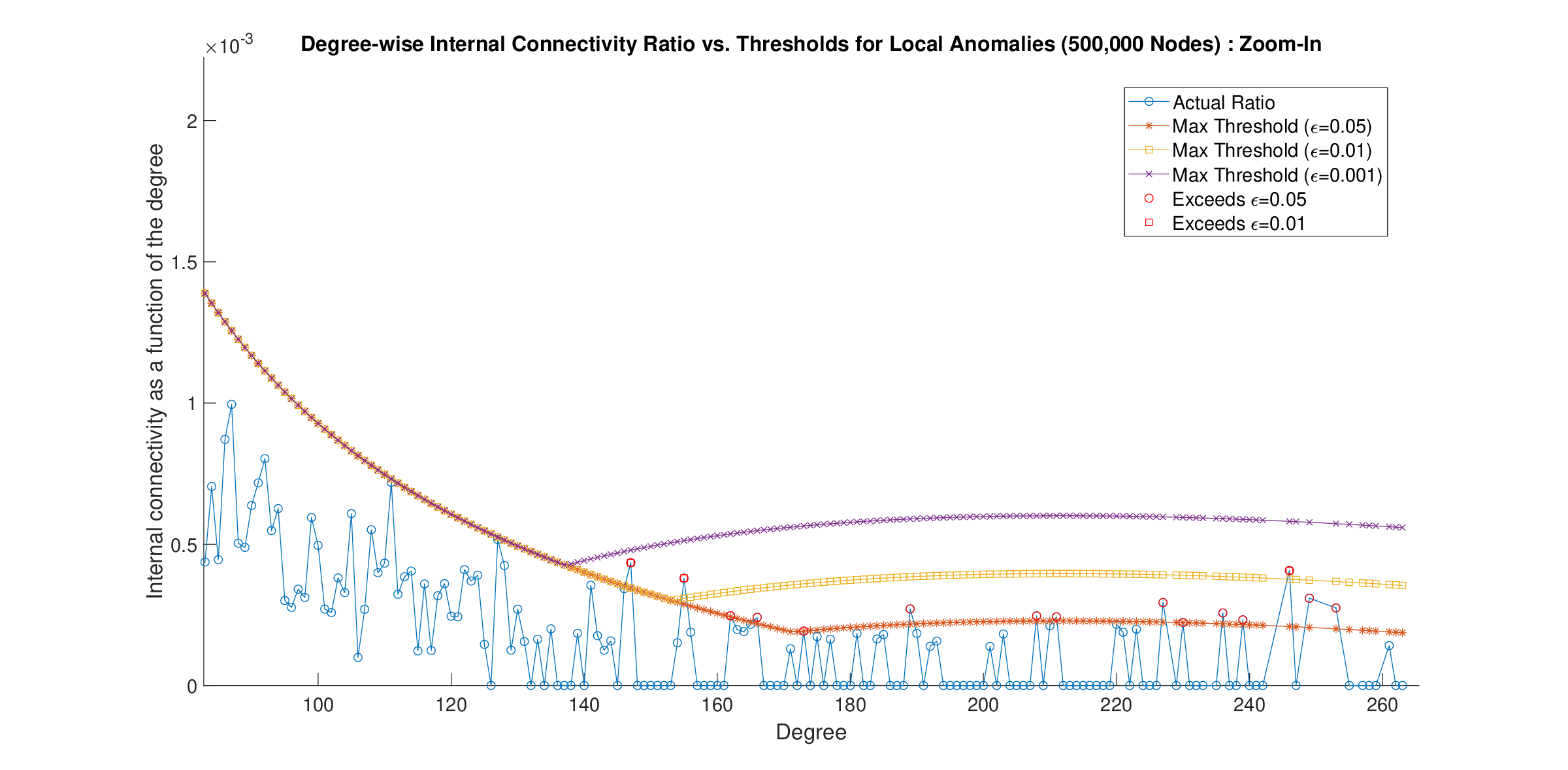}
    \caption{An illustration of Theorem \ref{thm.beta} contraststing the probabilistic boundary regarding the appearance of node clusters of the same degree with elevated internal connectivity, with the observed connectivity in a graph \( G \) of 500,000 nodes, generated via the Preferential Attachment method \cite{newman2001clustering}. The top chart displays the graph's degree distribution, reinforcing its alignment with power-law behavior. The middle chart delineates the theoretical limits on the internal connectivity for nodes of degree \( d \) throughout all present degrees in \( G \), set against distinct statistical confidence levels \( \epsilon = 0.05, 0.01, 0.001 \). As expected, a tighter \( \epsilon \) narrows the scope of degrees where node clusters exceed the anticipated connectivity values. The bottom chart offers a magnified perspective on the data from the middle chart.}
    \label{fig:thm5_500000}
\end{figure*}

Theorem \ref{thm.beta} is illustrated in Figures \ref{fig:thm5_10000}, \ref{fig:thm5_100000}, and \ref{fig:thm5_500000}. These figures contrast the thresholds predicted by the theorem for various confidence levels, \(\epsilon\), with graphs of differing sizes that conform to the power-law model. For each degree \(d\) that has at least 5 nodes, the internal connectivity of these nodes is contrasted with the ``unlikelihood thresholds'' outlined by Theorem \ref{thm.beta}. This highlights that while power-law graphs may inherently display certain patterns, arbitrary graphs -- without specific embedded signals -- are rarely expected to exhibit such local anomalies.

Drawing from Theorem \ref{thm.beta}, we possess a powerful tool to systematically probe k-mer networks, enabling us to swiftly identify and ``extract'' groups of nodes, representing specific k-mers. These k-mers are analytically established to be linked to distinct phenotypic attributes. The intriguing aspect of this linkage is that it might be directly or indirectly correlated with the biological state or condition we aim to elucidate. Several noteworthy implications emerge from this:

\begin{itemize}
  \item \textbf{Relevancy Determination:} It is rather straightforward to assess the significance of a k-mer cluster concerning a particular phenotypic trait. This can be achieved by juxtaposing the frequency of its manifestation across various samples against the intensity or prevalence of the attribute in question.
  \item \textbf{Conditional Dynamics:} The pertinence of k-mer clusters is not static. For instance, a cluster deemed irrelevant to a specific property during certain periods (like summer) might emerge as highly pertinent in different settings (say, during winter, when the biological functionality associated with it becomes more relevant).
  \item \textbf{Evolution of the Repository:} As we continuously enrich our database with k-mer clusters, each endorsed by empirical evidence to bear some functional significance, the value of this repository amplifies. This augmentation manifests in two dimensions: first, the broad spectrum of biological attributes it can shed light on, and second, the precision with which it can make such predictions.
\end{itemize}

In essence, as we cultivate this burgeoning collection of functionally significant k-mer clusters, we enhance our capacity to decipher an array of biological phenomena with increased accuracy and breadth.

\begin{figure*}[htbp]
    \centering
    \includegraphics[width=\textwidth]{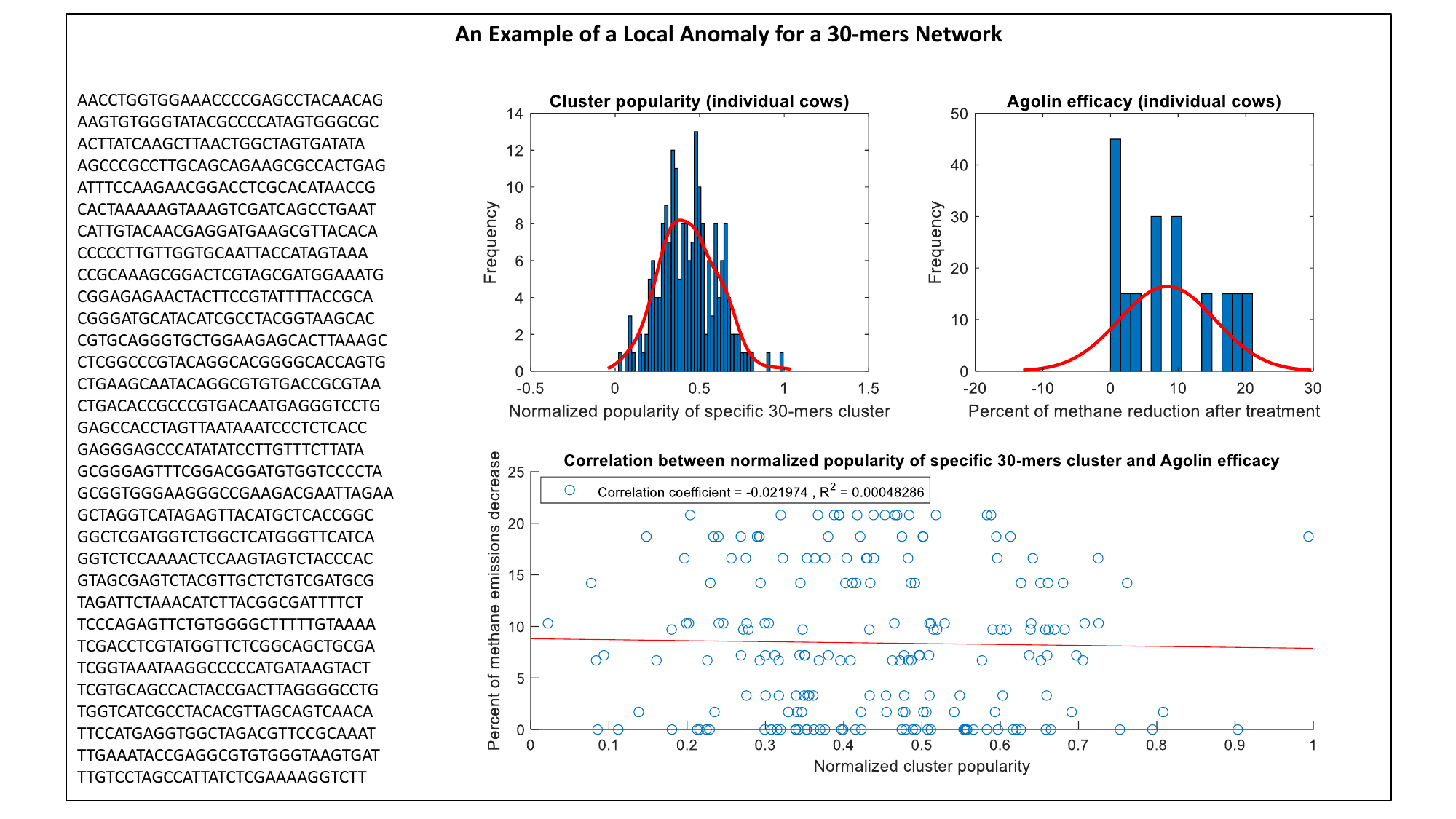}
    \caption{This illustration showcases a cluster of 32 k-mers (with \(k = 30\)) derived from combined microbiome samples of all collected samples, integrated into a single network. In this network, built using edges that represent the co-occurrence of k-mers in shared reads, all these k-mers exhibit the same degree. Their internal connectivity surpasses the threshold stipulated by Theorem \ref{thm.beta}, which suggests such a pattern is improbable to arise spontaneously in a power-law network.
    The chart on the top-left visualizes the cluster's prevalence among the Microbiome Train cows, assessed by the frequency of k-mer appearances in their samples, normalized against the average frequency. In contrast, the top-right chart represents the efficacy of the Agolin additive, determined for the Methane Train cows that originate from the same farms as the Microbiome Train cows. Here, each Microbiome Train cow's efficacy is linked to its farm's average efficacy, calculated from the corresponding Methane Train cows.
    The bottom chart juxtaposes these two datasets in a scatter plot, seeking a potential linear relationship. Notably, it reveals an absence of linear correlation between the abundance of this k-mer cluster and Agolin's effectiveness.}
    \label{fig:clusters1}
\end{figure*}

\begin{figure*}[htbp]
    \centering
    \includegraphics[width=\textwidth]{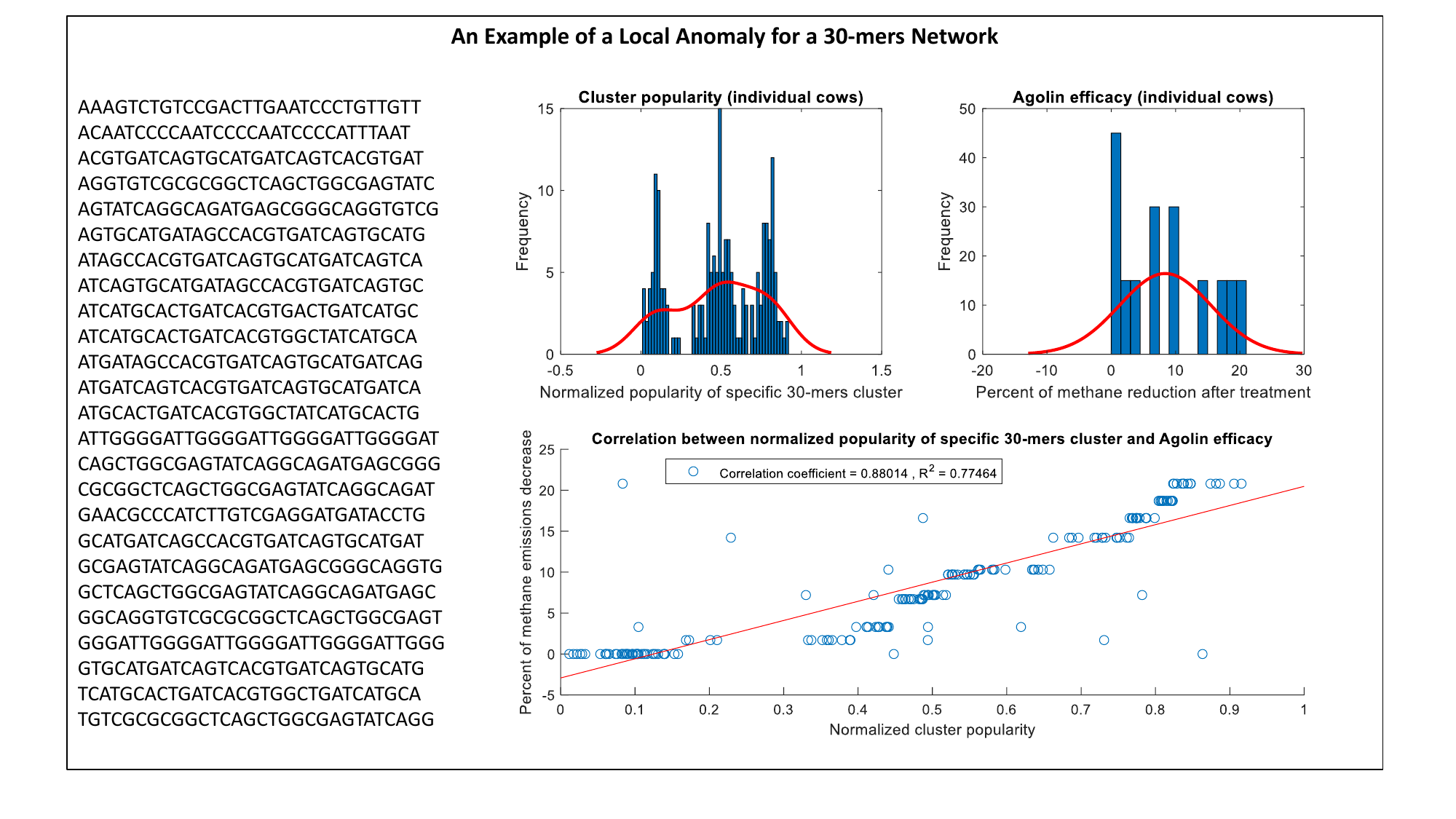}
    \caption{This illustration showcases a cluster of 25 k-mers (with \(k = 30\)) derived from combined microbiome samples of all collected samples, integrated into a single network. In this network, built using edges that represent the co-occurrence of k-mers in shared reads, all these k-mers exhibit the same degree. Their internal connectivity surpasses the threshold stipulated by Theorem \ref{thm.beta}, which suggests such a pattern is improbable to arise spontaneously in a power-law network.
    The chart on the top-left visualizes the cluster's prevalence among the Microbiome Train cows, assessed by the frequency of k-mer appearances in their samples, normalized against the average frequency. In contrast, the top-right chart represents the efficacy of the Agolin additive, determined for the Methane Train cows that originate from the same farms as the Microbiome Train cows. Here, each Microbiome Train cow's efficacy is linked to its farm's average efficacy, calculated from the corresponding Methane Train cows.
    The bottom chart contrasts these two datasets in a scatter plot, revealing a discernible statistical association. Specifically, it illustrates that the efficacy of Agolin in farms exhibits a linear rise -- with significant statistical backing -- as the prevalence of this cluster's k-mers increases among the cows of these farms.}
    \label{fig:clusters2}
\end{figure*}

Illustrated in Figures \ref{fig:clusters1} and \ref{fig:clusters2} are two distinct k-mer clusters extracted from microbiome samples, using $\epsilon = 0.001$. These clusters present anomalies within the microbiome k-mer network, indicating a high likelihood of their association with a particular biological trait. While one cluster does not show a significant statistical relationship with the performance of the feed additive Agolin, the other emerges as a compelling correlate. This strong association enables us to employ its presence in microbiome samples as a reliable marker for anticipating the enhanced effectiveness of the additive.

It's worth noting that the cluster depicted in Figure \ref{fig:clusters1}, which lacks correlation with Agolin's efficacy (as do many other clusters we extracted from the microbiome data), might still prove crucial in relation to other feed additives. Alternatively, this cluster could be relevant to other properties, unrelated to methane, that we might be keen on predicting in the future.

For the creation of the markers used in this study we have produced the following networks:
\begin{itemize}
  \item A dedicated network for each cow.
  \item A dedicated network for each farm, created as the superposition of the networks of the cows of this farm.
  \item A network of the entire data, created as the superposition of the dedicated cow networks.
\end{itemize}

For each of the constructed networks, we employed Theorem \ref{thm.beta} to identify and extract all anomalous 30-mer clusters at a statistical confidence level of \( \epsilon = 0.001 \). This comprehensive extraction yielded a substantial collection of over 20,000 k-mer clusters that exhibited notable statistical significance.

To pinpoint clusters relevant to each feed additive, we undertook a rigorous filtering process. We sought out clusters that exhibited a strong linear correlation with the additive's efficacy, demanding both a correlation coefficient of at least 0.75 in absolute value, and an \( R^2 \) value surpassing 0.7. Through this meticulous filtering, we discerned between 17 to 31 distinct clusters that met our criteria for each specific additive.

We synthesized our results into two markers for each additive. Clusters that demonstrated a positive linear relationship with the additive's efficacy were combined to form what we term the ``top marker''. Conversely, clusters that exhibited a negative linear association were amalgamated into the ``bottom marker''. During this merging process, we prioritized k-mers that were consistently recurrent, retaining only those that appeared in a minimum of two clusters of the same typology.

\section{Discussion}

The study implemented a novel methodology that utilizes the cow microbiome as a basis for predicting the efficacy of methane-reducing additives. The structure of the trial followed a two-step approach: initially, we collected microbiome samples from cattle across an array of farms. These farms were chosen to represent a broad spectrum of conditions, including geographical location, environmental context, herd size, and differing management methods. The genetic data obtained from these samples was then analyzed via unsupervised learning to discern meaningful microbial genetic patterns. In the subsequent stage, a selected subset of cows from these farms was subjected to methane emission measurements under both control and experimental conditions (the latter involving the feed additive). This stage utilized supervised learning, building upon the established microbial baseline from the first step.

The design of our study, combined with the unsupervised genetic material detection methodology, can be extended as a predictive framework for various other properties, given appropriate labeling for the supervised learning stage. This work highlights the positive network effect of this method: as more data are collected and incorporated into the unsupervised learning stage, the accuracy of the predictions improves. When new use-cases are introduced, they contribute additional microbiome samples, further improving the predictive capability across all use-cases.

The division of cows into different groups (e.g. ``control methane cows'', ``train methane cows'', and ``test methane cows'') was a critical part of our study design. It was done randomly, but with meticulous attention to mitigating any potential biases. Several key factors were taken into account, including the age of cows, their days in lactation, and average milk yield. By using these factors to inform our random selection, we were able to create groups that were representative of the general population of the herd. This ensured that the microbiome samples and methane measurements taken from each group were not skewed by any one particular attribute, enhancing the generalizability and applicability of our findings.

\paragraph*{\textbf{Macroscopic Microbial Data Properties:}}

Our analysis revealed a striking and noteworthy pattern in the distribution of the repetition of the reads and the k-mers. Specifically, we observed that the popularity of these repetitions follows a power-law distribution. This observation was consistent and sustained, evident in the data extracted from multiple cows, and across both read and k-mer repetitions.

It is crucial to underscore that the emergence of this power-law distribution is data-agnostic. This means that it is not an artifact of our data acquisition process or specific to the type of data being studied. Indeed, the power-law distribution we observed suggests the existence of underlying principles governing these distributions, which extend beyond the specifics of our study.

One might suggest that the `shotgun' metagenomic sequencing method, which we employed to sequence the DNA samples, could somehow skew the data towards this power-law distribution. However, this is highly unlikely. While the sequencing method can indeed introduce certain biases into the data, there is no reason to suspect that it would give rise to a power-law distribution. This is because sequencing, by its nature, is a pseudo random process and, as such, is not predisposed towards creating any specific pattern, much less one as distinctive as a power-law distribution.

Moreover, when we pooled the data from several cows together and then analyzed the distribution of popularity, we consistently observed the power-law distribution. This further strengthens our assertion that the power-law distribution is not an artifact of data acquisition. If it were, the distribution would be different for each cow, reflecting the individual variances in data collection. The consistent manifestation of the power-law distribution across different cows suggests a more profound, universal principle at work.

Interestingly, we observed the power-law distribution in the repetition of k-mers, as well as whole reads. This consistency across different levels of granularity in the data adds another layer of validation to our findings. The emergence of a power-law distribution at different scales strongly suggests the existence of scale-free, fractal-like patterns in the data, which is a key characteristic of power-law distributions.

Taken together, our findings strongly suggest that the observed power-law distribution is not a mere artifact of our methodology, but rather points to fundamental biological phenomena. This could have profound implications for our understanding of the cow microbiome and the factors influencing the efficacy of methane-reducing additives. Further research is required to fully understand the implications of these findings, but they present promising avenues for future investigations.

\paragraph*{\textbf{Limitations:}}

Despite the promising results of our study, we must also recognize its limitations. Firstly, while our methodology attempted to mitigate the potential for bias, with the division of cows into ``control methane cows'', ``train methane cows'', and ``test methane cows'' carefully executed to take into account factors such as the age of cows, their days in lactation, and average milk yield, some bias could still remain. Individual differences in cow physiology, diet, and environmental factors could potentially affect both the composition of the microbiome and the methane production. Future research should consider these factors in more depth, perhaps by introducing stratification or covariate balancing.

Moreover, while our study design can be extended as a predictive framework for various other properties, it is also reliant on the availability of accurate labeling data. Inadequate or inaccurately labeled data during the supervised learning stage could affect the model's predictive capability. The accuracy of our predictions is likely to improve with more data, highlighting the need for continued and broad-scale data collection efforts.

Lastly, our analysis primarily focused on the methane reduction potential of additives. It would be worthwhile to consider other potential effects of these additives on cow health, productivity, and the quality of dairy products.

Recognizing these limitations is vital to guide future research. While our study provides a significant step forward in the use of the cow microbiome to predict the efficacy of methane-reducing additives, there is still much work to be done to refine our understanding and improve our predictive capabilities.

\paragraph*{\textbf{Impact:}}

Our results were promising, demonstrating a correlation between certain microbiome patterns and the efficacy of methane-reducing additives. This allows us to predict the impact of additives based on a cow's microbiome, enhancing the average effect by using the additive only when it is expected to be effective.

The potential impact of these findings is substantial. Given the ongoing need to reduce methane emissions for environmental reasons, the ability to predict the effectiveness of methane-reducing additives at an individual cow level could lead to significant improvements in emission reduction strategies.

Moreover, it is important to emphasize a few additional points. Even though methane measurements were only conducted at some farms, the broad-scale collection of microbiome data enriched the initial unsupervised learning phase, enhancing the detection of significant microbial genetic patterns. There was no overlap between the cow groups from which microbiomes were sampled and those from which methane was measured, affirming that a small microbiome sample is representative of the entire herd's microbiome.

Our approach rests on the assumption of strong microbiome homogeneity across the cows within a single herd. This means that while our microbiome samples were collected from one group of cows, we consider these samples as indicative of the microbiome composition of the entire herd. The validity of this assumption is confirmed when we successfully use these microbiome samples to predict the efficacy of the methane-reducing additives in a distinct group of cows within the same herd. This suggests that despite no overlap in the cows from which the microbiome and methane measurements were taken, our microbiome sample effectively represents the herd's overall microbiome. It's this feature that makes our method both robust and scalable, as it allows for a comprehensive prediction model without requiring invasive and extensive sampling from every cow.

Finally, while the initial implementation of our methodology was conducted in Israel, its design and underlying principles suggest a wide-ranging applicability. This makes it a valuable tool for dairy farming operations worldwide, particularly in the context of sustainable agriculture and environmental management.

\paragraph*{\textbf{Future Work:}}

Our study opens up several intriguing avenues for future research. While we established a novel methodology for predicting the efficacy of methane-reducing additives, the potential of using the cow microbiome as a predictive tool extends beyond this singular application. In the future, our methodology could be employed to predict other biologically significant outcomes based on microbiome composition, such as disease susceptibility or productivity measures like milk yield and growth rate. Such predictive models could aid in early intervention strategies and improve the overall health and productivity of livestock.

Further, our research indicated that a small microbiome sample is representative of the entire herd's microbiome. This paves the way for non-invasive, rapid and cost-effective microbiome sampling techniques that could be used at scale. Developing these techniques and validating their efficacy would be a valuable next step.

Finally, while our study focused primarily on the methane reduction potential of additives, it would be worthwhile to investigate other potential effects of these additives. Are there any unintended consequences of their use, such as changes to cow health or milk quality? Understanding the holistic impact of these additives will provide a more comprehensive view and aid in making informed decisions regarding their widespread use.

Overall, this research has laid the groundwork for an exciting new direction in livestock management. Through continued refinement and expansion of this work, we envision a future where personalized additive regimens of the farm level, informed by each herd's dynamic microbiome, become a standard practice, optimizing both environmental impact and productivity outcomes.

\bibliographystyle{plain}

\begin{thebibliography}{10}

\bibitem{human2012framework}
A framework for human microbiome research.
\newblock {\em nature}, 486(7402):215--221, 2012.

\bibitem{human2012structure}
Structure, function and diversity of the healthy human microbiome.
\newblock {\em nature}, 486(7402):207--214, 2012.

\bibitem{milkProduction2023}
Milk production.
\newblock {\em National Agricultural Statistics Service}, 2023.

\bibitem{adesogan2013mitigation}
T~Adesogan, W~Yang, C~Lee, PJ~Gerber, B~Henderson, and JM~Tricarico.
\newblock Mitigation of methane and nitrous oxide emissions from
  animal—special topics.
\newblock {\em J. Anim. Sci}, 91:5045--5069, 2013.

\bibitem{ahn1effect}
JS~Ahn, JS~Shin, GH~Son, SS~Jang, and BK~Park.
\newblock Effect of allicin and illite supplementation on the methane
  production and growth performance of the beef cattle.
\newblock {\em Indian Journal of Animal Research}, 1:6.

\bibitem{altshuler2019modeling}
Tal Altshuler, Yaniv Altshuler, Rachel Katoshevski, and Yoram Shiftan.
\newblock Modeling and prediction of ride-sharing utilization dynamics.
\newblock {\em Journal of Advanced Transportation}, 2019, 2019.

\bibitem{SR-IIS}
Y.~Altshuler, N.~Aharony, A.~Pentland, Y.~Elovici, and M.~Cebrian.
\newblock Stealing reality: When criminals become data scientists (or vice
  versa).
\newblock {\em Intelligent Systems, IEEE}, 26(6):22--30, nov.-dec. 2011.

\bibitem{appuhamy2013anti}
JAD Ranga~Niroshan Appuhamy, AB~Strathe, S~Jayasundara, C~Wagner-Riddle,
  J~Dijkstra, J~France, and E~Kebreab.
\newblock Anti-methanogenic effects of monensin in dairy and beef cattle: A
  meta-analysis.
\newblock {\em Journal of Dairy Science}, 96(8):5161--5173, 2013.

\bibitem{barabasi2004network}
Albert-Laszlo Barabasi and Zoltan~N Oltvai.
\newblock Network biology: understanding the cell's functional organization.
\newblock {\em Nature reviews genetics}, 5(2):101--113, 2004.

\bibitem{beauchemin2008nutritional}
KA~Beauchemin, M~Kreuzer, F~O’mara, and TA~McAllister.
\newblock Nutritional management for enteric methane abatement: a review.
\newblock {\em Australian Journal of Experimental Agriculture}, 48(2):21--27,
  2008.

\bibitem{bhatta2007measurement}
Raghavendra Bhatta and Osamu Enishi.
\newblock Measurement of methane production from ruminants.
\newblock {\em Asian-australasian journal of animal sciences},
  20(8):1305--1318, 2007.

\bibitem{bishop2016intra}
Gregory~J Bishop-Hurley, David Paull, Philip Valencia, Leslie Overs, Kourosh
  Kalantar-zadeh, Andr{\'e}-Denis~G Wright, and Chris McSweeney.
\newblock Intra-ruminal gas-sensing in real time: a proof-of-concept.
\newblock {\em Animal Production Science}, 56(3):204--212, 2016.

\bibitem{bland1986statistical}
J~Martin Bland and DouglasG Altman.
\newblock Statistical methods for assessing agreement between two methods of
  clinical measurement.
\newblock {\em The lancet}, 327(8476):307--310, 1986.

\bibitem{boland2020feed}
Tommy~M Boland, Karina~M Pierce, Alan~K Kelly, David~A Kenny, Mary~B Lynch,
  Sin{\'e}ad~M Waters, Stephen~J Whelan, and Zoe~C McKay.
\newblock Feed intake, methane emissions, milk production and rumen methanogen
  populations of grazing dairy cows supplemented with various c 18 fatty acid
  sources.
\newblock {\em Animals}, 10(12):2380, 2020.

\bibitem{borlinghaus2014allicin}
Jan Borlinghaus, Frank Albrecht, Martin~CH Gruhlke, Ifeanyi~D Nwachukwu, and
  Alan~J Slusarenko.
\newblock Allicin: chemistry and biological properties.
\newblock {\em Molecules}, 19(8):12591--12618, 2014.

\bibitem{brambila2022evaluation}
Rosalio Brambila and Jorge Noricumbo-Saenz.
\newblock Evaluation of agolin{\textregistered} ruminant, an essential oil
  blend, as a feed additive for cows at two levels of production.
\newblock {\em Open Journal of Animal Sciences}, 12(3):380--389, 2022.

\bibitem{carrazco2020impact}
Angelica~V Carrazco, Carlyn~B Peterson, Yongjing Zhao, Yuee Pan, John~J
  McGlone, Edward~J DePeters, and Frank~M Mitloehner.
\newblock The impact of essential oil feed supplementation on enteric gas
  emissions and production parameters from dairy cattle.
\newblock {\em Sustainability}, 12(24):10347, 2020.

\bibitem{chikhi2014informed}
Rayan Chikhi and Paul Medvedev.
\newblock Informed and automated k-mer size selection for genome assembly.
\newblock {\em Bioinformatics}, 30(1):31--37, 2014.

\bibitem{clauset2009power}
Aaron Clauset, Cosma~Rohilla Shalizi, and Mark~EJ Newman.
\newblock Power-law distributions in empirical data.
\newblock {\em SIAM review}, 51(4):661--703, 2009.

\bibitem{danielsson2017methane}
Rebecca Danielsson, Johan Dicksved, Li~Sun, Horacio Gonda, Bettina M{\"u}ller,
  Anna Schn{\"u}rer, and Jan Bertilsson.
\newblock Methane production in dairy cows correlates with rumen methanogenic
  and bacterial community structure.
\newblock {\em Frontiers in microbiology}, 8:226, 2017.

\bibitem{de2020comparison}
Camila~Flavia de~Assis~Lage, Susanna~Elizabeth R{\"a}is{\"a}nen, Audino Melgar,
  Krum Nedelkov, Xianjiang Chen, Joonpyo Oh, Molly~Elizabeth Fetter, Nagaraju
  Indugu, Joseph~Samuel Bender, Bonnie Vecchiarelli, et~al.
\newblock Comparison of two sampling techniques for evaluating ruminal
  fermentation and microbiota in the planktonic phase of rumen digesta in dairy
  cows.
\newblock {\em Frontiers in Microbiology}, 11:618032, 2020.

\bibitem{dengel2011methane}
Sigrid Dengel, Peter~E Levy, John Grace, Stephanie~K Jones, and Ute~M Skiba.
\newblock Methane emissions from sheep pasture, measured with an open-path eddy
  covariance system.
\newblock {\em Global Change Biology}, 17(12):3524--3533, 2011.

\bibitem{metha-intro6}
Gareth~Frank Difford, Damian~Rafal Plichta, Peter L{\o}vendahl, Jan Lassen,
  Samantha~Joan Noel, Ole H{\o}jberg, Andr{\'e}-Denis~G Wright, Zhigang Zhu,
  Lise Kristensen, Henrik~Bj{\o}rn Nielsen, et~al.
\newblock Host genetics and the rumen microbiome jointly associate with methane
  emissions in dairy cows.
\newblock {\em PLoS genetics}, 14(10):e1007580, 2018.

\bibitem{duffield2004comparison}
T~Duffield, JC~Plaizier, A~Fairfield, R~Bagg, G~Vessie, P~Dick, J~Wilson,
  J~Aramini, and B~McBride.
\newblock Comparison of techniques for measurement of rumen ph in lactating
  dairy cows.
\newblock {\em Journal of dairy science}, 87(1):59--66, 2004.

\bibitem{eggleston20062006}
HS~Eggleston, Leandro Buendia, Kyoko Miwa, Todd Ngara, and Kiyoto Tanabe.
\newblock 2006 ipcc guidelines for national greenhouse gas inventories.
\newblock 2006.

\bibitem{kexxtone13}
{European Medicine Agency}.
\newblock Kexxtone, 2013.

\bibitem{fouts2022enteric}
Julia~Q Fouts, Mallory~C Honan, Breanna~M Roque, Juan~M Tricarico, and Ermias
  Kebreab.
\newblock Enteric methane mitigation interventions.
\newblock {\em Translational Animal Science}, 6(2):txac041, 2022.

\bibitem{garnsworthy2019comparison}
Philip~C Garnsworthy, Gareth~F Difford, Matthew~J Bell, Ali~R Bayat, Pekka
  Huhtanen, Bj{\"o}rn Kuhla, Jan Lassen, Nico Peiren, Marcin Pszczola, Diana
  Sorg, et~al.
\newblock Comparison of methods to measure methane for use in genetic
  evaluation of dairy cattle.
\newblock {\em Animals}, 9(10):837, 2019.

\bibitem{geishauser1996comparison}
T~Geishauser and A~Gitzel.
\newblock A comparison of rumen fluid sampled by oro-ruminal probe versus rumen
  fistula.
\newblock {\em Small Ruminant Research}, 21(1):63--69, 1996.

\bibitem{giavarina2015understanding}
Davide Giavarina.
\newblock Understanding bland altman analysis.
\newblock {\em Biochemia medica}, 25(2):141--151, 2015.

\bibitem{gonzalez:2008nature}
Marta~C. Gonzalez, Cesar~A. Hidalgo, and Albert-Laszlo Barabasi.
\newblock Understanding individual human mobility patterns.
\newblock {\em Nature}, 453(7196):779--782, 06 2008.

\bibitem{goodrich1984influence}
RD~Goodrich, JE~Garrett, DR~Gast, MA~Kirick, DA~Larson, and JC~Meiske.
\newblock Influence of monensin on the performance of cattle.
\newblock {\em Journal of animal science}, 58(6):1484--1498, 1984.

\bibitem{hagey2022rumen}
Jill~V Hagey, Maia Laabs, Elizabeth~A Maga, and Edward~J DePeters.
\newblock Rumen sampling methods bias bacterial communities observed.
\newblock {\em PLoS One}, 17(5):e0258176, 2022.

\bibitem{hammond2015methane}
KJ~Hammond, DJ~Humphries, LA~Crompton, Colin Green, and CK~Reynolds.
\newblock Methane emissions from cattle: Estimates from short-term measurements
  using a greenfeed system compared with measurements obtained using
  respiration chambers or sulphur hexafluoride tracer.
\newblock {\em Animal Feed Science and Technology}, 203:41--52, 2015.

\bibitem{hammond2016greenfeed}
KJ~Hammond, GC~Waghorn, and RS~Hegarty.
\newblock The greenfeed system for measurement of enteric methane emission from
  cattle.
\newblock {\em Animal Production Science}, 56(3):181--189, 2016.

\bibitem{hedges1981distribution}
Larry~V Hedges.
\newblock Distribution theory for glass's estimator of effect size and related
  estimators.
\newblock {\em journal of Educational Statistics}, 6(2):107--128, 1981.

\bibitem{hess2004robust}
Melinda~R Hess and Jeffrey~D Kromrey.
\newblock Robust confidence intervals for effect sizes: A comparative study of
  cohen’sd and cliff’s delta under non-normality and heterogeneous
  variances.
\newblock In {\em annual meeting of the American Educational Research
  Association}, volume~1. Citeseer, 2004.

\bibitem{honan2021feed}
M~Honan, X~Feng, JM~Tricarico, and E~Kebreab.
\newblock Feed additives as a strategic approach to reduce enteric methane
  production in cattle: Modes of action, effectiveness and safety.
\newblock {\em Animal Production Science}, 2021.

\bibitem{hristov2015use}
Alexander~N Hristov, Joonpyo Oh, Fabio Giallongo, Tyler Frederick, Holley
  Weeks, Patrick~R Zimmerman, Michael~T Harper, Rada~A Hristova, R~Scott
  Zimmerman, and Antonio~F Branco.
\newblock The use of an automated system (greenfeed) to monitor enteric methane
  and carbon dioxide emissions from ruminant animals.
\newblock {\em JoVE (Journal of Visualized Experiments)}, (103):e52904, 2015.

\bibitem{huhtanen2019enteric}
P~Huhtanen, M~Ramin, and AN~Hristov.
\newblock Enteric methane emission can be reliably measured by the greenfeed
  monitoring unit.
\newblock {\em Livestock science}, 222:31--40, 2019.

\bibitem{huss2007currency}
Mikael Huss and Petter Holme.
\newblock Currency and commodity metabolites: their identification and relation
  to the modularity of metabolic networks.
\newblock {\em IET systems biology}, 1(5):280--285, 2007.

\bibitem{ito2000toward}
Takashi Ito, Kosuke Tashiro, Shigeru Muta, Ritsuko Ozawa, Tomoko Chiba, Mayumi
  Nishizawa, Kiyoshi Yamamoto, Satoru Kuhara, and Yoshiyuki Sakaki.
\newblock Toward a protein--protein interaction map of the budding yeast: a
  comprehensive system to examine two-hybrid interactions in all possible
  combinations between the yeast proteins.
\newblock {\em Proceedings of the National Academy of Sciences},
  97(3):1143--1147, 2000.

\bibitem{metha-intro3}
Elie Jami, Adi Israel, Assaf Kotser, and Itzhak Mizrahi.
\newblock Exploring the bovine rumen bacterial community from birth to
  adulthood.
\newblock {\em The ISME journal}, 7(6):1069--1079, 2013.

\bibitem{jeong2000large}
Hawoong Jeong, B{\'a}lint Tombor, R{\'e}ka Albert, Zoltan~N Oltvai, and A-L
  Barab{\'a}si.
\newblock The large-scale organization of metabolic networks.
\newblock {\em Nature}, 407(6804):651--654, 2000.

\bibitem{johannesson2022forest}
Carl-Fredrik Johannesson, Klaus Steenberg~Larsen, and Jenni Nord{\'e}n.
\newblock Forest soil and deadwood ch4 fluxes in response to climate change and
  forest management.
\newblock In {\em EGU General Assembly Conference Abstracts}, pages
  EGU22--7061, 2022.

\bibitem{johnson1995methane}
Kristen~A Johnson and Donald~E Johnson.
\newblock Methane emissions from cattle.
\newblock {\em Journal of animal science}, 73(8):2483--2492, 1995.

\bibitem{kebreab2006methane}
E~Kebreab, K~Clark, C~Wagner-Riddle, and J~France.
\newblock Methane and nitrous oxide emissions from canadian animal agriculture:
  A review.
\newblock {\em Canadian Journal of Animal Science}, 86(2):135--157, 2006.

\bibitem{kekana2021effects}
MR~Kekana, D~Luseba, and MC~Muyu.
\newblock Effects of garlic supplementation on in vitro nutrient digestibility,
  rumen fermentation, and gas production.
\newblock {\em South African Journal of Animal Science}, 51(2):271--279, 2021.

\bibitem{korben2021estimation}
Piotr Korben.
\newblock {\em Estimation of methane emissions and investigation of isotopic
  composition of methane from selected sources in Germany, Poland and Romania}.
\newblock PhD thesis, 2021.

\bibitem{layeghifard2017disentangling}
Mehdi Layeghifard, David~M Hwang, and David~S Guttman.
\newblock Disentangling interactions in the microbiome: a network perspective.
\newblock {\em Trends in microbiology}, 25(3):217--228, 2017.

\bibitem{CSS-Lazer-Science-2009}
David Lazer, Alex Pentland, Lada Adamic, Sinan Aral, Albert-Laszlo Barabasi,
  Devon Brewer, Nicholas Christakis, Noshir Contractor, James Fowler, Myron
  Gutmann, Tony Jebara, Gary King, Michael Macy, Deb Roy, and Marshall~Van
  Alstyne.
\newblock Social science: Computational social science.
\newblock {\em Science}, 323(5915):721--723, 2009.

\bibitem{ma2016effect}
Tao Ma, Dandan Chen, Yan Tu, Naifeng Zhang, Bingwen Si, Kaidong Deng, and Qiyu
  Diao.
\newblock Effect of supplementation of allicin on methanogenesis and ruminal
  microbial flora in dorper crossbred ewes.
\newblock {\em Journal of animal science and biotechnology}, 7(1):1--7, 2016.

\bibitem{marumo2023enteric}
Joyce~L Marumo, P~Andrew LaPierre, and Michael~E Van~Amburgh.
\newblock Enteric methane emissions prediction in dairy cattle and effects of
  monensin on methane emissions: A meta-analysis.
\newblock {\em Animals}, 13(8):1392, 2023.

\bibitem{mcdermitt2011new}
D~McDermitt, G~Burba, L~Xu, T~Anderson, A~Komissarov, B~Riensche,
  J~Schedlbauer, G~Starr, D~Zona, W~Oechel, et~al.
\newblock A new low-power, open-path instrument for measuring methane flux by
  eddy covariance.
\newblock {\em Applied Physics B}, 102:391--405, 2011.

\bibitem{melgar2021enteric}
A~Melgar, CFA Lage, K~Nedelkov, SE~R{\"a}is{\"a}nen, H~Stefenoni, ME~Fetter,
  X~Chen, J~Oh, S~Duval, M~Kindermann, et~al.
\newblock Enteric methane emission, milk production, and composition of dairy
  cows fed 3-nitrooxypropanol.
\newblock {\em Journal of dairy science}, 104(1):357--366, 2021.

\bibitem{mercadante2015relationship}
Maria Eug{\^e}nia~Zerlotti Mercadante, Ana Paula de~Melo Caliman,
  Roberta~Carrilho Canesin, Sarah Figueiredo~Martins Bonilha, Alexandre Berndt,
  Rosa Toyoko~Shiraishi Frighetto, Elaine Magnani, and Renata~Helena Branco.
\newblock Relationship between residual feed intake and enteric methane
  emission in nellore cattle.
\newblock {\em Revista Brasileira de Zootecnia}, 44:255--262, 2015.

\bibitem{metha-intro2}
Itzhak Mizrahi, R~John Wallace, and Sarah Mora{\"\i}s.
\newblock The rumen microbiome: balancing food security and environmental
  impacts.
\newblock {\em Nature Reviews Microbiology}, 19(9):553--566, 2021.

\bibitem{metha-intro1}
Sarah Moraïs and Itzhak Mizrahi.
\newblock {Islands in the stream: from individual to communal fiber degradation
  in the rumen ecosystem}.
\newblock {\em FEMS Microbiology Reviews}, 43(4):362--379, 04 2019.

\bibitem{muizelaar2020rumen}
Wouter Muizelaar, Paolo Bani, Bj{\"o}rn Kuhla, Mogens Larsen, Ilma Tapio, David
  Y{\'a}{\~n}ez-Ruiz, and Sanne van Gastelen.
\newblock Rumen fluid sampling via oral stomach tubing method.
\newblock 2020.

\bibitem{nachar2008mann}
Nadim Nachar et~al.
\newblock The mann-whitney u: A test for assessing whether two independent
  samples come from the same distribution.
\newblock {\em Tutorials in quantitative Methods for Psychology}, 4(1):13--20,
  2008.

\bibitem{newman2001clustering}
Mark~EJ Newman.
\newblock Clustering and preferential attachment in growing networks.
\newblock {\em Physical review E}, 64(2):025102, 2001.

\bibitem{nocek1997bovine}
James~E Nocek.
\newblock Bovine acidosis: Implications on laminitis.
\newblock {\em Journal of dairy science}, 80(5):1005--1028, 1997.

\bibitem{odai2010estimation}
Masaharu Odai, Itoko Nonaka, Witthaya Sumamal, Rumphrai Narmsilee, Pimpaporn
  Pholsen, Taweesak Chuenprecha, Arun Phromloungsri, Nuttanart Khotprom,
  Ittiphon Phaowphaisal, Somchit Indramanee, et~al.
\newblock Estimation of methane production by lactating and dry crossbred
  holstein cows in thailand.
\newblock {\em Japan Agricultural Research Quarterly: JARQ}, 44(4):429--434,
  2010.

\bibitem{socialphysicsandcybercrime2018}
A.~Pentland and Y.~Altshuler.
\newblock {\em Social Physics and Cybercrime}, chapter New Solutions for
  Cybersecurity, pages 351--364.
\newblock MIT Press, 2018.

\bibitem{phibro2021}
{Phibro}.
\newblock {Phibro data 2021}.
\newblock Available on request, 2021.

\bibitem{place2011construction}
Sara~E Place, Yuee Pan, Yongjing Zhao, and Frank~M Mitloehner.
\newblock Construction and operation of a ventilated hood system for measuring
  greenhouse gas and volatile organic compound emissions from cattle.
\newblock {\em Animals}, 1(4):433--446, 2011.

\bibitem{qi2012study}
Zhu Qi, Yin Yi, Wang Qiaohua, Wang Zhihao, and Li~Zhe.
\newblock Study on the online dissolved gas analysis monitor based on the
  photoacoustic spectroscopy.
\newblock In {\em 2012 IEEE International Conference on Condition Monitoring
  and Diagnosis}, pages 433--436. IEEE, 2012.

\bibitem{metha-intro7}
Yuliaxis Ramayo-Caldas, Laura Zingaretti, Milka Popova, Jordi Estell{\'e},
  Aurelien Bernard, Nicolas Pons, Pau Bellot, N{\'u}ria Mach, Andrea Rau, Hugo
  Roume, et~al.
\newblock Identification of rumen microbial biomarkers linked to methane
  emission in holstein dairy cows.
\newblock {\em Journal of Animal Breeding and Genetics}, 137(1):49--59, 2020.

\bibitem{ramos2014use}
Eva Ramos-Morales, Ana Arco-P{\'e}rez, A~Ignacio Mart{\'\i}n-Garc{\'\i}a,
  David~Rafael Y{\'a}{\~n}ez-Ruiz, Pilar Frutos, and Gonzalo Herv{\'a}s.
\newblock Use of stomach tubing as an alternative to rumen cannulation to study
  ruminal fermentation and microbiota in sheep and goats.
\newblock {\em Animal Feed Science and Technology}, 198:57--66, 2014.

\bibitem{rey2019comparison}
Jagoba Rey, Raquel Atxaerandio, Roberto Ruiz, Eva Ugarte, Oscar
  Gonz{\'a}lez-Recio, Aser Garcia-Rodriguez, and Idoia Goiri.
\newblock Comparison between non-invasive methane measurement techniques in
  cattle.
\newblock {\em Animals}, 9(8):563, 2019.

\bibitem{ribeiro2020comparison}
Angelita~Alecchandra Ribeiro, Lerner~Ar{\'e}valo Pinedo, Luciane
  da~Cunha~Codognoto, Jucilene Cavali, Marlos~Oliveira Porto, Betina
  Raquel~Cunha dos Santos, Palloma Vit{\'o}ria~Carlos de~Oliveira, Dayana~Souza
  Amorim, Salvador~Gonz{\'a}lez Chac{\'o}n, and Salenilda~Soares Firmino.
\newblock Comparison of methods to measure enteric methane emissions from
  ruminants: an integrative review.
\newblock {\em Research, Society and Development},
  9(11):e8259118143--e8259118143, 2020.

\bibitem{rosenthal1994parametric}
Robert Rosenthal, Harris Cooper, Larry Hedges, et~al.
\newblock Parametric measures of effect size.
\newblock {\em The handbook of research synthesis}, 621(2):231--244, 1994.

\bibitem{roulston2018emissions}
Christopher~Thomas Roulston.
\newblock Emissions of fine particulate matter from tropical peat fires.
\newblock 2018.

\bibitem{VM41}
Mootral SA.
\newblock Vm0041 methodology for the reduction of enteric methane emissions
  from ruminants through the use of feed ingredients.
\newblock {\em Verra}, 2021.

\bibitem{metha-intro5}
Goor Sasson, Sheerli~Kruger Ben-Shabat, Eyal Seroussi, Adi Doron-Faigenboim,
  Naama Shterzer, Shamay Yaacoby, Margret E.~Berg Miller, Bryan~A. White, Eran
  Halperin, and Itzhak Mizrahi.
\newblock Heritable bovine rumen bacteria are phylogenetically related and
  correlated with the cow’s capacity to harvest energy from its feed.
\newblock {\em mBio}, 8(4):10.1128/mbio.00703--17, 2017.

\bibitem{metha-intro4}
Yoav Shaani, Tamar Zehavi, Stav Eyal, Joshuah Miron, and Itzhak Mizrahi.
\newblock Microbiome niche modification drives diurnal rumen community
  assembly, overpowering individual variability and diet effects.
\newblock {\em The ISME journal}, 12(10):2446--2457, 2018.

\bibitem{shah2023characterising}
Adil Shah, Olivier Laurent, Luc Lienhardt, Gr{\'e}goire Broquet, Rodrigo
  Rivera~Martinez, Elisa Allegrini, and Philippe Ciais.
\newblock Characterising the methane gas and environmental response of the
  figaro taguchi gas sensor (tgs) 2611-e00.
\newblock {\em Atmospheric Measurement Techniques}, 16(13):3391--3419, 2023.

\bibitem{shen2012insertion}
JS~Shen, Z~Chai, LJ~Song, JX~Liu, and YM~Wu.
\newblock Insertion depth of oral stomach tubes may affect the fermentation
  parameters of ruminal fluid collected in dairy cows.
\newblock {\em Journal of dairy science}, 95(10):5978--5984, 2012.

\bibitem{shmueli2015ride}
Erez Shmueli, Itzik Mazeh, Laura Radaelli, Alex~Sandy Pentland, and Yaniv
  Altshuler.
\newblock Ride sharing: A network perspective.
\newblock In {\em Social Computing, Behavioral-Cultural Modeling, and
  Prediction}, pages 434--439. Springer, 2015.

\bibitem{somin2020network}
Shahar Somin, Yaniv Altshuler, Goren Gordon, Alex Pentland, and Erez Shmueli.
\newblock network dynamics of a financial ecosystem.
\newblock {\em Scientific reports}, 10(1):1--10, 2020.

\bibitem{somin2022beyond}
Shahar Somin, Yaniv Altshuler, Alex ‘Sandy’Pentland, and Erez Shmueli.
\newblock Beyond preferential attachment: falling of stars and survival of
  superstars.
\newblock {\em Royal Society Open Science}, 9(8):220899, 2022.

\bibitem{somin2022remaining}
Shahar Somin, Yaniv Altshuler, Erez Shmueli, et~al.
\newblock Remaining popular: power-law regularities in network dynamics.
\newblock {\em EPJ Data Science}, 11(1):61, 2022.

\bibitem{storm2012methods}
Ida~MLD Storm, Anne Louise~F Hellwing, Nicolaj~I Nielsen, and J{\o}rgen Madsen.
\newblock Methods for measuring and estimating methane emission from ruminants.
\newblock {\em Animals}, 2(2):160--183, 2012.

\bibitem{takano2021spatial}
Tsugumi Takano and Masahito Ueyama.
\newblock Spatial variations in daytime methane and carbon dioxide emissions in
  two urban landscapes, sakai, japan.
\newblock {\em Urban Climate}, 36:100798, 2021.

\bibitem{tao2021understanding}
Ziletao Tao, Can Chen, Qi~Yang, Zhenyu Zhong, Yong Wan, Shengjie Chen, Fubing
  Yao, Zhoujie Pi, Xiaoming Li, and Dongbo Wang.
\newblock Understanding the impact of allicin for organic matter release and
  microorganism community in anaerobic co-digestion of food waste and waste
  activated sludge.
\newblock {\em Science of The Total Environment}, 776:145598, 2021.

\bibitem{van1977effect}
C~J\_ Van~Nevel and D\_I Demeyer.
\newblock Effect of monensin on rumen metabolism in vitro.
\newblock {\em Applied and Environmental Microbiology}, 34(3):251--257, 1977.

\bibitem{vranken2019reduction}
Hilde Vranken, Maria Suenkel, Paul~R Hargreaves, Lynette Chew, and Edward
  Towers.
\newblock Reduction of enteric methane emission in a commercial dairy farm by a
  novel feed supplement.
\newblock {\em Open Journal of Animal Sciences}, 9:286--296, 2019.

\bibitem{metha-intro8}
R~John Wallace, John~A Rooke, Nest McKain, Carol-Anne Duthie, Jimmy~J Hyslop,
  David~W Ross, Anthony Waterhouse, Mick Watson, and Rainer Roehe.
\newblock The rumen microbial metagenome associated with high methane
  production in cattle.
\newblock {\em BMC genomics}, 16(1):1--14, 2015.

\bibitem{wang2016sampling}
M~Wang, R~Wang, PH~Janssen, XM~Zhang, XZ~Sun, D~Pacheco, and ZL~Tan.
\newblock Sampling procedure for the measurement of dissolved hydrogen and
  volatile fatty acids in the rumen of dairy cows.
\newblock {\em Journal of Animal Science}, 94(3):1159--1169, 2016.

\bibitem{weller2023genetic}
Joel~Ira Weller, Ephraim Ezra, Eyal Seroussi, and Moran Gershoni.
\newblock Genetic and genomic analysis of cow mortality in the israeli holstein
  population.
\newblock {\em Genes}, 14(3):588, 2023.

\bibitem{wolfram2023chatgpt}
Stephen Wolfram.
\newblock {\em What Is ChatGPT Doing... and Why Does It Work?}
\newblock Stephen Wolfram, 2023.

\bibitem{xu2018chamber}
Liukang Xu and Richard Vath.
\newblock Chamber-based soil methane flux measurement system.
\newblock In {\em AGU Fall Meeting Abstracts}, volume 2018, pages B41H--2824,
  2018.

\bibitem{zwang2018detecting}
Morit Zwang, Shahar Somin, Alex~'Sandy' Pentland, and Yaniv Altshuler.
\newblock Detecting bot activity in the ethereum blockchain network, 2018.

\end{thebibliography}

\end{document}